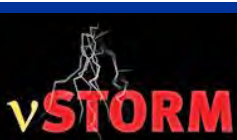

# nuSTORM Conventional Facilities

**PDR**

*Project Definition Report for the conventional facilities to house the nuSTORM Facilities.*

**ENERGY**
Office of Science

Fermi Research Alliance LLC

Fermilab

This Project Design Report (PDR) is intended to be a self-consistent basis for the development of a planning level cost estimate for the conventional facilities to support the programmatic requirements. This report has not answered every technical design question and as such, the current level of contingency is believed to be consistent with the degree of technical confidence in the design at this stage. It is recognized that some basic construction concerns will be reviewed and optimized during the remaining stages of the project.

Authors of this document:

T. Lackowski, FESS/Engineering
S. Dixon, FESS/Engineering
R. Jedziniak, LG Associates
M. Blewitt, Holabird & Root
L. Fink, Holabird & Root

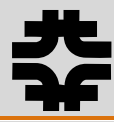

The objective of this Project Definition Report (PDR) is to document one (1) possible solution for the conventional facilities required to house the nuSTORM experiment on the Fermilab site. It is recognized that this effort is done in support of obtaining DOE approval for Critical Decision Zero and that the programmatic requirements are still considered preliminary at this stage of the project.

The nuSTORM Conventional Facilities is anticipated to consist of six (6) functional areas consisting of the Primary Beamline, Target Station, Transport Line/Muon Decay Ring, Near Detector, Far Detector and the Site Work.

The nuSTORM Conventional Facilities will be located on the Fermilab site, south of the existing Main Injector and west of Kautz Road.

**Project Costs**
The Total Project Cost (TPC) for the Conventional Facilities portion of this project is estimated at $148,350,000 for the construction and associated design services. The Other Project Costs (OPC) are estimated at $2,526,000.

The TPC includes Construction, EDIA (Engineering, Design, Inspection and Administration), Contingency and Indirect Costs. The TPC has been estimated in FY13 dollars. The Indirect Costs associated with this project, are based on current laboratory rates, dated October 2012.

**Schedule**
| | |
|---|---|
| CD-0 Approval | Month 0 |
| CD-1 Approval | Month 12 |
| CD-2 Approval | Month 24 |
| CD-3 Approval | Month 36 |
| Start Conventional Facilities Construction | Month 39 |
| Complete Conventional Facilities Construction | Month 57 |

The schedule is based on technically driven parameters and does not incorporate lags for DOE approvals or funding restrictions.

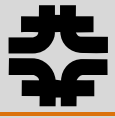

The nuSTORM Conventional Facilities design described in this Project Definition Report is based in information collected during discussions, meetings and workshops with the design leads of the functional areas.    Listed below are the functional areas and the design leads for each area:

| | |
|---|---|
| Primary Beamline | Michael Geelhoed |
| Target Station | Kris Anderson |
| Transport Line/Muon Decay Ring | Ao Liu |
| Near Detector | Alan Bross |
| Far Detector | Herman Cease |
| Site Work | FESS/Engineering |

The following assumptions were incorporated into the development of this Project Definition Report.

- It is assumed that DOE Order 413.3A, "*Program and Project Management for the Acquisition of Capital Assets*," will be followed during the execution of the subsequent phases of this project.
- The existing Main Injector accelerator will be reconfigured for use as an extraction beamline to support the nuSTORM facilities.  This reconfiguration will result in a beamline that is extracted from the existing MI-40 absorber eastward to the nuSTORM facilities.  It is assumed that the reconfiguration of the beamline and related devices is not part of the nuSTORM Conventional Facilities scope of work.
- While it is recognized that significant work associated with reconfiguring the cryogenic systems of the TeVatron is required to accommodate the nuSTORM Experiment, those costs are not included in the nuSTORM Conventional Facilities scope of work.
- For the purposes of this Project Definition Report, a shielding depth of 21 feet earth equivalent has been used for areas that contain extracted beam and twelve (12) feet for the Transport Line Enclosure, Muon Decay Ring Enclosure and those portions of the project associated with the secondary beam.
- It is assumed that the nuSTORM facilities will not be normally occupied.
- The guiding principals of high performance building design will be incorporated into the design of the nuSTORM Conventional Facilities. However, based on the type and use of the facilities, is not intended that the nuSTORM Conventional Facilities will become a LEED certified building.  The project processes and each project element will be evaluated during design to reduce their impact on natural resources without sacrificing program objectives.  The project design will incorporate maintainability, aesthetics,

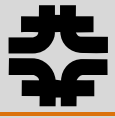

**Section II**

environmental justice and program requirements to deliver a well-balanced project.

The drawings contained in the Appendix B contain detailed descriptions of additional requirements for the nuSTORM Conventional Facilities.

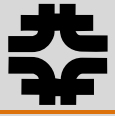

The nuSTORM Conventional Facilities will be located in an area south of the existing Main Injector accelerator and west of Kautz Road on the Fermilab site. In general terms, a proton beam will be extracted from the existing Main Injector at the MI-40 absorber, directed east towards a new below grade target station, muon decay ring. This extracted beam will be directed at a Near Detector east of the muon decay ring and towards the Far Detector located at the existing DZero Assembly Hall. Figure 1, below, shows a site photo depicting the nuSTORM Conventional Facilities.

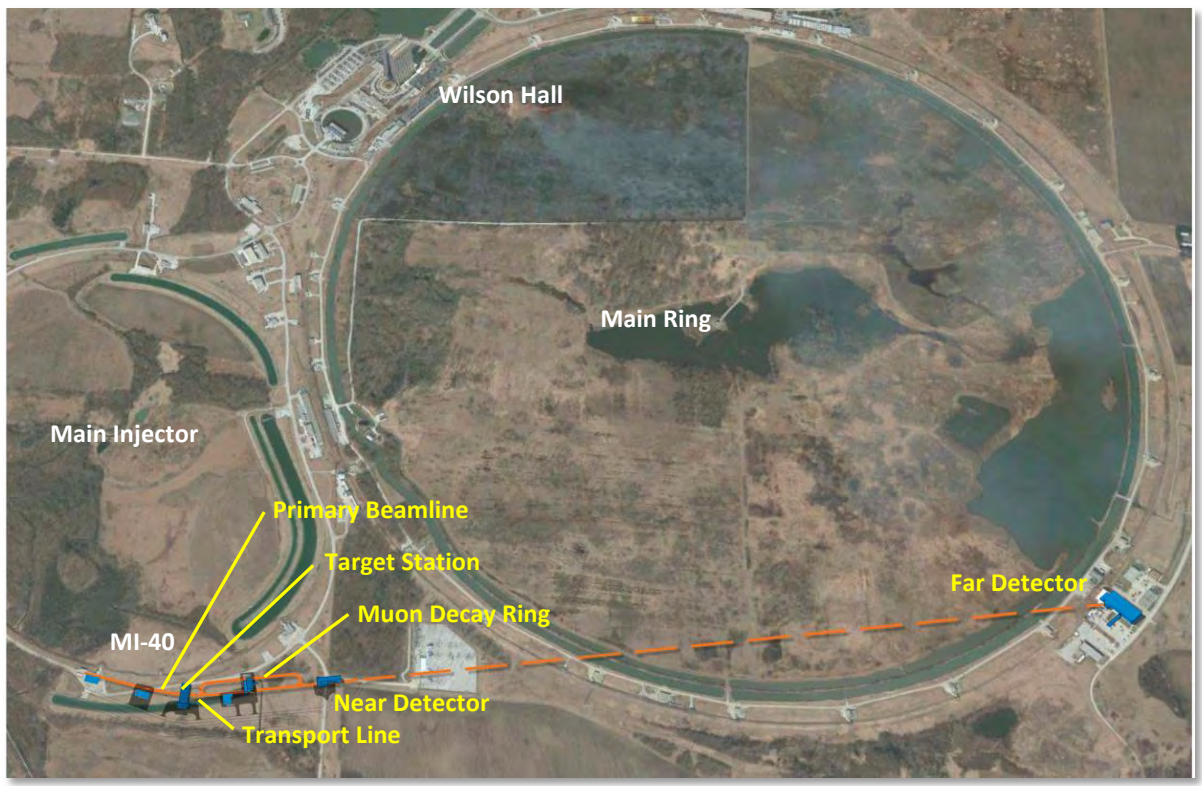

*Figure 1 – View Looking North of Fermilab with the nuSTORM Facilities*

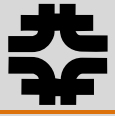

## FUNCTIONAL AREAS

The nuSTORM Conventional Facilities are envisioned to include seven (7) functional areas.  This is intended to provide a logical and constructible sequence to reduce the construction period to a minimum.   Further design iterations will be required to determine how to optimize the construction packages based on programmatic and funding driven limitations.

## Area 1 - Primary Beamline Enclosure (WBS 1.0)

The first functional area is the Primary Beamline Enclosure, consisting of the work involved with construction of a below-grade, concrete enclosure for the beamline components that will be required to transport the proton beam from the existing Main Injector enclosure into the Target Station.  Figure 2, below, shows the plan view of the beamline enclosure with the nuSTORM beamline indicated in orange.

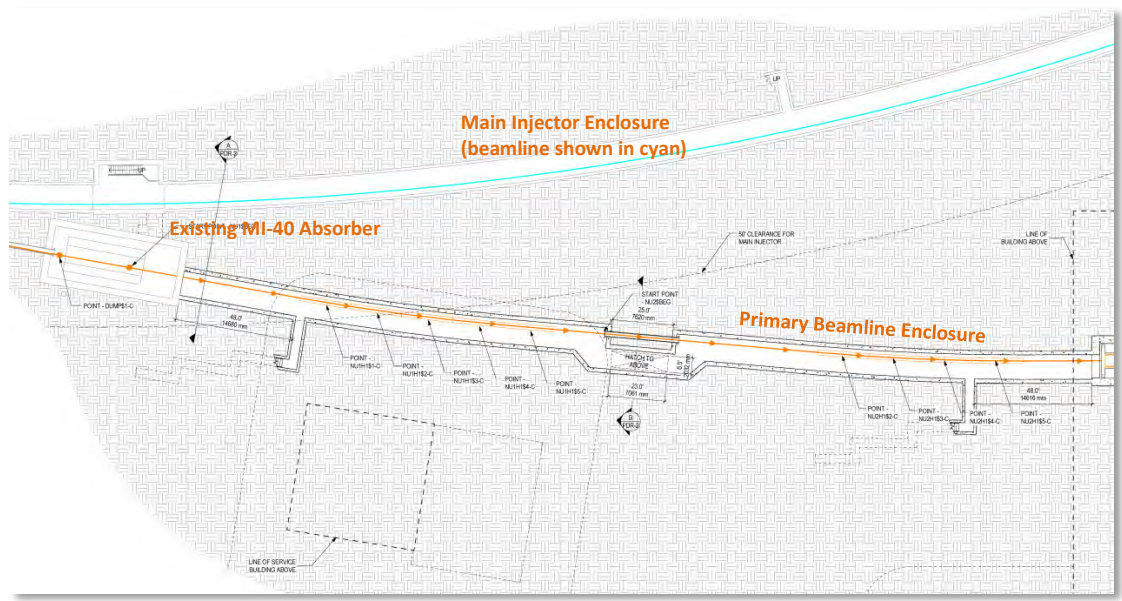

*Figure 2 – Plan View of Primary Beamline Enclosure*

At the upper portion of the figure is the existing Main Injector with the Main Injector beamline indicated in cyan.  The nuSTORM Primary Beamline Enclosure (PBE) is shown at the lower right of the existing MI-40 Absorber in orange.

The nuSTORM PBE will begin at the existing east wall of the MI-40 Absorber and will house components similar to the conventional beamline components in the Main Injector.  The PBE will include two (2) code compliance exit stairways provided at each end of the enclosure to ensure proper exiting.  These exit stairs will be similar in style and construction to stairs installed in the Main Injector Project.

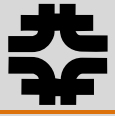

In addition, the PBE will include a shielded drop hatch to accommodate installation of beamline components located at the approximate mid-point of the enclosure. This hatch is intended to remain open during installation of the beamline components.  Once assembly is complete, the hatch will be filled with precast concrete shielding blocks to a depth of 21 feet in order to provide the required shielding for a primary beamline.

Portions of the PBE will be constructed during a Main Injector shutdown due to shielding requirements of the operating accelerator.

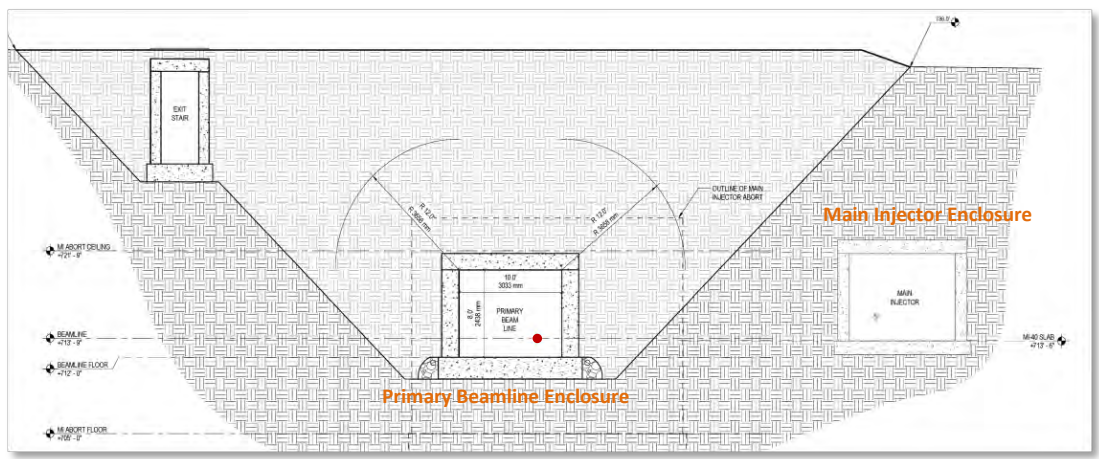

*Figure 3 –Typical Section of Primary Beamline Enclosure*

The PBE is based on the dimensions of a Main Injector style enclosure.  These enclosures provide a ten (10) foot interior width and an eight (8) foot interior height. Figure 3, above, depicts a typical enclosure cross section.  The beamline is shown as a red dot and is expected to be located three feet away from the north wall of the enclosure.  The height of the beamline components will vary with the position of the beam.

Utilities and access to the PBE will be provided through a new Service Building located at grade level approximately halfway along the length of the PBE.  This new 50'x50' service building will be based on a typical Main Injector style service building and associated electrical substation.

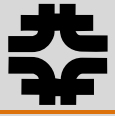

**Section III**

## Area 2 – Target Station (WBS 2.0)

The second functional area is the nuSTORM Target Station consisting of below-grade, cast-in-place concrete enclosure for the beamline and detector components and an above grade Target Support Building.

The nuSTORM Target Station below grade detector enclosure consists of the spaces required to house the nuSTORM target and beamline components, accommodate installation and safe operation of the nuSTORM target. Technical requirements and access information for the nuSTORM Target Station are based on the nuSTORM Target Hall Preliminary Layout drawing dated 11-MAR-2013 and subsequent discussions.

Figure 4, below, is a plan view of the below grade portion of the nuSTORM Target Station at the same elevation of the Primary Beamline Enclosure. The target station will be an isolated concrete enclosure that provides a shielded environment for the target components. The shielding requirements include 3 feet of concrete on the floor and walls, a 6 inch air gap and 4 feet of steel shielding. The shielding above the target station provides space for eight feet of steel shielding and 3 feet of concrete shielding. The conventional construction work provides the concrete for the floor and walls of the target chase, but the steel shielding for the walls and floor as well as the shielding above the target is provided by the target group.

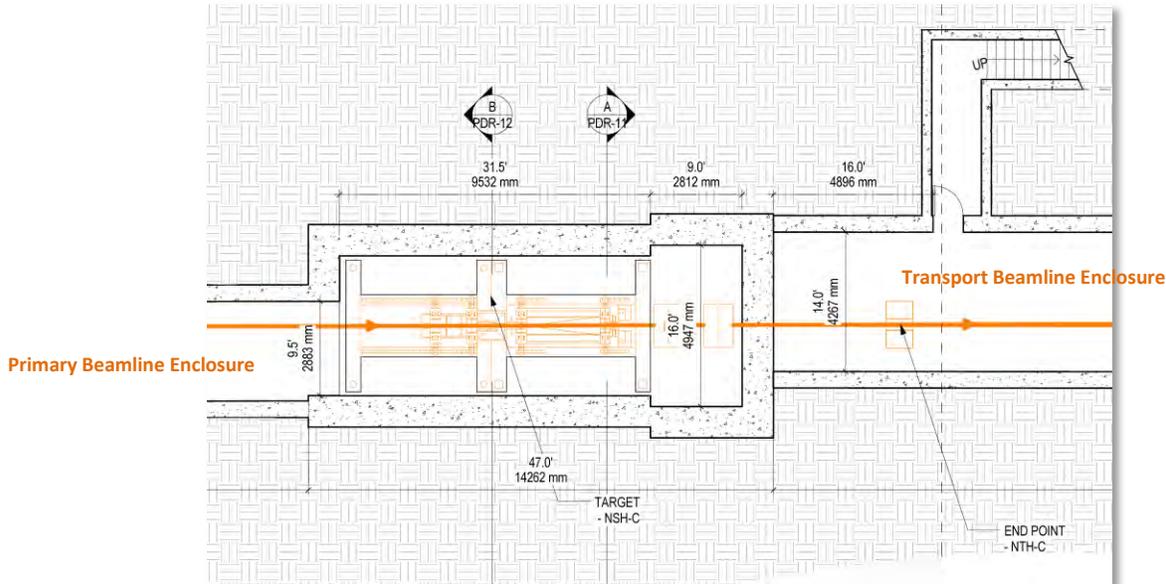

*Figure 4 – Plan View of the nuSTORM Target Chase*

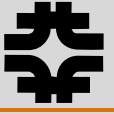

Figure 5, below, provides a plan view of the lower level of the nuSTORM Target Station. This level, approximately 14 feet below existing grade, provides space for the support functions associated with the nuSTORM target including a work cell for target components, shielded storage of activated components, staging areas for shield blocks as well as the mechanical and electrical systems required to operate the nuSTORM target.

**Section III**

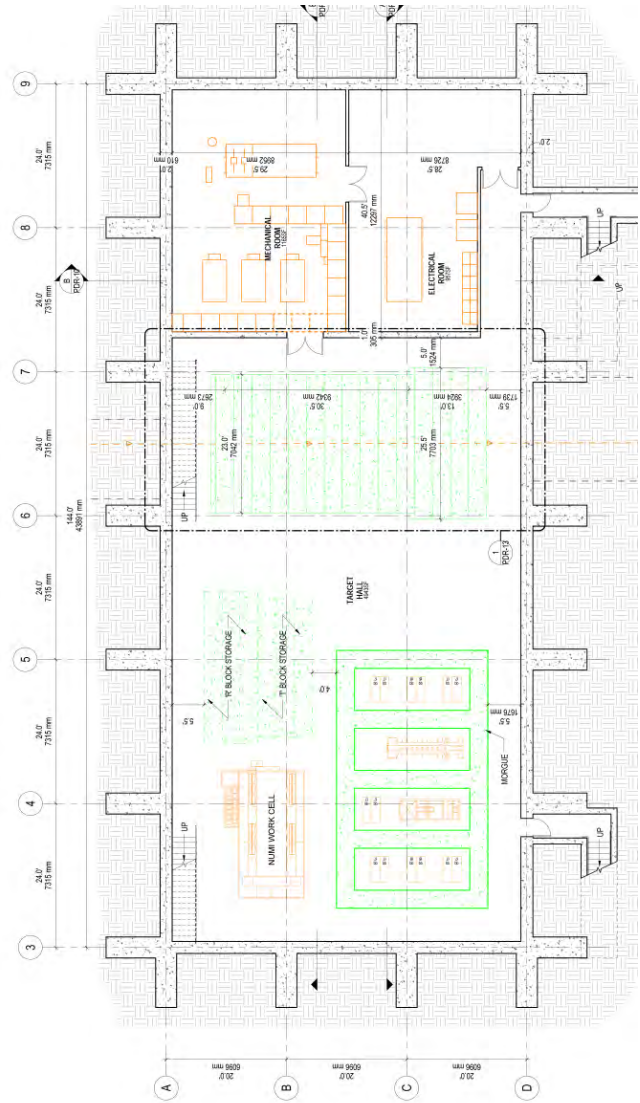

*Figure 5 – Plan View of the Below Grade portion of the nuSTORM Target Station*

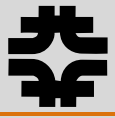

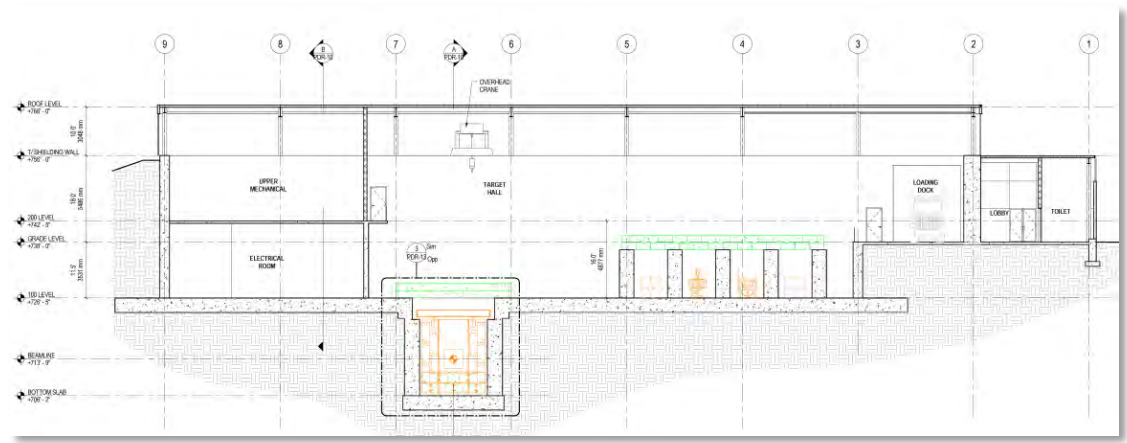

*Figure 6 – Section View of the nuSTORM Target Station*

Figure 6, above, depicts a section view of the nuSTORM Target Station showing the relationship of the target chase to the below grade areas and the above grade target service building. The above grade portion of the nuSTORM Target Station consists of a service building that includes a loading dock, toilet facilities, mechanical and electrical rooms to support the operation of the building. The above grade building has been designed to accommodate the movement and handling of activated target components. As a result, the walls of the above grade building provide 3 feet of shielding to a height of 10 feet above grade.

The below grade portion of the nuSTORM Target Station will be constructed of cast-in-place concrete. The above grade portion will be constructed of cast-in-place concrete walls and a structural steel frame to support the roof and the overhead bridge crane. The building will be equipped with a 50 ton capacity overhead bridge crane, similar in style and construction to cranes installed with the Main Injector project.

The arrangement of the spaces for the below grade portion of the nuSTORM Target Station include the provision for code compliant egress. This is accomplished with two (2) exit stairs and connecting smoke-proof corridors positioned to allow for two (2) means of egress for the below grade spaces.

The nuSTORM Target Service building contains the Mechanical Room, Electrical Room, toilets and janitor's closet to support the operation of the facility. These spaces have been sized based on experience with the Main Injector and NuMI projects and have not been optimized based on specific equipment or processes.

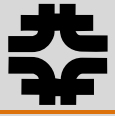

The mechanical systems for the nuSTORM Target Station include the following:

- Target Chase system to provide cooling for the target;
- Building system to condition the space and provide a pressure gradient towards the Target Chase;
- Service Building system to condition the support spaces;
- A tritium recovery system similar to the existing system used at NuMI to condense the moisture in the Target Chase, transport it to a boiler in the above grade building;
- A Target Chase underdrain system to intercept groundwater at the Target Chase.  This system will include a geo-membrane system capable of isolating the Target Chase from the surrounding area with appropriate witness zones to verify the operation of the system;
- An interior water collection within the Target Chase to collect water from the interior of the chase.  This water will be directed to a dedicated sump basin for testing.  Contaminated water will be collected and disposed of properly utilizing standard Laboratory procedures;
- A building underdrain system to intercept ground water and pump it to the surface.  This system will include redundant sump basins with duplex pumps.

**Area 3 – Transport Line Enclosure (WBS 3.0)**

The third functional area is the consists of the below grade beamline enclosure to house the Transport Beamline components required to transport the secondary beam from the Target Station to the Muon Decay Ring and the enclosure to house the Primary Beam Absorber.

The Transport Beamline Enclosure will begin at the east wall of the Target Station and will house both conventional and cryogenic beamline components.  The Transport Beamline Enclosure will include code compliance exits that will be similar in style and construction to stairs installed in the Main Injector.

The Transport Beamline Enclosure is based on the dimensions of a Main Injector style concrete enclosure.  These enclosures provide a ten (10) foot interior width and an eight (8) foot interior height.

Figure 7, below, depicts a plan view of the Transport Beamline Enclosure with the Target Station shown on the left side, the Muon Decay Ring Enclosure shown at the upper right corner and the Beam Absorber Enclosure at the lower right corner.

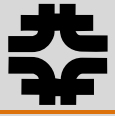

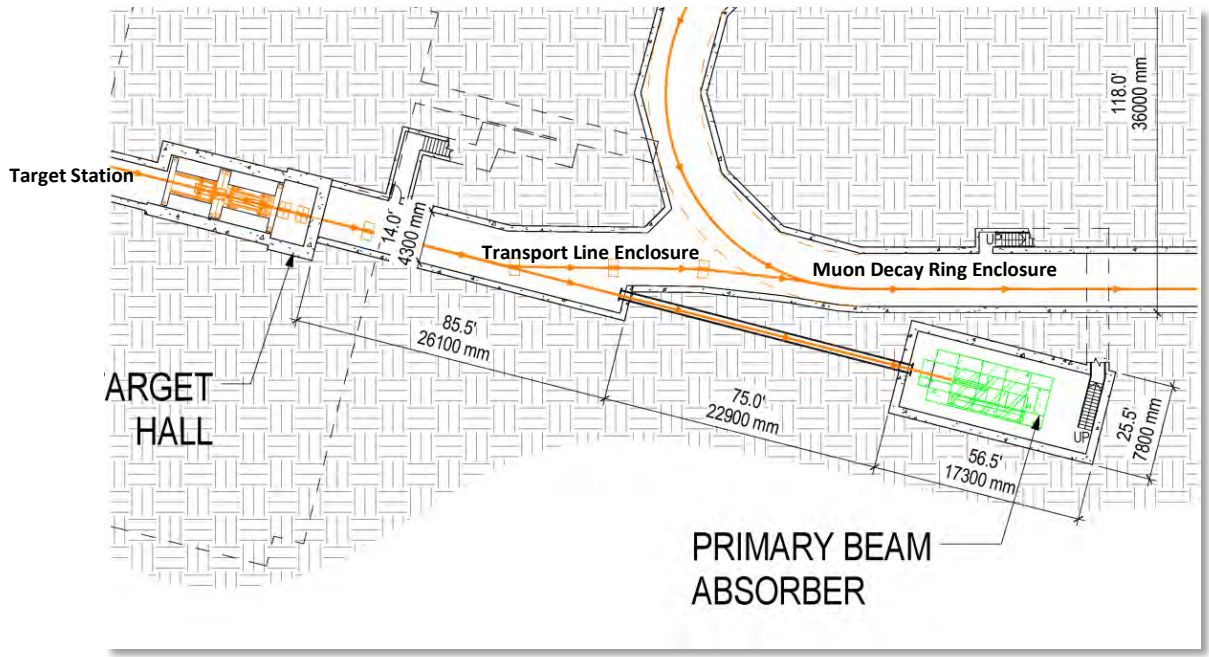

*Figure 7 – Plan View of Transport Line Enclosure*

The Primary Beam Absorber Enclosure has been designed to accommodate a steel and concrete beam absorber similar in size, type and construction to the existing MI-40 absorber. For the purposes of this report, the conventional construction is assumed to provide the cast-in-place concrete enclosure, code compliant exiting and the buried beam pipe that connects the Beam Absorber Enclosure to the Transport Beamline Enclosure. The shielding steel is assumed to be provided by the technical group.

Utilities and access for the Transport Beamline Enclosure will be provided through the nuSTORM Target Service Building located west of the Transport Beamline Enclosure

### Area 4 – Muon Decay Ring Enclosure (WBS 4.0)

The fourth functional area consists of the enclosures, buildings and infrastructure to support the installation and operation of the Muon Decay Ring beamline components including the following:

- The nuSTORM Muon Decay Ring (MDR) Enclosure consisting of a below grade beamline enclosure to house the beamline components for the muon decay ring;
- The Pion Absorber enclosure;

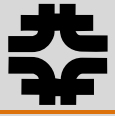

- The above grade Muon Decay Ring Service Building;
- The above grade Cryogenic Building.

Figure 8, below, depicts the below grade portion of this functional area with the existing Main Injector Enclosure shown at the upper left corner of the figure, the nuSTORM Target Station at the left hand side, and the Muon Decay Ring Enclosure shown at the center.

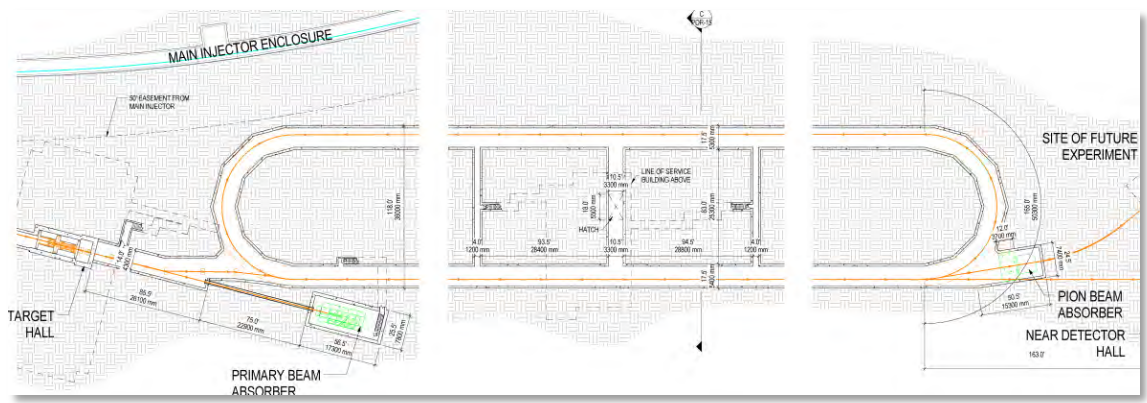

*Figure 8 – Plan View of Muon Decay Ring Enclosure*

The Muon Decay Ring Enclosure consists of a concrete enclosure with a 13 foot height and a 14 foot width. The enclosure is assumed to be outfitted in a similar manner as the typical Main Injector enclosure. The Muon Decay Ring Enclosure will be located approximately 27 feet below existing grade and provide more than the minimum of 12 feet of earth shielding. Figure 9, below depicts a cross section through the Muon Decay Ring Enclosure which shows the relative depth in relation to existing grade.

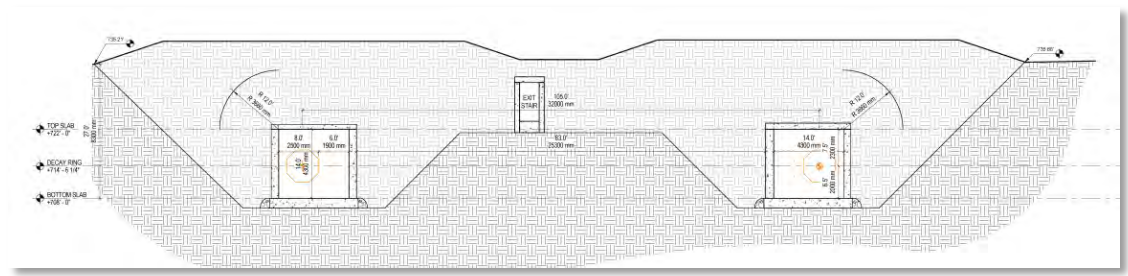

*Figure 9 – Section View of Muon Decay Ring Enclosure*

The Muon Decay Ring Enclosure will include code compliance exits similar in style and construction to exit stairs installed in the Main Injector.

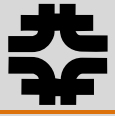

A

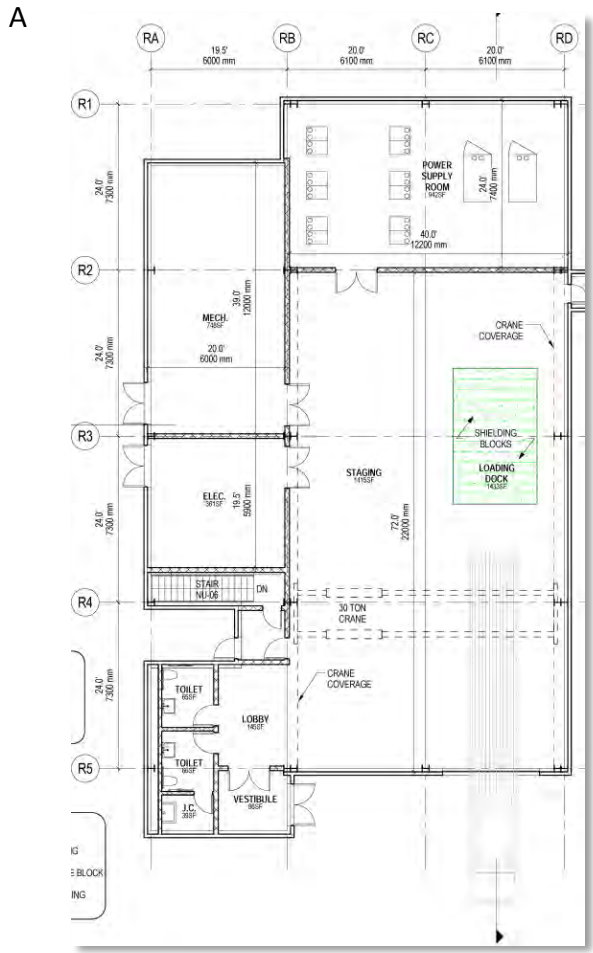

*Figure 10 – Plan View of Muon Decay Ring Service Building*

21 foot long by 12 foot wide shielded hatch located at the approximate center point of the Muon Decay Ring Enclosure will accommodate the vertical transport of beamline components.

The Pion Absorber Enclosure has been designed to accommodate a steel and concrete beam absorber that is 14 feet wide, by 14 feet high by 14 feet long. For the purposes of this report, the conventional construction is assumed to provide the concrete enclosure. The shielding steel and precast concrete shielding blocks are assumed to be provided by the technical group.

Utilities and access to the Muon Decay Ring Enclosure will be provided through the Muon Decay Ring Service Building located at grade at the approximate midpoint of the below grade Muon Decay Ring Enclosure.

The nuSTORM Muon Decay Ring Service Building, shown in figure 10 above, will house the functions to support the installation and operation of the Muon Decay Ring, including a loading dock, toilet facilities, janitor's closet and Mechanical Room and Electrical Room. These spaces have been sized based on experience with the Main Injector and NuMI projects and have not been optimized based on specific equipment or processes.

The high bay of the nuSTORM Muon Decay Ring Service Building will provide space for staging and installing the nuSTORM Muon Decay Ring. The high bay includes a shield block filled hatch that allow access to the below grade Muon Decay Ring Enclosure. This hatch is intended to remain open during installation of the beamline components. Once assembly is complete, the hatch will be filled with precast concrete shielding blocks to a depth of twelve (12) feet in order to provide the required shielding.

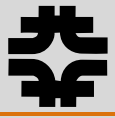

The high bay will be provided with a thirty (30) ton capacity overhead bridge crane that will be used for component staging and installations as well as shield block handling. The high bay contains space for the temporary staging of portions of the shield blocks should removal be required for maintenance and/or repair of the technical components.

The Structural Requirements for this functional area include:
- The below grade portions of the facility and building foundations will be constructed of cast-in-place concrete;
- The nuSTORM Muon Decay Ring Service Building will be metal sided, structural steel frame on a cast-in-place concrete foundation. The building will contain a 30 ton capacity overhead bridge crane to accommodate the unloading, staging and vertical transport of the beamline components;

The Mechanical Requirements for this function area include the following:
- The Mechanical Room will house the mechanical equipment that supports the assembly and operation of the nuSTORM Muon Decay Ring Service Building as well as the connection to the below grade Muon Decay Ring Enclosure;
- The HVAC systems for the nuSTORM Muon Decay Ring Service Building will conform to ASHRAE 90.1, ASHRAE 62 and applicable NFPA requirements and applicable sections of the Fermilab Engineering Standards Manual;
- Mechanical systems and Metasys controls will be further investigated during subsequent phases in accordance with ASHRAE 90.1 and Federal Life Cycle costing analysis;
- Heating, Ventilation and Air Conditioning Design Parameters:
  - Temperature: 68 degrees Fahrenheit to 78 degrees Fahrenheit
  - Humidity: 50% Maximum Relative Humidity, no minimum
- All plumbing work to be installed in accordance with Illinois Plumbing Code and Standard Specifications for Water & Sewer Main Construction in Illinois;
- The nuSTORM Muon Decay Ring Service Building will be conditioned with a roof mounted packaged HVAC system;
- A duplex sump pump system will be installed to collect subsurface water from around the nuSTORM Muon Decay Ring Service Building;

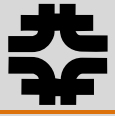

- Fire Alarm/Fire Suppression systems for the nuSTORM Muon Decay Ring Service Building shall be designed in accordance with the applicable sections of the Fermilab Engineering Standards Manual;
- Automatic sprinkler systems shall be designed to a minimum of an Ordinary Hazard Group 1 classification, in accordance with National Fire Protection Association (NFPA) latest edition. The most commonly used NFPA standards relative to automatic sprinkler systems are: 13, 20, 25, 231, 231C, 318, and 750;
- Fire alarm systems shall be designed with a minimum standby power (battery) capacity. These batteries shall be capable of maintaining the entire system in a non-alarm condition for 24 hours, in addition to 15 minutes in full load alarm condition. The most commonly used NFPA standards relative to fire alarm systems are: 70, 72, 90A, and 318;
- The facility will be equipped with a hard-wired, zoned, general evacuation fire alarm system consisting of:
  - Manual fire alarm stations at the building exits
  - Sprinkler system water flow and valve supervisory devices
  - Combination fire alarm horn/strobe located throughout the building
  - A 24 volt hard wire extension from the existing control panel
  - Connection to the site wide FIRUS monitoring system
  - Smoke detection as required.

The Electrical Requirements for this functional area include the following:
- The Electrical Equipment Room will house the electrical equipment that supports the assembly and operation of the nuSTORM Muon Decay Ring beamline components. This includes the incoming electrical service switchgear, panelboards, and related power supplies;
- The electrical power for the nuSTORM Muon Decay Ring Service Building will be provided by a new 1,500 kVA transformer. The transformer has been sized to accommodate the anticipated electrical power for both the conventional facilities and the programmatic equipment;
- The electrical substation will be located adjacent to the nuSTORM Muon Decay Ring Service Building. The substation will be designed to accommodate both the new transformer and associated 4-way air switch;
- The new electrical substation will be connected to the nuSTORM Muon Decay Ring Service Building via a new concrete encased duct bank and will be routed to new electrical service switchgear inside the Muon Decay Ring Service Building. The switchgear will be sized to accommodate the entire 1,500 kVA power from the new transformer;

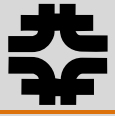

- From the switchgear the electrical power will serve conventional facilities equipment and programmatic equipment for the nuSTORM components;
- The power for the conventional HVAC equipment will be provided from the new electrical switchgear, utilizing 480v power;
- New electrical panels serving the lights, outlets and general house power will be included in the electrical power distribution system.

Adjacent to the Muon Decay Ring Service Building will be the nuSTORM Cryogenic Building. This building will consist of the work involved with construction of an at-grade building and related facilities required to house cryogenic equipment for the nuSTORM Experiment.

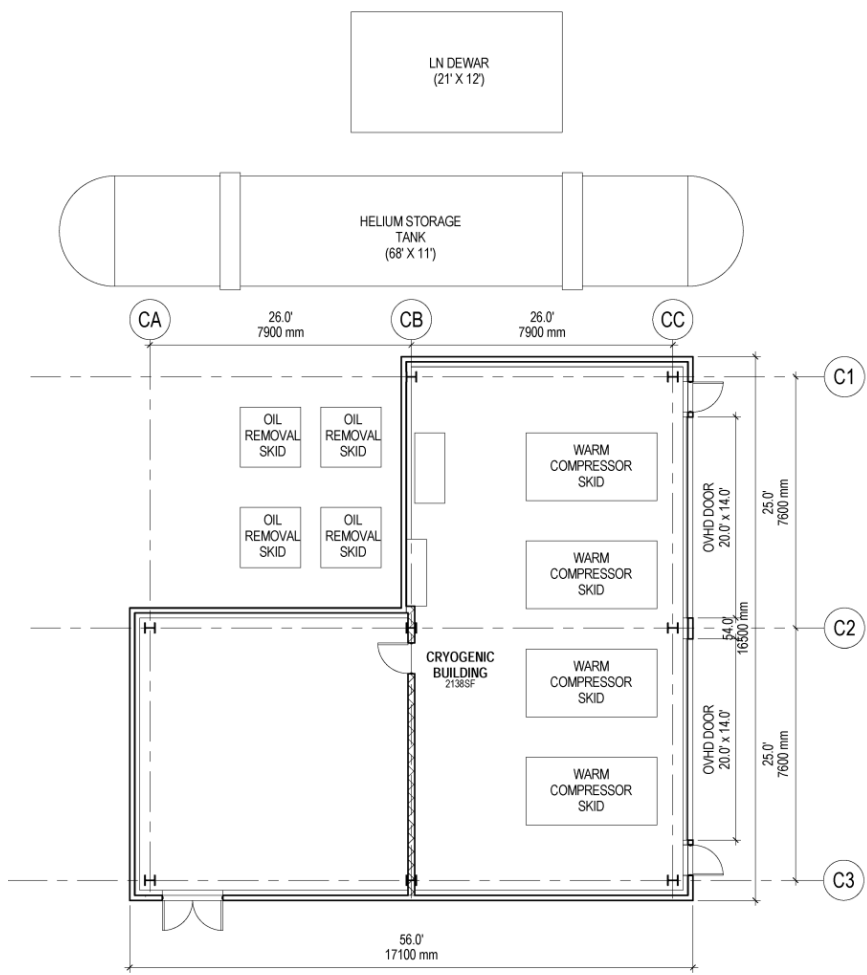

*Figure 11 – Plan View of Cryogenic Building*

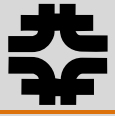

The Cryogenic Building, shown in figure 11 above, is sized based on existing buildings in the Main Ring that serves a similar function. Adjacent to the Cryogenic Building is a 60 foot wide by 100 foot long hardstand area intended for trailer parking to support the cryogenic operations.

The Cryogenic Building will be connected to the nuSTORM Muon Decay Ring Service Building via a below grade trench to house the refrigerant piping and associated control equipment to support operations. The trench will terminate in the drop shaft to the below grade Muon Decay Ring Enclosure.

**Area 5 – Near Detector (WBS 5.0)**

The fifth functional area is the nuSTORM Near Detector Hall consisting of a below grade enclosure to house the nuSTORM Near Detectors and the above grade nuSTORM Near Detector Service Building.

Figure 12, below, depicts the below grade portion of the Near Detector Service Building. This space has been designed to accommodate the nuSTORM Near Detector as well as space for future experiments.

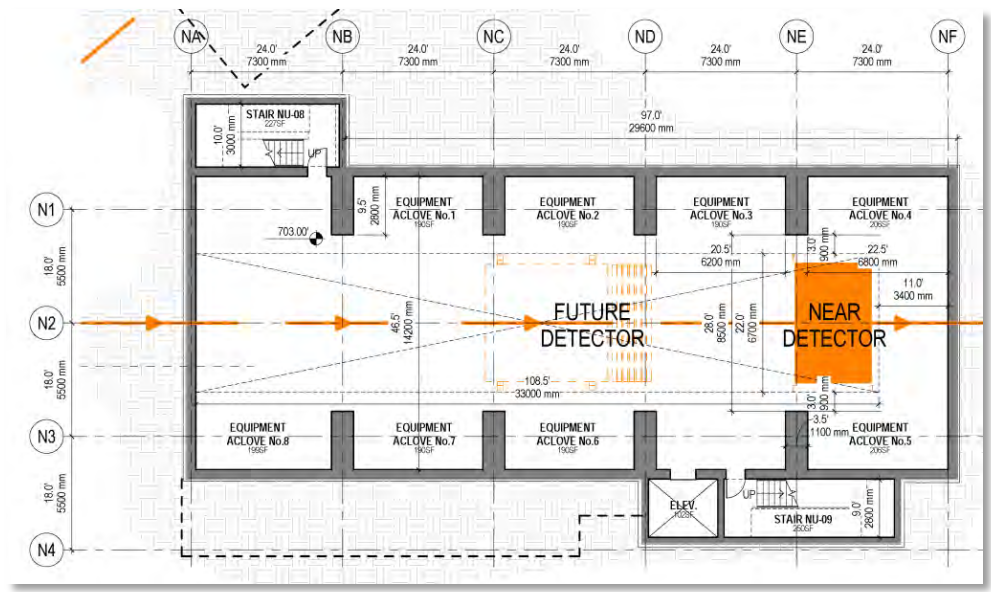

*Figure 12 – Plan View of the Near Detector Enclosure.*

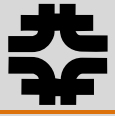

The below grade portions of the facility and building foundations will be constructed of cast-in-place concrete and has been designed to accommodate six (6) feet of precast concrete shielding blocks over the detector.

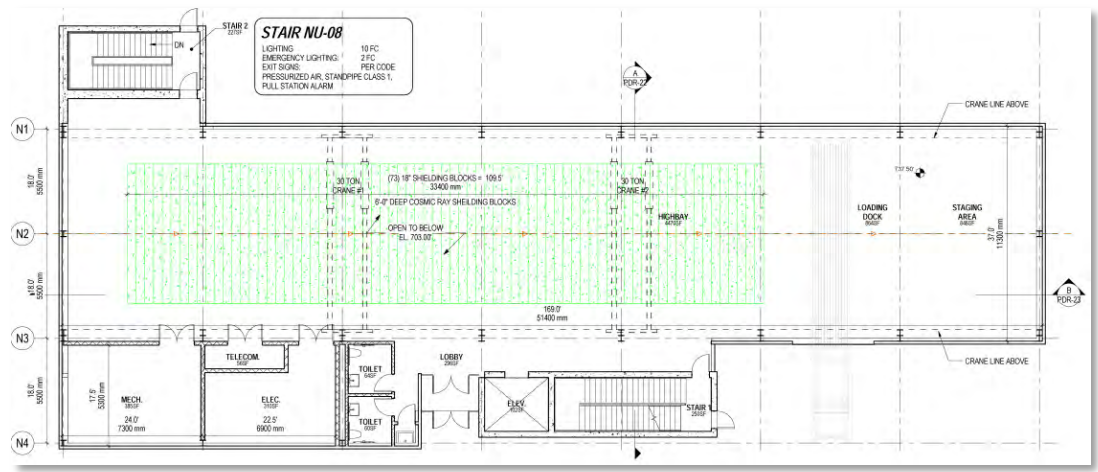

*Figure 13 – Plan View of the Near Detector Service Building*

The above grade portion, shown in Figure 13 above, of the nuSTORM Near Detector Service Building will be a braced steel frame with pre-finished metal siding. The high bay will provide space for staging and assembling the detectors.

The space is dominated by a shield block filled hatch that allow access to the below grade detector enclosure. This hatch is intended to remain open during assembly of the detector. Once assembly is complete, the hatch will be filled with precast concrete shielding blocks to a depth of six (6) feet in order to provide the required shielding.

The high bay will be provided with two (2) thirty (30) ton capacity overhead bridge cranes that will be used for detector assembly as well as shield block handling. The high bay contains space for the temporary staging of portions of the shield blocks should removal be required for maintenance and/or repair of the detector components.

The design provides space adjacent to the nuSTORM Near Detector for the operation of the nuSTORM Near Detector including the associated computer hardware. It is recognized that this space will be minimally occupied as the day to day operation of the detector will likely occur at a central facilities located elsewhere. As such, this space will serve as the front end to remote systems and will be used mainly during assembly, initial startup and maintenance/repair activities. This space will be designed in a similar manner to existing control rooms found on the Fermilab site.

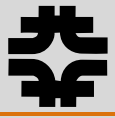

The Structural Requirements for this functional area include:

- The below grade portions of the facility and building foundations will be constructed of cast-in-place concrete and has been designed to accommodate six (6) feet of precast concrete shielding blocks over the detector;
- The Muon Decay Ring Service Building will be metal sided, structural steel frame on a cast-in-place concrete foundation. The building will contain a 30 ton capacity overhead bridge crane to accommodate the unloading, staging and vertical transport of the beamline components;
- The above grade portion, nuSTORM Near Detector Service Building will be a braced steel frame with pre-finished metal siding;

The Mechanical Requirements for this functional area include the following:

- The Mechanical Room will house the mechanical equipment that supports the assembly and operation of the nuSTORM Near Detector Service Building as well as the connection to the below grade portion of the building;
- The HVAC systems for the nuSTORM Near Detector Ring Service Building will conform to ASHRAE 90.1, ASHRAE 62 and applicable NFPA requirements and applicable sections of the Fermilab Engineering Standards Manual.
- Mechanical systems and Metasys controls will be further investigated during subsequent phases in accordance with ASHRAE 90.1 and Federal Life Cycle costing analysis;
- Heating, Ventilation and Air Conditioning Design Parameters:
  - Temperature: 68 degrees Fahrenheit to 78 degrees Fahrenheit
  - Humidity: 50% Maximum Relative Humidity, no minimum
- All plumbing work will be installed in accordance with Illinois Plumbing Code and Standard Specifications for Water & Sewer Main Construction in Illinois.
- The nuSTORM Near Detector Service Building will be conditioned with a roof mounted packaged HVAC system.
- A duplex sump pump system will be installed to collect subsurface water from around the below grade portion of the nuSTORM Near Detector Service Building.
- Fire Alarm/Fire Suppression systems for the nuSTORM Near Detector Service Building shall be designed in accordance with the applicable sections of the Fermilab Engineering Standards Manual.
- Automatic sprinkler systems shall be designed to a minimum of an Ordinary Hazard Group 1 classification, in accordance with National Fire Protection Association (NFPA) latest edition. The most commonly used NFPA standards

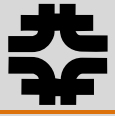

relative to automatic sprinkler systems are: 13, 20, 25, 231, 231C, 318, and 750.

- Fire alarm systems shall be designed with a minimum standby power (battery) capacity. These batteries shall be capable of maintaining the entire system in a non-alarm condition for 24 hours, in addition to 15 minutes in full load alarm condition. The most commonly used NFPA standards relative to fire alarm systems are: 70, 72, 90A, and 318.
- The facility will be equipped with a hard-wired, zoned, general evacuation fire alarm system consisting of:
  - Manual fire alarm stations at the building exits
  - Sprinkler system water flow and valve supervisory devices
  - Combination fire alarm horn/strobe located throughout the building
  - A 24 volt hard wire extension from the existing control panel
  - Connection to the site wide FIRUS monitoring system
  - Smoke detection as required.

The Electrical Requirements for this functional area include the following:

- The Electrical Equipment Room will house the electrical equipment that supports the assembly and operation of the technical components. This includes the incoming electrical service switchgear, panelboards, and related power supplies;
- The electrical power for the nuSTORM Near Detector Service Building will be provided by a new 1,500 kVA transformer. The transformer has been sized to accommodate the anticipated electrical power for both the conventional facilities and the programmatic equipment;
- The electrical substation will be located adjacent to the nuSTORM Near Detector Service Building. The substation will be designed to accommodate both the new transformer and associated 4-way air switch;
- The new electrical substation will be connected to the nuSTORM Near Detector Service Building via a new concrete encased duct bank and will be routed to new electrical service switchgear inside the Near Detector Service Building. The switchgear will be sized to accommodate the entire 1,500 kVA power from the new transformer;
- From the switchgear the electrical power will serve conventional facilities equipment and programmatic equipment for the nuSTORM components;
- The power for the conventional HVAC equipment will be provided from the new electrical switchgear, utilizing 480v power;
- New electrical panels serving the lights, outlets and general house power will be included in the electrical power distribution system.

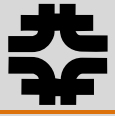

### Area 6 – Far Detector (WBS 6.0)

The sixth functional area is the nuSTORM Far Detector that will be housed in the existing DZero Assembly Hall.  The existing DZero facilities will be repurposed to provide the space and infrastructure to accommodate the nuSTORM Far Detector. Figure 14, below, depicts the nuSTORM Far Detector located in the existing DZero Assembly Hall.

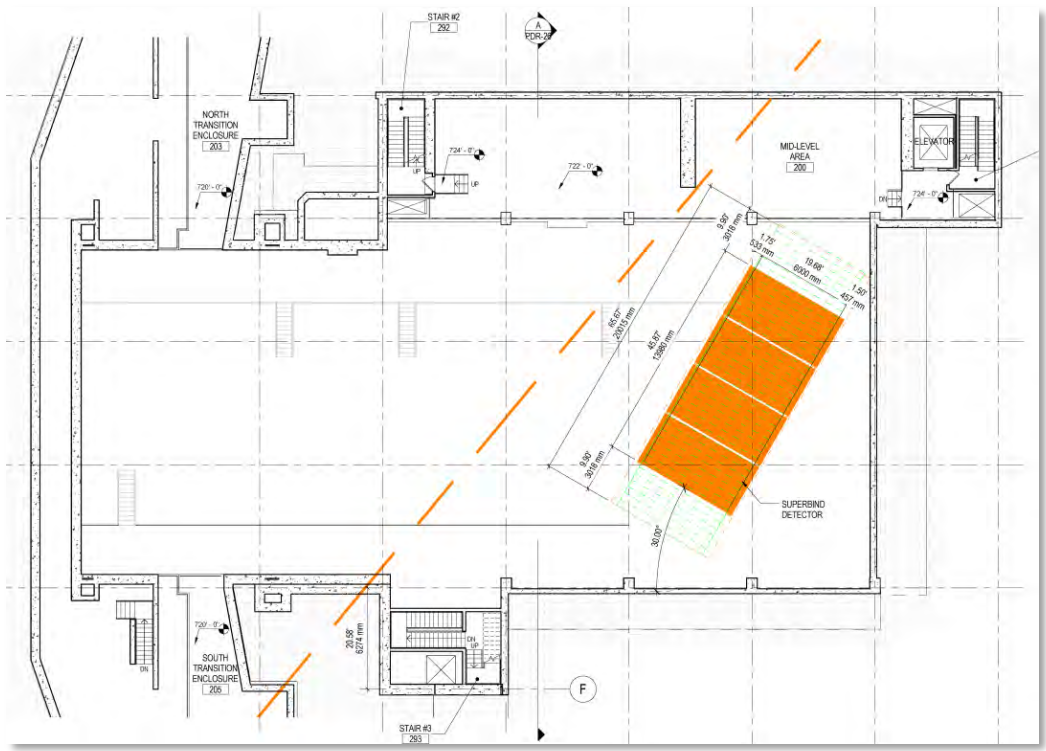

*Figure 14 – Plan View of the nuSTORM Far Detector in the DZero Assembly Hall*

For the purposes of this project definition report there is assumed to be minimal conventional facilities work required for the functional area.  This includes the removal of existing Counting House and Control Room within the building.  The technical components currently housed in these spaces are assumed to be removed by others prior to the start of the demolition work. Also included in this work scope is minor electrical rework associated with the demolition activities.

### Area 7 – Site Work (WBS 8.0)

The seventh functional area is the site work and utilities required to support the nuSTORM project.

The location for the nuSTORM Conventional Facilities was selected based on the programmatic requirement for extraction of a proton beam from the existing Main Injector at the MI-40 Absorber.   The location is adjacent to existing utilities serving the Main Injector including Industrial Cooling Water (ICW), Domestic Water Service (DWS), Low Conductivity Water (LCW), and electrical will be extended to serve the nuSTORM facilities.

Figure 15, below, depicts the overall site work required for the nuSTORM facilities. The existing Main Injector is at the upper portion of the figure (north is towards the top of the page), and the nuSTORM beamline is shown in orange.

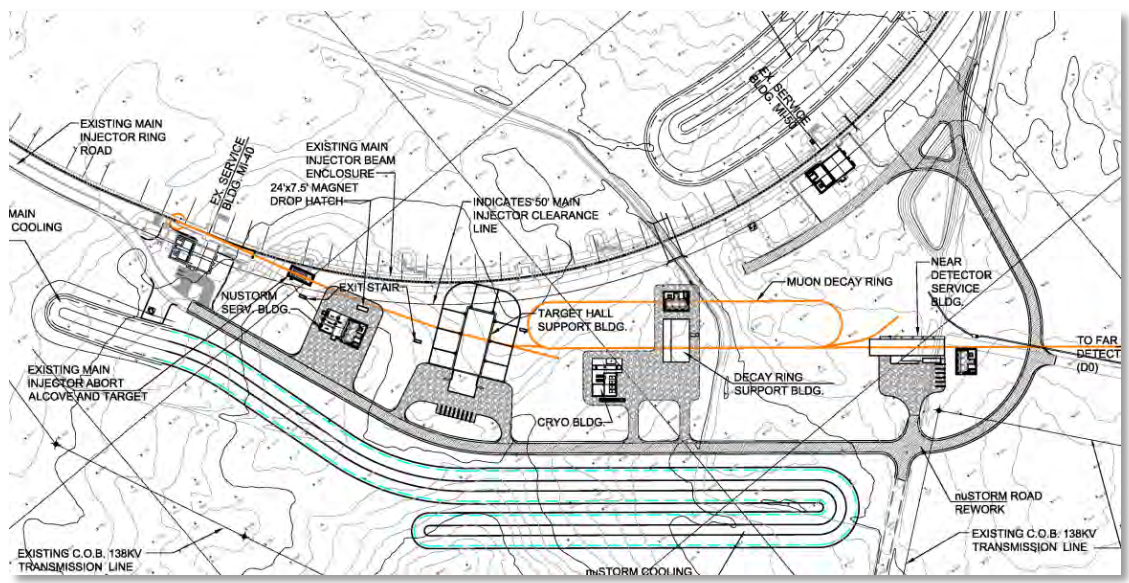

*Figure 15 – Site Plan for nuSTORM Far Detector*

The ICW, DWS and sanitary sewer services for the project will be extended from existing services originating at the MI-60 Service Building.  Additionally, data and communication services will be extended from sources at the Main Injector.  These services will be routed via a new utility corridor to the nuSTORM project site.

A new access road will provide access to the nuSTORM Conventional Facilities from existing Fermilab road network.  This road will be constructed in a similar manner to existing Fermilab roads and will be suitable for all weather access for assembly of the detector and normal operations.

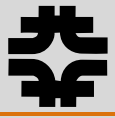

Paved parking will be provided for five (5) vehicles at the each of the nuSTORM Service Buildings along with an unpaved gravel hardstand to provide a staging area during detector assembly and tanker trailer parking for liquid helium trailers during operation. A paved approach to the at-grade loading dock with suitable truck maneuvering space will be provided.

Earth shielding will be provided at below grade beamline enclosures in order to provide the equivalent of 21 feet of shielding for the Primary Beamline Enclosure and 12 feet at enclosures with secondary beams. The earthen berm with maintainable side slopes will be used where possible.

A new concrete encased power duct bank will be installed to connect the nuSTORM facilities to the existing Fermilab 13.8 kV electrical infrastructure system. This connection extends the existing electrical infrastructure from the Kautz Road Substation to the nuSTORM facilities.

The existing Pond C will be impacted by the construction of the nuSTORM facilities. A new pond will be constructed that restores the capacity of Pond C as well as increases the cooling capacity to accommodate the expected heat loads from the nuSTORM facility. This pond will be constructed in a similar style and manner to existing Main Injector ponds.

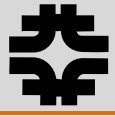

**SUGGESTED PROJECT BUDGET**        **nuSTORM Conventional Facilities**

The nuSTORM Conventional Facilities project team developed a suggested project budget based on the preliminary design described in the drawings and Section III. The design drawings, detailed descriptions and resulting suggested project budget should be considered preliminary and based on the assumptions and requirements contained in Section II.  As the programmatic requirements are refined the cost and schedule implications will also require refinement and adjustment.

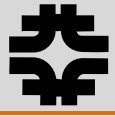

**PART 1        COST SUMMARY**
Listed below is a summary table of the costs associated with the scope of work described in this project definition report.

| | |
|---|---|
| Base Cost | $68,594,000 |
| Engineering Design Inspection and Administration | $20,579,000 |
| Contingency/Management Reserve | $47,200,000 |
| Indirect Costs | $11,977,000 |
| **TOTAL PROJECT COST** | **$148,350,000** |
| Demolition/Space Bank Offset | $1,250,000 |
| Engineering Design Inspection and Administration | $375,000 |
| Contingency/Management Reserve | $486,000 |
| Indirect Costs | $415,000 |
| **OTHER PROJECT COSTS** | **$2,526,000** |

The identified Other Project Costs include the costs associated with the space management costs required to demolish an equal amount of space to balance the new space created.

A further breakdown of the costs by the functional areas is listed below:

| WBS | Functional Area | Base Cost | EDIA 30% | Contingency % | Contingency $ | Indirects | Totals |
|---|---|---|---|---|---|---|---|
| 1.0 | Primary Beamline Enclosure | $7,013,000 | $2,104,000 | 40% | $3,647,000 | $1,266,000 | $14,030,000 |
| 2.0 | Target Station | $8,993,000 | $2,698,000 | 55% | $6,430,000 | $1,662,000 | $19,783,000 |
| 3.0 | Transport Line Enclosure | $1,883,000 | $565,000 | 60% | $1,469,000 | $504,000 | $4,421,000 |
| 4.0 | Muon Decay Ring Enclosure | $26,002,000 | $7,801,000 | 60% | $20,282,000 | $4,215,000 | $58,300,000 |
| 5.0 | Near Detector | $11,750,000 | $3,525,000 | 40% | $6,110,000 | $1,882,000 | $23,267,000 |
| 6.0 | Far Detector | $720,000 | $216,000 | 55% | $515,000 | $333,000 | $1,784,000 |
| 8.0 | Site Work | $12,233,000 | $3,670,000 | 55% | $8,747,000 | $2,115,000 | $26,765,000 |
| | **TOTALS** | **$68,594,000** | **$20,579,000** | | **$47,200,000** | **$11,977,000** | **$148,350,000** |

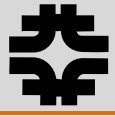

Listed below are descriptions of the basic components used to develop the suggested project budget.

### PART 2　　　　METHODOLOGY

DOE Order 413.3-21 - DOE Guide 430.1-1X, *DOE Cost Estimating Guide for Program and Project Management*" was used as the basis for the development of the ICS.

Fiscal Year Dollars - The costs contained in this Project Definition Report are based on FY2013 dollars.  Escalation adjustment to the above costs will need to be applied once a funding profile has been determined.

Cost Estimate Basis - The suggested project budget is based on cost data taken from Means Cost Estimating Guides, historical data and recent construction history at Fermilab.  While the suggested project budget can provide input for the feasibility of the project, further design refinement will affect the final cost of the project.

High Performance Sustainable Building - Fermilab incorporates sustainable design principals into the planning, design and construction of projects. This direction is taken from the Fermilab Director's Policy 3.  The suggested project costs include the compliance with the DOE order of high performance (sustainable) building.  In some instances, new facilities within the DOE complex achieve a "gold" level certification from the United States Green Building Council (USGBC) under the Leadership in Energy and Environmental Design (LEED) standards. It is recognized that the nuSTORM Conventional Facilities do not fall within the LEED criteria.  While this project is not intended to become a certified building, the project processes and each project element will be evaluated during design to reduce their impact on natural resources without sacrificing program objectives.  The project design will incorporate maintainability, aesthetics, environmental justice and program requirements to deliver a well-balanced project.

Space Management Requirements - Beginning in FY 2003, all new DOE Office of Science funded construction projects, including line items, GPP and IGPP, which provide new space, must have an equivalent amount of excess space allocated from the DOE Office of Science Space Bank.  The identified Other Project Costs include the costs associated with the space management costs required to demolish an equal amount of space to balance the new space created.

Engineering, Design, Inspection and Administration - Project management, design, construction management, cost and schedule for the conventional construction portion of projects are typically the responsibility of the Facilities Engineering

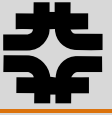

Services Section (FESS).  This effort will be accomplished using the resources of the FESS Engineering Department supplemented with appropriate consultant resources.

Engineering Design and Inspection (ED&I) activities are included in the suggested project budgets.  ED&I activities include the engineering and design activities in Preliminary and Final Design, the inspection activities associated with Construction Phase Support.  The descriptions are based on DOE Directive G413.3-21, Chapter 6.  Past historical data and DOE Directive G413.3-21, Section 5.4.3 indicates that 15%-25% of the construction cost is an appropriate range.  Non-traditional, first of a kind projects may be higher, while simple construction such as buildings will be lower than this range.  Generally, the more safety and regulatory intervention is involved, the higher the percentage.

**Section IV**

Administration (A) activities includes those defined by DOE Directive G413.3-21, Section 5.4.3 as Project Management (PM) and Construction Management (CM).  These costs range from 5%-15% of the other estimated project costs for most DOE projects, depending on the nature of the project and the scope of what is covered under project management.

For the purposes of this project definition report, the EDI and A functions have been combined.   Listed below is a range of EDIA estimates for the life of the project:

| | |
|---|---|
| 6%-8% | Preliminary Design |
| 8%-10% | Final Design |
| 10%-12% | Construction Phase |
| 2%-3% | Close Out Phase |
| ====== | |
| 26%-33% | Total Engineering, Design, Inspection and Administration |

For the nuSTORM Conventional Facilities, a value of 30% was used in the suggested project budget calculations.

<u>Management Reserve/Contingency</u> - The application of risk-based contingency has been considered in the development of this project definition report.  The contingency has been derived from a preliminary risk analysis of various aspects of the scope being estimated.  This analysis takes into consideration budget, schedule, and technical risks as they apply to the project effort, underscoring the uncertainties that exist in each of the elements.  The magnitude of the contingency estimate will depend on the status of planning, design, procurement, and construction, and the complexities and uncertainties of the operation or component parts of the project element.  Part 4 below contains the preliminary risks identified with the conventional facilities at this stage of the project.

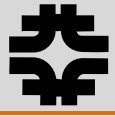

<u>Indirect Costs</u> - Indirect Costs rates are defined by DOE Order 4700.1 that states indirect costs are "...costs incurred by an organization for common or joint objectives and which cannot be identified specifically with a particular activity or project.  ". The suggested costs above include Indirect Costs calculated at the current rate for projects as of October 30, 2012. As the project progresses, the Directorate Budget Office should be consulted for updated rates and procedures.

**Section IV**

**PART 3          COST ESTIMATE CLASSIFICATION**

DOE Guide 430.1-1X, *DOE Cost Estimating Guide for Program and Project Management*" classifies cost estimates into one (1) of five (5) categories.  These classifications are listed below:

| Cost Estimate Classification | Primary Characteristics | |
| --- | --- | --- |
| | Level of Definition (% of Complete Definition) | Cost Estimating Description (Techniques) |
| Class 5, Order of Magnitude | 0% to 2% | Stochastic, most parametric, judgment (parametric, specific analogy, expert opinion, trend analysis) |
| Class 4, Intermediate | 1% to 15% | Various, more parametric (parametric, specific analogy, expert opinion, trend analysis) |
| Class 3, Preliminary | 10% to 40% | Various, including combinations (detailed, unit-cost, or activity-based; parametric; specific analogy; expert opinion; trend analysis) |
| Class 2, Intermediate | 30% to 70% | Various, more definitive (detailed, unit-cost, or activity-based; expert opinion; learning curve) |
| Class 1, Definitive | 50% to 100% | Deterministic, most definitive (detailed, unit-cost, or activity-based; expert opinion; learning curve) |

These classifications are based on the Association for the Advanced of Cost Engineering (AACE) Recommended Practice No. 18R-97. These classifications help ensure that the quality of the cost estimate is appropriately considered when applying escalation and contingency.

The level of detail and accuracy of the budget becomes more definitive as the project's scope is refined. In a project's earliest phases, the Initiation, or Pre-Conceptual Phase (before Critical Decision [CD] -0, an Order-of-Magnitude (or Parametric) Estimate is usually required.  When a capital asset acquisition project has completed the Conceptual Design Phase, a Preliminary Budget Range is required to establish the Budget Baseline at CD-1. Budget refinements shall be based on a Definitive Estimate for every element in the WBS and is required for CD-2.

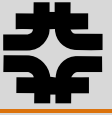

The classification for the nuSTORM Conventional Facilities PDR is considered a Class 4 (Intermediate) based on the preliminary nature and level of definition of the programmatic requirements.

**Section IV**

### PART 4          PRELIMINARY RISK IDENTIFICATION

Contingency estimates are generally applied as a percentage of a particular cost or category of work.  The PDR cost estimate is based on using the best available information to develop the expected cost, and then a risk analysis is performed to develop the required contingency budget based on risk probability and consequence. It is recognized that higher or lower contingency amounts are appropriate throughout the project based on an analysis of project complexity, technical characteristics, and associated risks.  As part of the development of the suggested project budget, the project team has developed a preliminary list of high impact risks for the nuSTORM Conventional Facilities.

Risk #1 – Programmatic Requirements
The early stage of the project and the likely evolution of the programmatic requirements have the ability to significantly impact the cost and schedule for the conventional facilities.

Risk #2 – Subsurface Characteristics
The nuSTORM Conventional Facilities project is intended to be located on a portion of the Fermilab site with limited historic subsurface data.  Experience at similar adjacent projects has been used as a basis of assumption for the development of this PDR. However, unknown subsurface conditions have the ability to impact the cost and schedule for the conventional facilities.

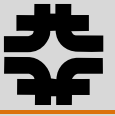

**Cost Details**

- **Functional Area 1 – Primary Beamline Enclosure Cost Estimate**
- **Functional Area 2 – Target Station Cost Estimate**
- **Functional Area 3 – Transport Beamline Enclosure Cost Estimate**
- **Functional Area 4 – Muon Decay Ring Cost Estimate**
- **Functional Area 5 – Near Detector Cost Estimate**
- **Functional Area 6 – Far Detector Cost Estimate**
- **Functional Area 7 – Site Work Cost Estimate**
- **FY13 Provisional Labor, Indirect and Shop Rates**
- **Indirect Burden Allocation Policy and Methodology**

# FERMILAB FESS  COST ESTIMATE

| | | ESTIMATED SUBCONTRACT AWARD AMOUNT | | | | | $7,013,000 |
|---|---|---|---|---|---|---|---|
| Escalation | | | | | | | $0 |
| Subcontractor's Overhead and Profit | | 20.0% | | | | | $1,169,000 |
| Difficult Conditions (Construction at Main Injector) | | 10.0% | | | | | $531,000 |
| Subcontract Base Estimate | | | | | | | $5,313,000 |

| nuSTORM Conventional Facilities | | | | Project No. | Status: | Date: | Rev Date |
|---|---|---|---|---|---|---|---|
| *Fixed Priced Construction - Primary Beamline Enclosure* | | | | 6-13-1 | PDR | 13-May-13 | 28-May-13 |
| ITEM NO. | | DESCRIPTION OF WORK: | | QUANTITY | UNITS | UNIT COST | AMOUNT |
| **00** | | **General** | **$10,000** | | | | |
| | 01 | Mobilize | | 1 | lot | $10,000.00 | $10,000 |
| | | | | | | | |
| | | | | | | | |
| **02** | | **Site Construction** | **$1,695,000** | | | | |
| | 01 | Silt Fence and Erosion control | | 1 | Lot | $50,000 | $50,000 |
| | 02 | Strip Topsoil | | 1630 | CY | $6 | $9,778 |
| | 03 | Haul Topsoil | | 2119 | CY | $7 | $14,830 |
| | 04 | Excavate clays for Base Slab | | 32725 | CY | $15 | $490,875 |
| | 05 | Haul Clays | | 42543 | CY | $7 | $297,798 |
| | 06 | Stone Backfill | | 640 | CY | $32 | $20,480 |
| | 07 | Haul and Backfill Clays | | 30918 | CY | $25 | $772,940 |
| | 08 | Haul and spread topsoil | | 1630 | CY | $13 | $21,185 |
| | 09 | Final Grading and seeding | | 1 | Acre | $23,800 | $17,484 |
| | 10 | Support MI Duct Bank across excavation | | 0 | Lot | $45,000 | $0 |
| | 11 | Support ICW piping across excavation | | 0 | Lot | $7,500 | $0 |
| | | | | | | | |
| **03** | | **Concrete** | **$1,971,000** | | | | |
| | 01 | Mud slab | *Enclosure* | 71 | CY | $250 | $17,824.07 |
| | 02 | Base Slab | *Enclosure* | 570 | CY | $600 | $342,222.22 |
| | 03 | Walls | *Enclosure* | 596 | CY | $850 | $506,977.78 |
| | 04 | Top Slab | *Enclosure* | 463 | CY | $900 | $417,083.33 |
| | 05 | Hatch Walls (7.5 x 24) | *Enclosure* | 145 | CY | $900 | $130,784.00 |
| | 06 | Hatch Shielding blocks | *Enclosure* | 160 | CY | $1,100 | $176,000.00 |
| | 07 | Stairs | *Enclosure* | 2 | Each | $175,000 | $350,000.00 |
| | 08 | FBP Footings | | | | | |
| | | *Foundation - Walls* | *Service Bldg* | 16 | CY | $325 | $5,200 |
| | | *Column Pier - 24" - T/Pier 12" below grade* | *Service Bldg* | 3 | CY | $325 | $975 |
| | | *Footing - Continuous Wall - 12" x 2'-6"* | *Service Bldg* | 16 | CY | $300 | $4,800 |
| | | *Footing - Column Pier - 12" x 4'-0"* | *Service Bldg* | 5 | CY | $300 | $1,500 |
| | 09 | FBP Slab | | | | | |
| | | *Granular Fill at Slab on Grade - 4"* | *Service Bldg* | 24 | CY | $28 | $672 |
| | | *2" Extruded Polystyrene Insulation Under Slab* | *Service Bldg* | 1945 | SF | $1 | $2,334 |
| | | *15 Mil Vapor Barrier Under Slab* | *Service Bldg* | 1945 | SF | $0 | $584 |
| | | *6" Concrete Slab w 6x6 WWF* | *Service Bldg* | 1945 | SF | $5 | $9,628 |
| | | *8" Concrete Housekeeping Slabs* | *Service Bldg* | 1 | LS | $4,500 | $4,500 |
| | | | | | | | |
| | | | | | | | |
| **04** | | **Masonry** | **$6,000** | | | | |
| | 01 | Exterior Masonry Walls | | 450 | SF | $13 | $5,850 |
| | | | | | | | |
| | | | | | | | |
| **05** | | **Metals** | **$72,000** | | | | |
| | 01 | Misc. Metals | | | | | |
| | | *Steel Girt Channels at Building Mid-Height* | | 4560 | lbs. | $3 | $11,400 |
| | | *Metal Panel Support Angles* | | 2692 | lbs. | $3 | $6,730 |
| | | *Steel Channel Opening Framing* | | 1560 | lbs. | $3 | $3,900 |
| | 02 | Erect Structural Steel | | | | | |
| | | *W12 Columns - Wide Flange* | | 6240 | lbs. | $3 | $15,600 |
| | | *Beams and Girts- Wide Flange* | | 11258 | lbs. | $3 | $28,145 |
| | 05 | Erect Roof Deck - 1 1/2" , 20 ga | | 2050 | SF | $3 | $6,458 |

# FERMILAB FESS COST ESTIMATE

| | ESTIMATED SUBCONTRACT AWARD AMOUNT | | $7,013,000 |
|---|---|---|---|
| Escalation | | | $0 |
| Subcontractor's Overhead and Profit | 20.0% | | $1,169,000 |
| Difficult Conditions (Construction at Main Injector) | 10.0% | | $531,000 |
| Subcontract Base Estimate | | | $5,313,000 |

| nuSTORM Conventional Facilities | | Project No. | Status: | Date: | Rev Date |
|---|---|---|---|---|---|
| *Fixed Priced Construction - Primary Beamline Enclosure* | | 6-13-1 | PDR | 13-May-13 | 28-May-13 |
| ITEM NO. | DESCRIPTION OF WORK: | *QUANTITY* | *UNITS* | *UNIT COST* | *AMOUNT* |
| **06** | **Wood and Plastic** | | | | **$0** |
| | | | | | $0 |
| **07** | **Thermal and Moisture Protection** $412,000 | | | | **$0** |
| | 01 Damp Proof enclosure | *Enclosure* | 18150 | SF | $8 | $145,200 |
| | 02 Install Metal Siding | | | | | $0 |
| | *3" Centrica Formawall* | *Service Bldg* | 3180 | SF | $46 | $146,280 |
| | 03 Install Built-Up Roofing | | | | | $0 |
| | *2 Layers 1 1/2" Insulation, 1/2" Fiberboard, 4 Ply Felt, White Marble Ballast* | *Service Bldg* | *2050* | *SF* | *$18* | *$36,900* |
| | *Tapered Insulation* | *Service Bldg* | *240* | *SF* | *$3* | *$600* |
| | 04 Install Aluminum Coping | *Service Bldg* | 220 | LF | $15 | $3,300 |
| | 05 Architectural Treatment (10%) | *Service Bldg* | 1 | Lot | $30,000 | $30,000 |
| | 06 Guiding Principals Compliance (15%) | *Service Bldg* | 1 | Lot | $50,000 | $50,000 |
| | | | | | | |
| **08** | **Doors and Windows** $46,000 | | | | **$0** |
| | 01 Enclosure Man doors 3 x 7 | *Enclosure* | 4 | Each | $1,400 | $5,600 |
| | 02 Install Exterior Man Doors | | | | | |
| | *Single Insulated Hollow Metal Door and Frame* | *Service Bldg* | *2* | *Each* | *$1,400* | *$2,800* |
| | *Double Insulated Hollow Metal Door and Frame (6'x7')* | *Service Bldg* | *1* | *Each* | *$1,800* | *$1,800* |
| | 03 Install Interior Man Doors | | | | | |
| | *Hollow Metal Door and Frame (3'x7')* | *Service Bldg* | *1* | *Each* | *$1,200* | *$1,200* |
| | 04 Install Overhead Door | | | | | |
| | *Insulated Coil Door - 18'X14'* | *Service Bldg* | *2* | *Each* | *$13,000* | *$26,000* |
| | 05 Install Skylight/ daylighting | *Service Bldg* | 6 | Each | $1,400 | $8,400 |
| | | | | | | $0 |
| **09** | **Finishes** $70,000 | | | | **$0** |
| | 01 Painting | *Enclosure* | 14300 | SF | $3 | $42,900 |
| | 02 Paint Exposed Ceiling | *Service Bldg* | 2200 | SF | $3 | $6,600 |
| | 03 Paint Structural Steel | *Service Bldg* | 1 | Lot | $10,000 | $10,000 |
| | 04 Paint Interior Walls | *Service Bldg* | 1 | Lot | $10,000 | $10,000 |
| | | | | | | $0 |
| **10** | **Specialties** | | | | **$0** |
| | | | | | | $0 |
| **14** | **Conveying Equipment** | | | | **$0** |
| | | | | | | $0 |
| **15** | **Mechanical** $311,000 | | | | **$0** |
| | 01 Install Fans | *Enclosure* | 1 | Lot | $13,988 | $13,988 |
| | 02 Install Dehumidifiers | *Enclosure* | 1 | Lot | $30,000 | $30,000 |
| | 03 Install Controls | *Enclosure* | 1 | Lot | $25,000 | $25,000 |
| | 04 Install Under drain | *Enclosure* | 1100 | LF | $28 | $30,800 |
| | 05 Install Sumps | *Enclosure* | 2 | Each | $50,000 | $100,000 |
| | 06 Install Sump Above Ground Piping | *Enclosure* | 2 | Lot | $16,963 | $33,926 |
| | 07 Install Sump Controls | *Enclosure* | 2 | Lot | $15,000 | $30,000 |
| | 08 Start Up Sump | *Enclosure* | 2 | Lot | $800 | $1,600 |
| | 09 Fire Riser | *Service Bldg* | 1 | Lot | $15,000 | $15,000 |
| | 10 Sprinklers | *Service Bldg* | 2200 | SF | $8 | $17,600 |
| | 11 Fire Detection | *Service Bldg* | 2200 | SF | $6 | $13,200 |
| | | | | | | |

## FERMILAB FESS COST ESTIMATE

| | | ESTIMATED SUBCONTRACT AWARD AMOUNT | $7,013,000 |
|---|---|---|---|
| Escalation | | | $0 |
| Subcontractor's Overhead and Profit | 20.0% | | $1,169,000 |
| Difficult Conditions (Construction at Main Injector) | 10.0% | | $531,000 |
| Subcontract Base Estimate | | | $5,313,000 |

| nuSTORM Conventional Facilities | | Project No. | Status: | Date: | Rev Date |
|---|---|---|---|---|---|
| *Fixed Priced Construction - Primary Beamline Enclosure* | | 6-13-1 | PDR | 13-May-13 | 28-May-13 |

| ITEM NO. | | DESCRIPTION OF WORK: | | QUANTITY | UNITS | UNIT COST | AMOUNT |
|---|---|---|---|---|---|---|---|
| 16 | | **Electrical** | **$720,000** | | | | |
| | 01 | Cable Tray | *Enclosure* | 1650 | LF | $74 | $122,100 |
| | 02 | Ground Wire | *Enclosure* | 550 | LF | $18 | $9,900 |
| | 03 | Ground Rods | *Enclosure* | 8 | ea. | $690 | $5,520 |
| | 04 | Unit heaters | *Enclosure* | 2 | ea. | $870 | $1,740 |
| | 05 | Panels | *Enclosure* | 1 | lot | $40,000 | $40,000 |
| | 06 | Fire Detection | *Enclosure* | 5500 | SF | $6 | $33,000 |
| | 07 | Lighting | *Enclosure* | 1 | Lot | $40,000 | $40,000 |
| | 08 | Electrical Distribution | *Enclosure* | 1 | Lot | $100,000 | $100,000 |
| | 09 | Exterior Lighting | *Service Bldg* | 1 | Lot | $25,000 | $25,000 |
| | 10 | Install Switchboard | *Service Bldg* | 1 | EA | $62,500 | $62,500 |
| | 11 | Install 480 V Ductbank | *Service Bldg* | 60 | LF | $563 | $33,750 |
| | 12 | Install Transformers | *Service Bldg* | 1 | EA | $14,375 | $14,375 |
| | 13 | Install Ground Grid Wire | *Service Bldg* | 550 | LF | $18 | $9,625 |
| | 14 | Install Ground Rods | *Service Bldg* | 10 | EA | $690 | $6,900 |
| | 15 | Install Building Bond Connections | *Service Bldg* | 140 | LF | $179 | $25,069 |
| | 16 | Install Panel Feeders | *Service Bldg* | 1 | LS | $12,769 | $12,769 |
| | 17 | Install Distribution Circuits | *Service Bldg* | 1 | LS | $4,460 | $4,460 |
| | 18 | Install PP-208Y/120 VAC | *Service Bldg* | 1 | EA | $5,750 | $5,750 |
| | 19 | Install LP - 480Y/277 VAC | *Service Bldg* | 1 | EA | $10,844 | $10,844 |
| | 20 | Install 45 kVA, 480 - 208Y/277 VAC | *Service Bldg* | 1 | EA | $13,375 | $13,375 |
| | 21 | Install Welding Receptacles | *Service Bldg* | 6 | EA | $955 | $5,730 |
| | 22 | Install 120/208 VAC Receptacles | *Service Bldg* | 16 | EA | $239 | $3,820 |
| | 23 | Install Disconnect Switches | *Service Bldg* | 6 | EA | $540 | $3,240 |
| | 24 | Install Unit Heaters | *Service Bldg* | 4 | EA | $2,000 | $8,000 |
| | 25 | Connect Other mechanical Equipment | *Service Bldg* | 4 | EA | $2,000 | $8,000 |
| | 26 | Install  Flour. Lighting | *Service Bldg* | 2200 | SF | $13 | $28,215 |
| | 27 | Install Emergency Lighting (UPS) | *Service Bldg* | 1 | LS | $13,875 | $13,875 |

| | | ESTIMATED SUBCONTRACT AWARD AMOUNT | | | | $8,993,000 |
|---|---|---|---|---|---|---|
| Escalation | | | | | | $0 |
| Subcontractor's Overhead and Profit | | 20.0% | | | | $1,499,000 |
| Difficult Conditions | | | | | | $0 |
| Subcontract Base Estimate | | | | | | $7,494,000 |

| nuSTORM Conventional Facilities | | | Project No. | Status: | Date: | Rev Date |
|---|---|---|---|---|---|---|
| *Fixed Priced Construction - Target Station* | | | **6-13-1** | **PDR** | **13-May-13** | **28-May-13** |
| ITEM NO. | | DESCRIPTION OF WORK: | QUANTITY | UNITS | UNIT COST | AMOUNT |
| **00** | | **General** | **$10,000** | | | |
| | 01 | Mobilize | 1 | lot | $10,000.00 | $10,000 |
| | | | | | | |
| | | | | | | |
| **02** | | **Site Construction** | **$760,000** | | | |
| | 01 | Silt Fence and Erosion control | 1 | Lot | $50,000 | $50,000 |
| | 02 | Strip Topsoil | 89 | CY | $6 | $534 |
| | 03 | Haul Topsoil | 124 | CY | $7 | $868 |
| | 04 | Excavate clays for Base Slab | 6098 | CY | $15 | $91,470 |
| | 05 | Excavate clays for Building | 8958 | CY | $15 | $134,370 |
| | 05 | Haul Clays | 21079 | CY | $7 | $147,553 |
| | 06 | Stone Backfill | 201 | CY | $32 | $6,432 |
| | 07 | Haul and Backfill Clays | 3500 | CY | $25 | $87,500 |
| | 08 | Haul and spread topsoil | 50 | CY | $13 | $650 |
| | 09 | Final Grading and seeding | 2 | Acre | $23,800 | $47,600 |
| | 10 | Backfill Along Building | 7704 | CY | $25 | $192,600 |
| | | | | | | |
| | | | | | | |
| **03** | | **Concrete** | **$2,032,000** | | | |
| | 01 | Target Pile Mud Slab | 9 | CY | $250 | $2,250 |
| | 02 | Target Pile Base Slab | 111 | CY | $600 | $66,600 |
| | 03 | Target Pile Walls | 302 | CY | $850 | $256,700 |
| | 04 | Lower Level Slabs | 933 | CY | $850 | $793,050 |
| | 05 | Lower Level Walls | 581 | CY | $600 | $348,600 |
| | 06 | Loading Dock Granular Base | 56 | CY | $35 | $1,960 |
| | 07 | Loading Dock Slab | 56 | CY | $500 | $28,000 |
| | 08 | Target Pile Shielding Blocks | 100 | CY | $1,100 | $110,000 |
| | 07 | Stairs | 2 | Each | $175,000 | $350,000 |
| | 08 | Foundations for At-Grade Building | 100 | CY | $500 | $50,000 |
| | 09 | Slab for At-Grade Building | 80 | CY | $250 | $20,000 |
| | 10 | 8" Concrete Housekeeping Slabs | 1 | LS | $4,500 | $4,500 |
| | | | | | | |
| | | | | | | |
| **04** | | **Masonry** | **$13,000** | | | |
| | 01 | Masonry Walls | 1000 | SF | $13 | $13,000 |
| | | | | | | |
| | | | | | | |
| **05** | | **Metals** | **$564,000** | | | |
| | 01 | Mezzinine Columns | 2400 | lbs | 3 | $7,200 |
| | 02 | Mezzinine beams | 20400 | lbs | 3 | $61,200 |
| | 03 | Mezzinine metal deck | 3000 | SF | 5 | $15,000 |
| | 04 | Crane Columns | 10880 | lbs | 3 | $32,640 |
| | 05 | High Bay girders | 64800 | lbs | 3 | $194,400 |
| | 06 | High Bay Roof beams | 46080 | lbs | 3 | $138,240 |
| | 07 | Roof Deck | 11520 | SF | 5 | $57,600 |
| | 08 | Crane Girders | 44160 | lbs | 5 | $57,600 |
| | | | | | | |
| **06** | | **Wood and Plastic** | | | | $0 |
| | | | | | | $0 |
| **07** | | **Thermal and Moisture Protection** | **$1,401,000** | | | |
| | 01 | Damp Proof enclosure | 15000 | SF | $8 | $120,000 |
| | 02 | Geo-Membrane at Target Chase | 1 | Lot | $100,000 | $100,000 |

# FERMILAB FESS  COST ESTIMATE

| | | ESTIMATED SUBCONTRACT AWARD AMOUNT | | | | $8,993,000 |
|---|---|---|---|---|---|---|
| Escalation | | | | | | $0 |
| Subcontractor's Overhead and Profit | | 20.0% | | | | $1,499,000 |
| Difficult Conditions | | | | | | $0 |
| Subcontract Base Estimate | | | | | | $7,494,000 |

| **nuSTORM Conventional Facilities** | | | Project No. | Status: | Date: | Rev Date |
|---|---|---|---|---|---|---|
| *Fixed Priced Construction - Target Station* | | | **6-13-1** | **PDR** | **13-May-13** | **28-May-13** |
| ITEM NO. | | DESCRIPTION OF WORK: | QUANTITY | UNITS | UNIT COST | AMOUNT |
| | 02 | Install Metal Siding | | | | |
| | | *3" Centrica Formawall* | 15120 | SF | $46 | $695,520 |
| | 03 | Install Built-Up Roofing | | | | $0 |
| | | *2 Layers 1 1/2" Insulation, 1/2" Fiberboard,    4 Ply Felt, White Marble Ballast* | 9000 | SF | $18 | $162,000 |
| | | *Tapered Insulation* | 9000 | SF | $3 | $22,500 |
| | 04 | Install Aluminum Coping | 420 | LF | $15 | $6,300 |
| | 05 | Architectural Treatment (10%) | 1 | Lot | $110,000 | $110,000 |
| | 06 | Guiding Principals Compliance (15%) | 1 | Lot | $185,000 | $185,000 |
| | | | | | | |
| **08** | | **Doors and Windows** | **$51,000** | | | |
| | 01 | Enclosure Man doors 3 x 7 | 4 | Each | $1,400 | $5,600 |
| | 02 | Install Exterior Man Doors | | | | |
| | | *Single Insulated Hollow Metal Door and Frame* | 2 | Each | $1,400 | $2,800 |
| | | *Double Insulated Hollow Metal Door and Frame (6'x7')* | 2 | Each | $1,800 | $3,600 |
| | 03 | Install Interior Man Doors | | | | |
| | | *Hollow Metal Door and Frame (3'x7')* | 4 | Each | $1,200 | $4,800 |
| | 04 | Install Overhead Door | | | | |
| | | Insulated Coil Door - 18'X14' | 2 | Each | $13,000 | $26,000 |
| | 05 | Install Skylight/ daylighting | 6 | Each | $1,400 | $8,400 |
| | | | | | | $0 |
| **09** | | **Finishes** | **$108,000** | | | **$0** |
| | 01 | Painting | 12000 | SF | $3 | $36,000 |
| | 02 | Paint Exposed Ceiling | 9000 | SF | $3 | $27,000 |
| | 03 | Paint Structural Steel | 1 | Lot | $20,000 | $20,000 |
| | 04 | Paint Interior Walls | 1 | Lot | $25,000 | $25,000 |
| | | | | | | $0 |
| **10** | | **Specialties** | | | | **$0** |
| | | | | | | $0 |
| **14** | | **Conveying Equipment** | **$440,000** | | | **$0** |
| | 01 | 40 Ton Overhead Bridge Crane | 2 | Each | $180,000 | $360,000 |
| | | *Standard Industrial Crane* | | | | |
| | 02 | Install Bridge Crane | 2 | Lot | $25,000 | $50,000 |
| | 03 | Load Test Bridge Crane | 2 | Lot | $15,000 | $30,000 |

# FERMILAB FESS COST ESTIMATE

| | | ESTIMATED SUBCONTRACT AWARD AMOUNT | | | | $8,993,000 |
|---|---|---|---|---|---|---|
| Escalation | | | | | | $0 |
| Subcontractor's Overhead and Profit | | 20.0% | | | | $1,499,000 |
| Difficult Conditions | | | | | | $0 |
| Subcontract Base Estimate | | | | | | $7,494,000 |

| **nuSTORM Conventional Facilities** | | | Project No. | Status: | Date: | Rev Date |
|---|---|---|---|---|---|---|
| *Fixed Priced Construction - Target Station* | | | 6-13-1 | PDR | 13-May-13 | 28-May-13 |
| ITEM NO. | | *DESCRIPTION OF WORK:* | *QUANTITY* | *UNITS* | *UNIT COST* | *AMOUNT* |
| **15** | | **Mechanical** | **$1,036,000** | | | **$0** |
| | 01 | Domestic Water | 1 | Lot | $29,171 | $29,171 |
| | 02 | Plumbing Fixtures & Setting | 1 | Lot | $7,452 | $7,452 |
| | 03 | Sanitary Waste & Vent | | | | |
| | | *Underground* | 1 | lot | $21,238 | $21,238 |
| | | *Aboveground* | 1 | lot | $25,143 | $25,143 |
| | 06 | Domestic Hot & Cold Water Insulation | 1 | lot | $5,038 | $5,038 |
| | 07 | Outside Air Units Air Handling Units, | 1 | lot | $212,845 | $212,845 |
| | 08 | Fans | 1 | lot | $60,803 | $60,803 |
| | 09 | Chilled Water Piping | 1 | lot | $95,493 | $95,493 |
| | 10 | Chilled Water Accessories | 1 | lot | $28,720 | $28,720 |
| | 11 | Process Cooling Water Piping (ICW) | 1 | lot | $34,203 | $34,203 |
| | 12 | Sheet Metal Ductwork | 1 | lot | $100,200 | $100,200 |
| | 13 | Duct Accessories | 1 | lot | $1,667 | $1,667 |
| | 14 | Supply, Return, Exhaust Registers | 1 | lot | $2,227 | $2,227 |
| | 15 | Insulation | 1 | lot | $77,639 | $77,639 |
| | 16 | Controls | 1 | lot | $185,000 | $185,000 |
| | 17 | Install Fire Riser | 1 | lot | $25,000 | $25,000 |
| | 18 | Install Pre Action Valve | 1 | lot | $15,000 | $15,000 |
| | 19 | Install High Bay Sprinklers | 9000 | SF | $8 | $72,000 |
| | 20 | Install Low Bay Sprinklers | 4000 | SF | $8 | $32,000 |
| | 20 | Test Sprinklers | 1 | lot | $5,000 | $5,000 |
| | | | | | | |
| **16** | | **Electrical** | **$1,079,000** | | | |
| | 01 | Cable Tray | 1500 | LF | $74 | $111,000 |
| | 02 | Ground Wire | 400 | LF | $18 | $7,200 |
| | 03 | Ground Rods | 8 | ea. | $690 | $5,520 |
| | 04 | Unit heaters | 2 | ea. | $870 | $1,740 |
| | 05 | Panels | 1 | lot | $40,000 | $40,000 |
| | 06 | Fire Detection | 9000 | SF | $6 | $54,000 |
| | 07 | Lighting | 1 | Lot | $40,000 | $40,000 |
| | 08 | Electrical Distribution | 9000 | Lot | $30 | $270,000 |
| | 09 | Exterior Lighting | 1 | Lot | $50,000 | $50,000 |
| | 10 | Install Switchboard | 2 | EA | $62,500 | $125,000 |
| | 11 | Install 480 V Ductbank | 100 | LF | $563 | $56,250 |
| | 12 | Install Transformers | 2 | EA | $14,375 | $28,750 |
| | 13 | Install Ground Grid Wire | 400 | LF | $18 | $7,000 |
| | 14 | Install Ground Rods | 8 | EA | $690 | $5,520 |
| | 15 | Install Building Bond Connections | 250 | LF | $179 | $44,766 |
| | 16 | Install Panel Feeders | 2 | LS | $12,769 | $25,538 |
| | 17 | Install Distribution Circuits | 2 | LS | $4,460 | $8,920 |
| | 18 | Install PP-208Y/120 VAC | 1 | EA | $5,750 | $5,750 |
| | 19 | Install LP - 480Y/277 VAC | 1 | EA | $10,844 | $10,844 |
| | 20 | Install 45 kVA, 480 - 208Y/277 VAC | 2 | EA | $13,375 | $26,750 |
| | 21 | Install Welding Receptacles | 6 | EA | $955 | $5,730 |
| | 22 | Install 120/208 VAC Receptacles | 16 | EA | $239 | $3,820 |
| | 23 | Install Disconnect Switches | 6 | EA | $540 | $3,240 |
| | 24 | Install Unit Heaters | 4 | EA | $2,000 | $8,000 |
| | 25 | Connect Other mechanical Equipment | 4 | EA | $2,000 | $8,000 |
| | 26 | Install Flour. Lighting | 2200 | SF | $13 | $28,215 |
| | 27 | Install Emergency Lighting (UPS) | 1 | LS | $25,000 | $25,000 |

# FERMILAB FESS COST ESTIMATE

| | | ESTIMATED SUBCONTRACT AWARD AMOUNT | | | | $1,883,000 |
|---|---|---|---|---|---|---|
| Escalation | | | | | | $0 |
| Subcontractor's Overhead and Profit | | 20.0% | | | | $314,000 |
| Difficult Conditions | | | | | | $0 |
| Subcontract Base Estimate | | | | | | $1,569,000 |

| **nuSTORM Conventional Facilities** | | | Project No. | Status: | Date: | Rev Date |
|---|---|---|---|---|---|---|
| *Fixed Priced Construction - Transport Line Enlcosure* | | | **6-13-1** | **PDR** | **13-May-13** | **28-May-13** |
| ITEM NO. | | *DESCRIPTION OF WORK:* | *QUANTITY* | *UNITS* | *UNIT COST* | *AMOUNT* |
| **00** | | **General** | **$9,000** | | | |
| | 01 | Mobilize | 1 | lot | $9,000.00 | $9,000 |
| | | | | | | |
| **02** | | **Site Construction** | **$272,000** | | | |
| | 01 | Silt Fence and Erosion control | 1 | Lot | $10,000 | $10,000 |
| | 02 | Strip Topsoil | 404 | CY | $6 | $2,422 |
| | 03 | Haul Topsoil | 565 | CY | $7 | $3,956 |
| | 04 | Excavate clays for Base Slab | 4998 | CY | $15 | $74,967 |
| | 05 | Haul Clays | 6997 | CY | $7 | $48,978 |
| | 06 | Stone Backfill | 179 | CY | $32 | $5,741 |
| | 07 | Haul and Backfill Clays | 3768 | CY | $25 | $94,189 |
| | 08 | Haul and spread topsoil | 135 | CY | $13 | $1,749 |
| | 09 | Final Grading and seeding | 1 | Acre | $23,800 | $23,800 |
| | 10 | 8' Topsoil Haul and Spread | 100 | CY | $13 | $1,300 |
| | 11 | Seed and Landscaping | 1 | LS | $5,000 | $5,000 |
| | | | | | | |
| **03** | | **Concrete** | **$944,000** | | | |
| | 01 | Target to Ring Base Slab | 174 | CY | $600 | $104,444 |
| | 02 | Target to Ring Walls | 202 | CY | $850 | $171,889 |
| | 03 | Target to Ring top Slab | 174 | CY | $900 | $156,667 |
| | 04 | Primary Abort Base Slab | 111 | CY | $600 | $66,667 |
| | 05 | Primary Abort Walls | 220 | CY | $850 | $187,000 |
| | 06 | Primary Abort Top Slab | 105 | CY | $900 | $94,500 |
| | 07 | Primary Absorber Base Slab | 90 | CY | $600 | $53,778 |
| | 08 | Primary Aborsber walls | 38 | CY | $850 | $32,262 |
| | 09 | Primary Absorber top Slab | 85 | CY | $900 | $76,500 |
| | | | | | | |
| **04** | | **Masonry** | **$0** | | | |
| | | | | | | |
| **05** | | **Metals** | **$0** | | | |
| | | | | | | |
| **06** | | **Wood and Plastic** | | | | $0 |
| | | | | | | $0 |
| **07** | | **Thermal and Moisture Protection** | **$83,000** | | | |
| | 01 | Damp Proof enclosure | 4152 | SF | $8 | $33,216 |
| | 02 | Geo-Membrane at Absorbers | 1 | Lot | $50,000 | $50,000 |
| | | | | | | |
| **08** | | **Doors and Windows** | **$3,000** | | | |
| | 01 | Enclosure Man doors 3 x 7 | 2 | Each | $1,400 | $2,800 |
| | | | | | | $0 |
| **09** | | **Finishes** | **$11,000** | | | $0 |
| | 01 | Painting | 3633 | SF | $3 | $10,899 |
| | | | | | | $0 |

# FERMILAB FESS  COST ESTIMATE

| | | | | | ESTIMATED SUBCONTRACT AWARD AMOUNT | $1,883,000 |
|---|---|---|---|---|---|---|
| Escalation | | | | | | $0 |
| Subcontractor's Overhead and Profit | | 20.0% | | | | $314,000 |
| Difficult Conditions | | | | | | $0 |
| Subcontract Base Estimate | | | | | | $1,569,000 |

| nuSTORM Conventional Facilities | | | Project No. | Status: | Date: | Rev Date |
|---|---|---|---|---|---|---|
| *Fixed Priced Construction - Transport Line Enlcosure* | | | **6-13-1** | **PDR** | **13-May-13** | **28-May-13** |
| ITEM NO. | | DESCRIPTION OF WORK: | QUANTITY | UNITS | UNIT COST | AMOUNT |
| **10** | | **Specialties** | | | | $0 |
| | | | | | | $0 |
| **14** | | **Conveying Equipment** | **$0** | | | $0 |
| | | | | | | |
| **15** | | **Mechanical** | **$44,000** | | | $0 |
| | 01 | Fans | 1 | Each | $13,988 | $13,988 |
| | 06 | Dehumidifiers | 1 | Each | $30,000 | $30,000 |
| | | | | | | |
| **16** | | **Electrical** | **$203,000** | | | |
| | 01 | Cable Tray | 519 | LF | $74 | $38,406 |
| | 02 | Unit heaters | 1 | Each | $870 | $870 |
| | 03 | Panels | 1 | Each | $40,000 | $40,000 |
| | 04 | Fire Detection | 1557 | SF | $6 | $9,342 |
| | 05 | Lighting - Enclosure | 1557 | SF | $13 | $20,241 |
| | 06 | Lighting - Surface Building | 425 | SF | $13 | $5,525 |
| | 07 | Electrical Distribution - Enclosure | 1557 | Lot | $10 | $15,570 |
| | | | | | | |

# FERMILAB FESS COST ESTIMATE

| | | | | | ESTIMATED SUBCONTRACT AWARD AMOUNT | $26,002,000 |
|---|---|---|---|---|---|---|
| Escalation | | | | | | $0 |
| Subcontractor's Overhead and Profit | | 20.0% | | | | $4,334,000 |
| Difficult Conditions | | | | | | $0 |
| Subcontract Base Estimate | | | | | | $21,668,000 |

| nuSTORM Conventional Facilities | | | Project No. | Status: | Date: | Rev Date |
|---|---|---|---|---|---|---|
| *Fixed Priced Construction - Muon Decay Ring Enclosure* | | | 6-13-1 | PDR | 13-May-13 | 28-May-13 |
| ITEM NO. | | DESCRIPTION OF WORK: | QUANTITY | UNITS | UNIT COST | AMOUNT |
| 00 | | **General** | **$10,000** | | | |
| | 01 | Mobilize | | 1 | Lot | $10,000.00 | $10,000 |
| | | | | | | | |
| 02 | | **Site Construction** | **$4,968,000** | | | |
| | 01 | Silt Fence and Erosion control | Enclosure | 1 | Lot | $90,000 | $90,000 |
| | 02 | Strip Topsoil | Enclosure | 7669 | CY | $6 | $46,016 |
| | 03 | Haul Topsoil | Enclosure | 10738 | CY | $7 | $75,165 |
| | 04 | Excavate clays for Base Slab | Enclosure | 94958 | CY | $15 | $1,424,373 |
| | 05 | Haul Clays | Enclosure | 132941 | CY | $7 | $930,588 |
| | 06 | Stone Backfill | Enclosure | 3409 | CY | $32 | $109,075 |
| | 07 | Haul and Backfill Clays | Enclosure | 71583 | CY | $25 | $1,789,586 |
| | 08 | Haul and spread topsoil | Enclosure | 2556 | CY | $13 | $33,234 |
| | 09 | Final Grading and seeding | Enclosure | 8 | Acre | $23,800 | $180,880 |
| | 10 | Backfill Along Building | Enclosure | 2500 | CY | $25 | $62,500 |
| | 11 | Transport Line Enclosure Excavation | Enclosure | 1 | Lot | $30,000 | $30,000 |
| | 12 | Primary Absorber Enclosure Excavation | Enclosure | 1 | Lot | $50,000 | $50,000 |
| | 13 | Pion Absorber Enclosure Excavation | Enclosure | 1 | Lot | $30,000 | $30,000 |
| | 14 | Soil and Erosion Control | Cyro Bldg | 1 | LS | $800 | $800 |
| | 15 | Paving, incl. stone base | Cyro Bldg | 1588 | SY | $40 | $63,520 |
| | 16 | Dewar Tank Foundations | Cyro Bldg | 1 | LS | $22,000 | $22,000 |
| | 17 | Pads for tanks | Cyro Bldg | 1 | LS | $7,500 | $7,500 |
| | 18 | Clear and Grub | Cyro Bldg | 1 | LS | $200 | $200 |
| | 19 | Excavate Topsoil | Cyro Bldg | 220 | CY | $6 | $1,320 |
| | 20 | Haul and Stockpile Topsoil | Cyro Bldg | 277 | CY | $7 | $1,939 |
| | 21 | Excavate for Footings | Cyro Bldg | 350 | CY | $22 | $7,700 |
| | 22 | Haul and Stockpile | Cyro Bldg | 438 | CY | $7 | $3,066 |
| | 23 | 8' Topsoil Haul and Spread | Cyro Bldg | 100 | CY | $13 | $1,300 |
| | 24 | Seed and Landscaping | Cyro Bldg | 1 | LS | $7,500 | $7,500 |
| | | | | | | | |
| 03 | | **Concrete** | **$7,952,000** | | | |
| | 01 | Mud Slab | Enclosure | 352 | CY | $250 | $88,000 |
| | 02 | Base Slab | Enclosure | 2819 | CY | $600 | $1,691,400 |
| | 03 | Walls | Enclosure | 3283 | CY | $850 | $2,790,550 |
| | 04 | Top Slab | Enclosure | 2462 | CY | $900 | $2,215,800 |
| | 05 | Hatch Walls | Enclosure | 61 | CY | $850 | $51,850 |
| | 06 | Hatch Shielding Blocks (1.5'x12'x15') | Enclosure | 120 | CY | $1,100 | $132,000 |
| | 07 | side passageways 86' lg x 2 | Enclosure | 239 | CY | $750 | $179,167 |
| | 08 | center passageway | Enclosure | 268 | CY | $750 | $200,667 |
| | 18 | Pion Absorber Base Slab | Enclosure | 45 | CY | $600 | $27,000 |
| | 19 | Pion Aborsber walls | Enclosure | 20 | CY | $850 | $17,000 |
| | 20 | Pion Absorber top Slab | Enclosure | 40 | CY | $900 | $36,000 |
| | 21 | Stairs | Enclosure | 2 | Each | $175,000 | $350,000 |
| | 22 | Foundations for At-Grade Building | Service Bldg | 200 | CY | $500 | $100,000 |
| | 23 | Slab for At-Grade Building | Service Bldg | 150 | CY | $250 | $37,500 |
| | 24 | 8" Concrete Housekeeping Slabs | Service Bldg | 1 | LS | $4,500 | $4,500 |
| | 25 | FBP Footings | | | | | $0 |
| | | *Foundation - Walls* | Cyro Bldg | 16 | cy | $325 | $5,200 |
| | | *Column Pier - 24" - T/Pier 12" below grade* | Cyro Bldg | 3 | cy | $325 | $975 |
| | | *Footing - Continuous Wall - 12" x 2'-6"* | Cyro Bldg | 16 | cy | $300 | $4,800 |
| | | *Footing - Column Pier - 12" x 4'-0"* | Cyro Bldg | 5 | cy | $300 | $1,500 |
| | 26 | FBP Slab | | | | | $0 |
| | | *Granular Fill at Slab on Grade - 4"* | Cyro Bldg | 24 | cy | $28 | $672 |
| | | *2" Extruded Polystyrene Insulation Under Slab* | Cyro Bldg | 1945 | sf | $1 | $2,334 |

# FERMILAB FESS COST ESTIMATE

| | | ESTIMATED SUBCONTRACT AWARD AMOUNT | | | | $26,002,000 |
|---|---|---|---|---|---|---|
| Escalation | | | | | | $0 |
| Subcontractor's Overhead and Profit | 20.0% | | | | | $4,334,000 |
| Difficult Conditions | | | | | | $0 |
| Subcontract Base Estimate | | | | | | $21,668,000 |

| nuSTORM Conventional Facilities | | | Project No. | Status: | Date: | Rev Date |
|---|---|---|---|---|---|---|
| *Fixed Priced Construction - Muon Decay Ring Enclosure* | | | 6-13-1 | PDR | 13-May-13 | 28-May-13 |
| ITEM NO. | | DESCRIPTION OF WORK: | QUANTITY | UNITS | UNIT COST | AMOUNT |
| | | *15 Mil Vapor Barrier Under Slab* | *Cyro Bldg* | 1945 | sf | $0 | $584 |
| | | *6" Concrete Slab w 6x6 WWF* | *Cyro Bldg* | 1945 | sf | $5 | $9,628 |
| | | *8" Concrete Housekeeping Slabs* | *Cyro Bldg* | 1 | LS | $4,500 | $4,500 |
| | | | | | | | |
| **04** | | **Masonry** | **$32,000** | | | | |
| | 01 | Masonry Walls | *Service Bldg* | 2000 | SF | $13 | $26,000 |
| | 02 | Masonry Walls | *Cyro Bldg* | 450 | SF | $13 | $5,850 |
| | | | | | | | |
| **05** | | **Metals** | **$3,643,000** | | | | |
| | 01 | High Bay Building (SF Cost) | *Service Bldg* | 7500 | SF | 400 | $3,000,000 |
| | 02 | Low Bay Building (SF Cost) | *Service Bldg* | 1000 | SF | 350 | $350,000 |
| | 03 | Crane Girders | *Service Bldg* | 44160 | lbs | 5 | $220,800 |
| | 04 | Misc. Metals | | | | | |
| | | *Steel Girt Channels at Building Mid-Height* | *Cyro Bldg* | 4560 | lbs. | $3 | $11,400 |
| | | *Metal Panel Support Angles* | *Cyro Bldg* | 2692 | lbs. | $3 | $6,730 |
| | | *Steel Channel Opening Framing* | *Cyro Bldg* | 1560 | lbs. | $3 | $3,900 |
| | 05 | Erect Structural Steel | | | | | |
| | | *W12 Columns - Wide Flange* | *Cyro Bldg* | 6240 | lbs. | $3 | $15,600 |
| | | *Beams and Girts- Wide Flange* | *Cyro Bldg* | 11258 | lbs. | $3 | $28,145 |
| | 06 | Erect Roof Deck | *Cyro Bldg* | 2050 | sf | $3 | $6,458 |
| | | | | | | | |
| **06** | | **Wood and Plastic** | | | | | $0 |
| | | | | | | | $0 |
| **07** | | **Thermal and Moisture Protection** | **$1,218,000** | | | | |
| | 01 | Damp Proof enclosure | *Enclosure* | 78888 | SF | $8 | $631,104 |
| | 02 | Geo-Membrane at Absorbers | *Enclosure* | 1 | Lot | $50,000 | $50,000 |
| | 03 | Install Metal Siding | | | | | |
| | | *3" Centrica Formawall* | *Cyro Bldg* | 3180 | sf | $46 | $146,280 |
| | 04 | Install Built-Up Roofing | | | | | $0 |
| | | 2 Layers 1 1/2" Insulation, 1/2" Fiberboard, 4 Ply Felt, White Marble Ballast | *Cyro Bldg* | 2050 | sf | $18 | $36,900 |
| | | *Tapered Insulation* | *Cyro Bldg* | 240 | sf | $3 | $600 |
| | 05 | Install Aluminum Coping | *Cyro Bldg* | 220 | lf | $15 | $3,300 |
| | 06 | Architectural Treatment (10%) | *Service Bldg* | 1 | Lot | $150,000 | $150,000 |
| | 07 | Guiding Principals Compliance (15%) | *Service Bldg* | 1 | Lot | $200,000 | $200,000 |
| | | | | | | | |
| **08** | | **Doors and Windows** | **$96,000** | | | | |
| | 01 | Enclosure Man doors 3 x 7 | *Enclosure* | 7 | Each | $1,400 | $9,800 |
| | 02 | Install Exterior Man Doors | *Service Bldg* | | | | |
| | | *Single Insulated Hollow Metal Door and Frame* | | *2* | *Each* | *$1,400* | *$2,800* |
| | | *Double Insulated Hollow Metal Door and Frame (6'x7')* | | *2* | *Each* | *$1,800* | *$3,600* |
| | 03 | Install Interior Man Doors | *Service Bldg* | | | | |
| | | *Hollow Metal Door and Frame (3'x7')* | | *4* | *Each* | *$1,200* | *$4,800* |
| | 04 | Install Overhead Door | *Service Bldg* | | | | |
| | | *Insulated Coil Door - 18'X14'* | | *2* | *Each* | *$13,000* | *$26,000* |
| | 05 | Install Skylight/ daylighting | *Service Bldg* | 6 | Each | $1,400 | $8,400 |
| | 06 | Install Exterior Man Doors | | | | | $0 |
| | | *Single Insulated Hollow Metal Door and Frame* | *Cyro Bldg* | *2* | *Each* | *$1,400* | *$2,800* |
| | | *Double Insulated Hollow Metal Door and Frame (6'x7')* | *Cyro Bldg* | *1* | *Each* | *$1,800* | *$1,800* |
| | 07 | Install Interior Man Doors | *Cyro Bldg* | 1 | Each | $1,200 | $1,200 |
| | | *Hollow Metal Door and Frame (3'x7')* | *Cyro Bldg* | | | | $0 |
| | 08 | Install Overhead Door | *Cyro Bldg* | 2 | Each | $13,000 | $26,000 |
| | | *Insulated Coil Door - 18'X14'* | *Cyro Bldg* | | | | $0 |
| | 09 | Install Skylight/ daylighting | *Cyro Bldg* | 6 | Each | $1,400 | $8,400 |
| | | | | | | | $0 |

# FERMILAB FESS  COST ESTIMATE

| | | ESTIMATED SUBCONTRACT AWARD AMOUNT | | | | $26,002,000 |
|---|---|---|---|---|---|---|
| Escalation | | | | | | $0 |
| Subcontractor's Overhead and Profit | 20.0% | | | | | $4,334,000 |
| Difficult Conditions | | | | | | $0 |
| Subcontract Base Estimate | | | | | | $21,668,000 |

| nuSTORM Conventional Facilities | | | Project No. | Status: | Date: | Rev Date |
|---|---|---|---|---|---|---|
| *Fixed Priced Construction - Muon Decay Ring Enclosure* | | | **6-13-1** | **PDR** | **13-May-13** | **28-May-13** |
| *ITEM NO.* | | *DESCRIPTION OF WORK:* | *QUANTITY* | *UNITS* | *UNIT COST* | *AMOUNT* |
| **09** | | **Finishes** | **$289,000** | | | **$0** |
| | 01 | Painting | *Enclosure* | 69027 | SF | $3 | $207,081 |
| | 02 | Paint Exposed Ceiling | *Service Bldg* | 8500 | SF | $3 | $25,500 |
| | 03 | Paint Structural Steel | *Service Bldg* | 1 | Lot | $20,000 | $20,000 |
| | 04 | Paint Interior Walls | *Service Bldg* | 1 | Lot | $25,000 | $25,000 |
| | 05 | Paint Exposed Ceiling | *Cyro Bldg* | 2200 | SF | $3 | $5,500 |
| | 06 | Paint Exposed Columns | *Cyro Bldg* | 15 | Each | $400 | $6,000 |
| | | | | | | | $0 |
| **10** | | **Specialties** | | | | **$0** |
| | | | | | | | $0 |
| **14** | | **Conveying Equipment** | **$187,000** | | | **$0** |
| | 01 | 30 Ton Overhead Bridge Crane | *Service Bldg* | 1 | Each | $140,000 | $140,000 |
| | | *Standard Industrial Crane* | *Service Bldg* | | | | |
| | 02 | Install Bridge Crane | *Service Bldg* | 1 | Lot | $25,000 | $25,000 |
| | 03 | Load Test Bridge Crane | *Service Bldg* | 1 | Lot | $15,000 | $15,000 |
| | 04 | Hoists and Trolleys | *Cyro Bldg* | 4 | Each | $1,850 | $7,400 |
| | | | | | | | |
| **15** | | **Mechanical** | **$1,136,000** | | | **$0** |
| | 01 | Domestic Water | *Service Bldg* | 1 | Lot | $30,000 | $30,000 |
| | 02 | Plumbing Fixtures & Setting | *Service Bldg* | 1 | Lot | $10,000 | $10,000 |
| | 03 | Sanitary Waste & Vent | *Service Bldg* | | | | |
| | | *Underground* | | 1 | lot | $21,238 | $21,238 |
| | | *Aboveground* | | 1 | lot | $25,143 | $25,143 |
| | 04 | Domestic Hot & Cold Water Insulation | *Service Bldg* | 1 | lot | $5,038 | $5,038 |
| | 05 | Outside Air Units Air Handling Units, | *Service Bldg* | 1 | lot | $212,845 | $212,845 |
| | 06 | Fans | *Service Bldg* | 4 | Each | $13,988 | $55,952 |
| | 07 | Chilled Water Piping | *Service Bldg* | 1 | lot | $95,493 | $95,493 |
| | 08 | Chilled Water Accessories | *Service Bldg* | 1 | lot | $28,720 | $28,720 |
| | 09 | Process Cooling Water Piping (ICW) | *Service Bldg* | 1 | lot | $34,203 | $34,203 |
| | 10 | Sheet Metal Ductwork | *Service Bldg* | 1 | lot | $100,200 | $100,200 |
| | 11 | Duct Accessories | *Service Bldg* | 1 | lot | $1,667 | $1,667 |
| | 12 | Supply, Return, Exhaust Registers | *Service Bldg* | 1 | lot | $2,227 | $2,227 |
| | 13 | Insulation | *Service Bldg* | 1 | lot | $77,639 | $77,639 |
| | 14 | Controls | *Service Bldg* | 1 | lot | $185,000 | $185,000 |
| | 15 | Install Fire Riser | *Service Bldg* | 1 | lot | $25,000 | $25,000 |
| | 16 | Install Pre Action Valve | *Service Bldg* | 1 | lot | $15,000 | $15,000 |
| | 17 | Install High Bay Sprinklers | *Service Bldg* | 7500 | SF | $8 | $60,000 |
| | 18 | Install Low Bay Sprinklers | *Service Bldg* | 1000 | SF | $8 | $8,000 |
| | 19 | Test Sprinklers | *Service Bldg* | 1 | lot | $5,000 | $5,000 |
| | 20 | Dehumidiefiers | *Enclosure* | 3 | Each | $30,000 | $90,000 |
| | 21 | Install Ventilation System | *Cyro Bldg* | 1 | lot | $3,200 | $3,200 |
| | 22 | Install Controls | *Cyro Bldg* | 1 | lot | $7,000 | $7,000 |
| | 23 | Start-up & Test Ventilation | *Cyro Bldg* | 1 | lot | $1,000 | $1,000 |
| ITEM NO. | 24 | Install Unit Heaters | *Cyro Bldg* | 1 | lot | $11,956 | $11,956 |
| | 25 | Install Fire Riser | *Cyro Bldg* | 1 | lot | $10,000 | $10,000 |
| | 26 | Install  Sprinklers | *Cyro Bldg* | 1950 | SF | $8 | $14,625 |

# FERMILAB FESS COST ESTIMATE

| | | | ESTIMATED SUBCONTRACT AWARD AMOUNT | | | $26,002,000 |
|---|---|---|---|---|---|---|
| Escalation | | | | | | $0 |
| Subcontractor's Overhead and Profit | 20.0% | | | | | $4,334,000 |
| Difficult Conditions | | | | | | $0 |
| Subcontract Base Estimate | | | | | | $21,668,000 |

| **nuSTORM Conventional Facilities** | | | Project No. | Status: | Date: | Rev Date |
|---|---|---|---|---|---|---|
| *Fixed Priced Construction - Muon Decay Ring Enclosure* | | | 6-13-1 | PDR | 13-May-13 | 28-May-13 |
| *ITEM NO.* | | *DESCRIPTION OF WORK:* | *QUANTITY* | *UNITS* | *UNIT COST* | *AMOUNT* |
| **16** | | **Electrical** | **$2,137,000** | | | |
| | 01 | Cable Tray | *Enclosure* | 4931 | LF | $74 | $364,857 |
| | 02 | Unit heaters | *Enclosure* | 9 | Each | $870 | $7,830 |
| | 03 | Panels | *Enclosure* | 2 | Each | $40,000 | $80,000 |
| | 04 | Fire Detection - Surface Building | *Service Bldg* | 9000 | SF | $6 | $54,000 |
| | 05 | Lighting - Enclosure | *Enclosure* | 31140 | SF | $13 | $404,820 |
| | 06 | Lighting - Surface Building | *Service Bldg* | 8500 | SF | $13 | $110,500 |
| | 07 | Electrical Distribution - Enclosure | *Enclosure* | 29583 | Lot | $10 | $295,830 |
| | 08 | Exterior Lighting | *Service Bldg* | 1 | Lot | $50,000 | $50,000 |
| | 09 | Install Switchboard | *Service Bldg* | 2 | EA | $62,500 | $125,000 |
| | 10 | Install 480 V Ductbank | *Service Bldg* | 100 | LF | $563 | $56,250 |
| | 11 | Install Transformers | *Service Bldg* | 2 | EA | $14,375 | $28,750 |
| | 12 | Install Generator | *Service Bldg* | 1 | Each | $20,000 | $20,000 |
| | 13 | Install Ground Grid Wire | *Service Bldg* | 400 | LF | $18 | $7,000 |
| | 14 | Install Ground Rods | *Service Bldg* | 8 | Each | $690 | $5,520 |
| | 15 | Install Building Bond Connections | *Service Bldg* | 250 | LF | $179 | $44,766 |
| | 16 | Install Panel Feeders | *Service Bldg* | 2 | LS | $12,769 | $25,538 |
| | 17 | Install Distribution Circuits | *Service Bldg* | 2 | LS | $4,460 | $8,920 |
| | 18 | Install PP-208Y/120 VAC | *Service Bldg* | 1 | Each | $5,750 | $5,750 |
| | 19 | Install LP - 480Y/277 VAC | *Service Bldg* | 1 | Each | $10,844 | $10,844 |
| | 20 | Install 45 kVA, 480 - 208Y/277 VAC | *Service Bldg* | 2 | Each | $13,375 | $26,750 |
| | 21 | Install Welding Receptacles | *Service Bldg* | 6 | Each | $955 | $5,730 |
| | 22 | Install 120/208 VAC Receptacles | *Service Bldg* | 16 | Each | $239 | $3,820 |
| | 23 | Install Disconnect Switches | *Service Bldg* | 6 | Each | $540 | $3,240 |
| | 24 | Install Unit Heaters | *Service Bldg* | 4 | Each | $2,000 | $8,000 |
| | 25 | Connect Other mechanical Equipment | *Service Bldg* | 4 | Each | $2,000 | $8,000 |
| | 26 | Install Emergency Lighting (UPS) | *Service Bldg* | 1 | LS | $25,000 | $25,000 |
| | 27 | Install Switchboard | *Cyro Bldg* | 1 | Each | $62,500 | $62,500 |
| | 28 | Install 480 V Ductbank | *Cyro Bldg* | 60 | LF | $563 | $33,750 |
| | 29 | Install Transformers | *Cyro Bldg* | 1 | Each | $14,375 | $14,375 |
| | 30 | Install Ground Grid Wire | *Cyro Bldg* | 400 | LF | $18 | $7,000 |
| | 31 | Install Ground Rods | *Cyro Bldg* | 8 | Each | $690 | $5,520 |
| | 32 | Install Building Bond Connections | *Cyro Bldg* | 140 | LF | $179 | $25,069 |
| | 33 | Install Panel Feeders | *Cyro Bldg* | 1 | LS | $12,769 | $12,769 |
| | 34 | Install Distribution Circuits | *Cyro Bldg* | 1 | LS | $4,460 | $4,460 |
| | 35 | Install PP-208Y/120 VAC | *Cyro Bldg* | 1 | Each | $5,750 | $5,750 |
| | 36 | Install LP - 480Y/277 VAC | *Cyro Bldg* | 1 | Each | $10,844 | $10,844 |
| | 37 | Install 45 kVA, 480 - 208Y/277 VAC | *Cyro Bldg* | 1 | Each | $13,375 | $13,375 |
| | 38 | Install Welding Receptacles | *Cyro Bldg* | 4 | Each | $955 | $3,820 |
| | 39 | Install 120/208 VAC Receptacles | *Cyro Bldg* | 16 | Each | $239 | $3,820 |
| | 40 | Install Disconnect Switches | *Cyro Bldg* | 6 | Each | $540 | $3,240 |
| | 41 | Install Unit Heaters | *Cyro Bldg* | 4 | Each | $863 | $3,450 |
| | 42 | Connect Other mechanical Equipment | *Cyro Bldg* | 4 | Each | $863 | $3,450 |
| | 43 | Install  Flour. Lighting | *Cyro Bldg* | 1950 | SF | $13 | $25,009 |
| | 44 | Install Emergency Lighting (UPS) | *Cyro Bldg* | 1 | LS | $13,875 | $13,875 |
| | 45 | Install Exterior Lighting | *Cyro Bldg* | 4 | Each | $656 | $2,625 |
| | 46 | Install Fire Detection | *Cyro Bldg* | 1950 | SF | $12 | $23,400 |

# FERMILAB FESS COST ESTIMATE

| | | ESTIMATED SUBCONTRACT AWARD AMOUNT | | | | $11,750,000 |
|---|---|---|---|---|---|---|
| Escalation | | | | | | $0 |
| Subcontractor's Overhead and Profit | | 20.0% | | | | $1,958,000 |
| Difficult Conditions | | | | | | $0 |
| Subcontract Base Estimate | | | | | | $9,792,000 |

| nuSTORM Conventional Facilities | | | Project No. | Status: | Date: | Rev Date |
|---|---|---|---|---|---|---|
| Fixed Priced Construction - Near Detector | | | 6-13-1 | PDR | 13-May-13 | 28-May-13 |
| ITEM NO. | | DESCRIPTION OF WORK: | QUANTITY | UNITS | UNIT COST | AMOUNT |
| 00 | | General | $10,000 | | | |
| | 01 | Mobilize | 1 | lot | $10,000.00 | $10,000 |
| | | | | | | |
| 02 | | Site Construction | $885,000 | | | |
| | 01 | Silt Fence and Erosion control | 1 | Lot | $50,000 | $50,000 |
| | 02 | Strip Topsoil | 556 | CY | $6 | $3,336 |
| | 03 | Haul Topsoil | 778 | CY | $7 | $5,446 |
| | 04 | Excavate clays for Below Grade Enclosure | 17630 | CY | $15 | $264,450 |
| | 05 | Haul Clays | 24681 | CY | $7 | $172,767 |
| | 06 | Stone Backfill | 3111 | CY | $32 | $99,552 |
| | 07 | Haul and Backfill Clays | 9528 | CY | $25 | $238,200 |
| | 08 | Haul and spread topsoil | 300 | CY | $13 | $3,900 |
| | 09 | Final Grading and seeding | 2 | Acre | $23,800 | $47,600 |
| | | | | | | |
| 03 | | Concrete | $3,246,000 | | | |
| | 01 | Mud Slab | 58 | CY | $250 | $14,500 |
| | 02 | Base Slab | 810 | CY | $600 | $486,000 |
| | 03 | Lower Walls | 718 | CY | $850 | $610,300 |
| | 04 | Upper Walls | 227 | CY | $850 | $226,667 |
| | 05 | Counterfort | 267 | CY | $900 | $240,000 |
| | 06 | Top Slab between counterforts | 267 | CY | $900 | $240,000 |
| | 07 | Hatch Shielding blocks | 584 | CY | $1,100 | $642,400 |
| | 08 | Stairs | 2 | Each | $175,000 | $350,000 |
| | 09 | Elevator Wall and pit | 57 | CY | $900 | $51,200 |
| | 10 | Slab on grade Loading dock | | | | |
| | | Granular Fill at Slab on Grade - 4" | 64 | cy | $28 | $1,792 |
| | | 2" Extruded Polystyrene Insulation Under Slab | 1728 | sf | $1 | $2,074 |
| | | 15 Mil Vapor Barrier Under Slab | 1728 | sf | $0 | $518 |
| | | 5" Concrete Slab w 6x6 WWF | 1728 | sf | $12 | $20,736 |
| | 16 | Slab on grade side bay | | | | |
| | | Granular Fill at Slab on Grade - 4" | 2 | cy | $28 | $47 |
| | | 2" Extruded Polystyrene Insulation Under Slab | 90 | sf | $1 | $108 |
| | | 15 Mil Vapor Barrier Under Slab | 90 | sf | $0 | $27 |
| | | 5" Concrete Slab w 6x6 WWF | 90 | sf | $12 | $1,080 |
| | 21 | Top Level Substructure | | | | |
| | | Spread Footings | 21 | cy | $700 | $14,933 |
| | | Drilled Piers | 7 | Ea | $1,800 | $12,600 |
| | | Strip Footings/Foundation Walls | 414 | cy | $800 | $331,200 |
| | | | | | | |
| 04 | | Masonry | $23,000 | | | |
| | 01 | Masonry Walls | 1800 | SF | $13 | $23,400 |
| | | | | | | |
| | | | | | | |
| 05 | | Metals | $1,121,000 | | | |
| | 01 | Crane Columns | 40960 | Lbs | $3 | $122,880 |
| | 02 | Side Columns | 140000 | Lbs | $3 | $420,000 |
| | 03 | Girts | 1 | Lot | $40,000 | $40,000 |
| | 04 | Second Floor Structure | 5000 | SF | $40 | $200,000 |
| | 05 | Premium girts for CA siding | 1 | Lot | $30,000 | $30,000 |
| | 06 | Roof Girders High Bay | 30400 | Lbs | $3 | $91,200 |

# FERMILAB FESS  COST ESTIMATE

| | ESTIMATED SUBCONTRACT AWARD AMOUNT | | | | $11,750,000 |
|---|---|---|---|---|---|
| Escalation | | | | | $0 |
| Subcontractor's Overhead and Profit | 20.0% | | | | $1,958,000 |
| Difficult Conditions | | | | | $0 |
| Subcontract Base Estimate | | | | | $9,792,000 |

| nuSTORM Conventional Facilities | | Project No. | Status: | Date: | Rev Date |
|---|---|---|---|---|---|
| *Fixed Priced Construction - Near Detector* | | 6-13-1 | PDR | 13-May-13 | 28-May-13 |
| ITEM NO. | DESCRIPTION OF WORK: | QUANTITY | UNITS | UNIT COST | AMOUNT |
| 07 | Roof Beam High Bay | 23520 | Lbs | $3 | $70,560 |
| 08 | Crane Beam | 38304 | Lbs | $3 | $114,912 |
| 09 | Erect Roof Deck | 8664 | SF | $3 | $25,992 |
| 10 | Guardrails | 268 | LF | $21 | $5,628 |
| | | | | | |
| **06** | **Wood and Plastic** | | | | $0 |
| | | | | | $0 |
| **07** | **Thermal and Moisture Protection** | $1,687,000 | | | |
| 01 | Damp Proof enclosure | 20000 | SF | $8 | $160,000 |
| 03 | Install Metal Siding | | | | |
| | 3" Centrica Formawall | 13696 | sf | $79 | $1,081,984 |
| 04 | Install Built-Up Roofing | | | | $0 |
| | 2 Layers 1 1/2" Insulation, 1/2" Fiberboard, 4 Ply Felt, White Marble Ballast | 8664 | sf | $18 | $155,952 |
| | Tapered Insulation | 2166 | sf | $3 | $5,415 |
| 05 | Install Aluminum Coping | 548 | lf | $15 | $8,220 |
| 06 | Architectural Treatment (10%) | 1 | Lot | $125,000 | $125,000 |
| 07 | Guiding Principals Compliance (15%) | 1 | Lot | $150,000 | $150,000 |
| | | | | | |
| **08** | **Doors and Windows** | $51,000 | | | |
| 01 | Enclosure Man doors 3 x 7 | 4 | Each | $1,400 | $5,600 |
| 02 | Install Exterior Man Doors | | | | |
| | *Single Insulated Hollow Metal Door and Frame* | *5* | *Each* | *$1,400* | *$7,000* |
| | *Double Insulated Hollow Metal Door and Frame (6'x7')* | *3* | *Each* | *$1,800* | *$5,400* |
| 03 | Install Interior Man Doors | | | | |
| | *Hollow Metal Door and Frame (3'x7')* | *10* | *Each* | *$1,200* | *$12,000* |
| 04 | Install Overhead Door | | | | |
| | *Insulated Coil Door - 18'X14'* | 1 | Each | $13,000 | $13,000 |
| 05 | Install Skylight/ daylighting | 6 | Each | $1,400 | $8,400 |
| | | | | | $0 |
| **09** | **Finishes** | $135,000 | | | $0 |
| 01 | Painting | *Enclosure* | 20000 | SF | $3 | $60,000 |
| 02 | Paint Exposed Ceiling | 10000 | SF | $3 | $30,000 |
| 03 | Paint Structural Steel | 1 | Lot | $20,000 | $20,000 |
| 04 | Paint Interior Walls | 1 | Lot | $25,000 | $25,000 |
| | | | | | $0 |
| **10** | **Specialties** | | | | $0 |
| | | | | | $0 |
| **14** | **Conveying Equipment** | $525,000 | | | $0 |
| 01 | 30 Ton Overhead Bridge Crane | 2 | Each | $140,000 | $280,000 |
| | *Standard Industrial Crane* | | | | |
| 02 | Install Bridge Crane | 2 | Lot | $25,000 | $50,000 |
| 03 | Load Test Bridge Crane | 2 | Lot | $15,000 | $30,000 |
| 04 | Elevator | 1 | Each | $165,000 | $165,000 |
| | | | | | |

# FERMILAB FESS COST ESTIMATE

| | | | ESTIMATED SUBCONTRACT AWARD AMOUNT | | | $11,750,000 |
|---|---|---|---|---|---|---|
| Escalation | | | | | | $0 |
| Subcontractor's Overhead and Profit | | 20.0% | | | | $1,958,000 |
| Difficult Conditions | | | | | | $0 |
| Subcontract Base Estimate | | | | | | $9,792,000 |

| nuSTORM Conventional Facilities | | | Project No. | Status: | Date: | Rev Date |
|---|---|---|---|---|---|---|
| *Fixed Priced Construction - Near Detector* | | | 6-13-1 | PDR | 13-May-13 | 28-May-13 |
| ITEM NO. | | DESCRIPTION OF WORK: | QUANTITY | UNITS | UNIT COST | AMOUNT |
| **15** | | **Mechanical** | **$1,143,000** | | | **$0** |
| | 01 | Domestic Water | 1 | Lot | $30,000 | $30,000 |
| | 02 | Plumbing Fixtures & Setting | 1 | Lot | $10,000 | $10,000 |
| | 03 | Sanitary Waste & Vent | | | | |
| | | *Underground* | 1 | lot | $21,238 | $21,238 |
| | | *Aboveground* | 1 | lot | $25,143 | $25,143 |
| | 04 | Domestic Hot & Cold Water Insulation | 1 | lot | $5,038 | $5,038 |
| | 05 | Outside Air Units Air Handling Units, | 1 | lot | $212,845 | $212,845 |
| | 06 | Fans | 4 | Each | $13,988 | $55,952 |
| | 07 | Chilled Water Piping | 1 | lot | $95,493 | $95,493 |
| | 08 | Chilled Water Accessories | 1 | lot | $28,720 | $28,720 |
| | 09 | Process Cooling Water Piping (ICW) | 1 | lot | $34,203 | $34,203 |
| | 10 | Sheet Metal Ductwork | 1 | lot | $100,200 | $100,200 |
| | 11 | Duct Accessories | 1 | lot | $1,667 | $1,667 |
| | 12 | Supply, Return, Exhaust Registers | 1 | lot | $2,227 | $2,227 |
| | 13 | Insulation | 1 | lot | $77,639 | $77,639 |
| | 14 | Controls | 1 | lot | $185,000 | $185,000 |
| | 15 | Install Fire Riser | 1 | lot | $50,000 | $50,000 |
| | 16 | Install Pre Action Valve | 1 | lot | $15,000 | $15,000 |
| | 17 | Install High Bay Sprinklers | 7500 | SF | $8 | $60,000 |
| | 18 | Install Low Bay Sprinklers | 1000 | SF | $8 | $8,000 |
| | 19 | Test Sprinklers | 1 | lot | $5,000 | $5,000 |
| | 20 | Dehumidiefiers | 4 | Each | $30,000 | $120,000 |
| | | | | | | |
| **16** | | **Electrical** | **$966,000** | | | |
| | 02 | Unit heaters | 10 | Each | $870 | $8,700 |
| | 03 | Panels | 2 | Each | $40,000 | $80,000 |
| | 04 | Fire Detection - Surface Building | 8000 | SF | $6 | $48,000 |
| | 05 | Lighting - Below Grade | 8000 | SF | $13 | $104,000 |
| | 06 | Lighting - Surface Building | 8000 | SF | $13 | $104,000 |
| | 07 | Electrical Distribution - Below Grade | 8000 | Lot | $10 | $80,000 |
| | 08 | Exterior Lighting | 1 | Lot | $50,000 | $50,000 |
| | 09 | Install Switchboard | 2 | EA | $62,500 | $125,000 |
| | 10 | Install 480 V Ductbank | 100 | LF | $563 | $56,250 |
| | 11 | Install Transformers | 2 | EA | $14,375 | $28,750 |
| | 12 | Install Generator | 1 | Each | $20,000 | $20,000 |
| | 13 | Install Ground Grid Wire | 400 | LF | $18 | $7,000 |
| | 14 | Install Ground Rods | 8 | Each | $690 | $5,520 |
| | 15 | Install Building Bond Connections | 250 | LF | $179 | $44,766 |
| | 16 | Install Panel Feeders | 2 | LS | $12,769 | $25,538 |
| | 17 | Install Distribution Circuits | 2 | LS | $4,460 | $8,920 |
| | 18 | Install PP-208Y/120 VAC | 1 | Each | $5,750 | $5,750 |
| | 19 | Install LP - 480Y/277 VAC | 1 | Each | $10,844 | $10,844 |
| | 20 | Install 45 kVA, 480 - 208Y/277 VAC | 2 | Each | $13,375 | $26,750 |
| | 21 | Install Welding Receptacles | 6 | Each | $955 | $5,730 |
| | 22 | Install 120/208 VAC Receptacles | 16 | Each | $239 | $3,820 |
| | 23 | Install Disconnect Switches | 6 | Each | $540 | $3,240 |
| | 24 | Install Unit Heaters | 4 | Each | $2,000 | $8,000 |
| | 25 | Connect Other mechanical Equipment | 4 | Each | $2,000 | $8,000 |
| | 26 | Install Emergency Lighting (UPS) | 1 | LS | $25,000 | $25,000 |

# FERMILAB FESS COST ESTIMATE

| | | ESTIMATED SUBCONTRACT AWARD AMOUNT | | | | $720,000 |
|---|---|---|---|---|---|---|
| Escalation | | | | | | $0 |
| Subcontractor's Overhead and Profit | | 20.0% | | | | $120,000 |
| Difficult Conditions | | 10.0% | | | | $55,000 |
| Subcontract Base Estimate | | | | | | $545,000 |

| nuSTORM Conventional Facilities | | | Project No. | Status: | Date: | Rev Date |
|---|---|---|---|---|---|---|
| *Fixed Priced Construction - Far Detector* | | | **6-13-1** | **PDR** | **13-May-13** | **28-May-13** |
| ITEM NO. | | *DESCRIPTION OF WORK:* | *QUANTITY* | *UNITS* | *UNIT COST* | *AMOUNT* |
| **00** | | **General** | **$120,000** | | | |
| | 01 | Mobilize | 1 | Lot | $5,000 | $5,000 |
| | 02 | Demolish/Remove Existing Counting House | 1 | Lot | $65,000 | $65,000 |
| | 03 | Demolish/Remove Existing Clean Room | 1 | Lot | $50,000 | $50,000 |
| | | | | | | $0 |
| **02** | | **Civil** | **$0** | | | |
| | | | | | | |
| **03** | | **Concrete** | **$0** | | | |
| | | | | | | $0 |
| **04** | | **Masonry** | **$0** | | | |
| | | | | | | $0 |
| **05** | | **Metals** | **$0** | | | |
| | | | | | | $0 |
| **06** | | **Wood and Plastic** | **$0** | | | |
| | | | | | | $0 |
| **07** | | **Thermal and Moisture Protection** | **$0** | | | |
| | | | | | | |
| **08** | | **Doors and Windows** | **$0** | | | |
| | | | | | | $0 |
| **09** | | **Finishes** | **$0** | | | |
| | | | | | | $0 |
| **10** | | **Specialties** | **$0** | | | |
| | | | | | | $0 |
| **14** | | **Conveying Equipment** | **$0** | | | |
| | | | | | | $0 |
| **15** | | **Mechanical** | **$350,000** | | | |
| | 01 | Cooling Tower System | 1 | Lot | $350,000 | $350,000 |
| | | *Sized for two cyro compressors* | | | | $0 |
| | | *Based on AZero System for mu2e* | | | | $0 |
| | | *Includes water treatment system* | | | | |
| | | | | | | |
| **16** | | **Electrical** | **$75,000** | | | |
| | 01 | Electrical Service Extension to Detector | 1 | Lot | $75,000 | $75,000 |
| | | | | | | $0 |

# FERMILAB FESS COST ESTIMATE

| | | | ESTIMATED SUBCONTRACT AWARD AMOUNT | | | $12,233,000 |
|---|---|---|---|---|---|---|
| Escalation | | | | | | $0 |
| Subcontractor's Overhead and Profit | | 20.0% | | | | $2,039,000 |
| Difficult Conditions | | | | | | $0 |
| Subcontract Base Estimate | | | | | | $10,194,000 |

| nuSTORM Conventional Facilities | | | Project No. | Status: | Date: | Rev Date |
|---|---|---|---|---|---|---|
| *Fixed Priced Construction - Site Work* | | | 6-13-1 | PDR | 13-May-13 | 28-May-13 |
| ITEM NO. | | *DESCRIPTION OF WORK:* | *QUANTITY* | *UNITS* | *UNIT COST* | *AMOUNT* |
| **00** | | **General** | **$260,000** | | | |
| | 01 | Mobilize | 1 | Lot | $10,000 | $10,000 |
| | 02 | Wetland Permits | 1 | Lot | $100,000 | $100,000 |
| | 03 | Wetland Mitigation | 1 | Lot | $150,000 | $150,000 |
| | | | | | | $0 |
| **02** | | **Civil** | **$6,778,000** | | | |
| | 01 | Silt Fence and Erosion Control Measures | 9,750 | LF | $8 | $78,000 |
| | 02 | Roads | | | | |
| | | *Strip and haul soils* | 5,556 | CY | $12 | $66,667 |
| | | *Aggregate Base* | 5000 | SF | $3 | $15,000 |
| | | *Binder and surface course* | 60000 | SF | $2 | $120,000 |
| | 03 | Drives and parking | | | | |
| | | *Strip and haul soils* | 9,574 | SF | $2 | $517,000 |
| | | *Aggregate Base* | 258500 | SF | $3 | $775,500 |
| | | *Binder and surface course* | 258500 | SF | $2 | $517,000 |
| | 04 | Hardstands | 1667 | CY | $30 | $50,000 |
| | 05 | Ponds | | | | |
| | | *Excavate for ponds* | 39000 | CY | $15 | $585,000 |
| | | *Liner* | 198900 | SF | $8 | $1,591,200 |
| | | *12" clay over liner* | 7540 | CY | $25 | $188,500 |
| | | *Stone banks* | 867 | CY | $85 | $73,667 |
| | | *Pumps and pump station* | 2 | Each | $400,000 | $800,000 |
| | | *Intake and discharge structures* | 4 | Each | $40,000 | $160,000 |
| | | *Pressure Piping 12' HDPE* | 1200 | LF | $140 | $168,000 |
| | | *Gravity Piping* | 750 | LF | $110 | $82,500 |
| | 06 | Creek Diversion | 1 | Lot | $50,000 | $50,000 |
| | 07 | Piped utilities | | | | |
| | | *ICW 10" Main* | 2700 | LF | $60 | $162,000 |
| | | *ICW 6" Laterals* | 240 | LF | $45 | $10,800 |
| | | *DWS* | 2700 | LF | $45 | $121,500 |
| | | *Sanitary Sewer* | 4600 | LF | $35 | $161,000 |
| | | *Sanitary Lift Stations* | 2 | Each | $40,000 | $80,000 |
| | | *Hydrants* | 10 | Each | $2,500 | $25,000 |
| | | *Valves and fittings* | 45 | Each | $1,000 | $45,000 |
| | 08 | Storm Drainage Structures | 1 | Lot | $100,000 | $100,000 |
| | 09 | Seeding and Landscaping | 23.5 | Acres | $10,000 | $235,000 |
| | | | | | | |
| **03** | | **Concrete** | **$350,000** | | | |
| | 01 | Substation Pads with Containment | 5 | Each | $50,000 | $250,000 |
| | 02 | Walks and Aprons | 200 | CY | $500 | $100,000 |
| | | | | | | $0 |

# FERMILAB FESS COST ESTIMATE

| | | ESTIMATED SUBCONTRACT AWARD AMOUNT | | | | $12,233,000 |
|---|---|---|---|---|---|---|
| Escalation | | | | | | $0 |
| Subcontractor's Overhead and Profit | 20.0% | | | | | $2,039,000 |
| Difficult Conditions | | | | | | $0 |
| Subcontract Base Estimate | | | | | | $10,194,000 |

| nuSTORM Conventional Facilities | | | Project No. | Status: | Date: | Rev Date |
|---|---|---|---|---|---|---|
| *Fixed Priced Construction - Site Work* | | | **6-13-1** | **PDR** | **13-May-13** | **28-May-13** |
| **04** | **Masonry** | **$0** | | | | |
| | | | | | | $0 |
| **05** | **Metals** | | | | | |
| | | | | | | $0 |
| **06** | **Wood and Plastic** | | | | | |
| | | | | | | $0 |
| **07** | **Thermal and Moisture Protection** | | | | | |
| | | | | | | $0 |
| **09** | **Finishes** | | | | | |
| | | | | | | $0 |
| **10** | **Specialties** | | | | | |
| | | | | | | $0 |
| **14** | **Conveying Equipment** | | | | | |
| | | | | | | $0 |
| **15** | **Mechanical** | **$0** | | | | |
| | | | | | | |
| **16** | **Electrical** | **$2,806,000** | | | | |
| | 01 | Temp. Power Pole Line | 1 | Lot | $75,000 | $75,000 |
| | 02 | Electrical Duct Banks | 2200 | LF | $250 | $550,000 |
| | 03 | 15kv feeder cable | 4400 | LF | $32 | $140,800 |
| | 04 | Termination and splices | 25 | Each | $250 | $6,250 |
| | 05 | Transformers 750 KVA | 4 | Each | $85,000 | $340,000 |
| | 06 | Transformers 1500 KVA | 4 | Each | $130,000 | $520,000 |
| | 07 | Secondary ducts and wires | 280 | LF | $600 | $168,000 |
| | 08 | 15 KV Switchgear | 5 | Each | $22,000 | $110,000 |
| | 09 | Generators | 4 | Each | $60,000 | $240,000 |
| | 10 | Kautz Rd Substation Upgrade | 1 | Lot | $300,000 | $300,000 |
| | 11 | Comm Duct Bank | 2200 | LF | $100 | $220,000 |
| | 12 | Electrical Manholes | 4 | Each | $25,000 | $100,000 |
| | 13 | Exterior road lighting | 20 | Each | $1,800 | $36,000 |
| | | | | | | $0 |

# Fermi National Accelerator Laboratory
## FY13 Provisional Labor, Indirect and Shop Rates

**Consult your Division/Section/Project Field Financial Manager for the proper application of these rates to costs.**

### Labor Burdens

| | Labor Burden Rate is applied to: | | |
|---|---|---|---|
| Vacation | *Monthly* Time Worked | 10.25% | |
| Vacation | *Weekly* Time Worked | 11.50% | |
| | | | |
| OPTO (Other Paid Time Off) | *Monthly* Time Worked | 6.50% | |
| OPTO (Other Paid Time Off) | *Weekly* Time Worked | 9.00% | |
| | | | |
| Fringe | Time Worked + Vacation + OPTO | 32.75% | |
| | | | |
| Effective Labor (Vacation, OPTO, Fringe) | *Monthly* Time Worked | *54.99%* | (1) |
| Effective Labor (Vacation, OPTO, Fringe) | *Weekly* Time Worked | *59.96%* | (1) |
| | | | |
| Summer/Temp Fringe | | 8.00% | |

### Indirect Rates

| | | | | Non HEP & Users Excludes TSCS |
|---|---|---|---|---|
| PS (Program Support) | *Accelerator Division* | 27.5% | | |
| PS (Program Support) | *Accelerator Division - Effective Rate (PS, CSS, TSCS & G&A)* | *99.29%* | (2) | *94.05%* |
| PS (Program Support) | *Computing Division* | 10.2% | | |
| PS (Program Support) | *Computing Division - Effective Rate (PS, CSS, TSCS & G&A)* | *72.25%* | (2) | *67.72%* |
| PS (Program Support) | *Particle Physics Division* | 13.0% | | |
| PS (Program Support) | *Particle Physics Division - Effective Rate (PS, CSS, TSCS & G&A)* | *76.63%* | (2) | *71.98%* |
| PS (Program Support) | *Technical Division* | 20.2% | | |
| PS (Program Support) | *Technical Division - Effective Rate (PS, CSS, TSCS & G&A)* | *87.88%* | (2) | *82.94%* |
| | | | | |
| PS (Program Support) | *Fully Loaded **Monthly** Time Worked - **Accelerator Division*** | *208.87%* | (3) | *200.75%* |
| PS (Program Support) | *Fully Loaded **Weekly** Time Worked - **Accelerator Division*** | *218.79%* | (3) | *210.41%* |
| PS (Program Support) | *Fully Loaded **Monthly** Time Worked - **Computing Division*** | *166.96%* | (3) | *159.94%* |
| PS (Program Support) | *Fully Loaded **Weekly** Time Worked - **Computing Division*** | *175.54%* | (3) | *168.29%* |
| PS (Program Support) | *Fully Loaded **Monthly** Time Worked - **Particle Physics Division*** | *173.75%* | (3) | *166.55%* |
| PS (Program Support) | *Fully Loaded **Weekly** Time Worked - **Particle Physics Division*** | *182.54%* | (3) | *175.11%* |
| PS (Program Support) | *Fully Loaded **Monthly** Time Worked - **Technical Division*** | *191.19%* | (3) | *183.53%* |
| PS (Program Support) | *Fully Loaded **Weekly** Time Worked - **Technical Division*** | *200.54%* | (3) | *192.64%* |
| | | | | |
| MSA (Material/Services Acquisition) | | 5.2% | | |
| MSA (Material/Services Acquisition) | *Effective Rate (MSA, TSCS & G&A)* | *22.90%* | (4) | *19.67%* |
| | | | | |
| CSS (Common Site Support) | | 33.80% | | |
| CSS (Common Site Support) | *Effective Rate - Non-Divisional (CSS, TSCS & G&A)* | *56.31%* | (5) | *52.20%* |
| CSS (Common Site Support) | *Fully Loaded **Monthly** Time Worked - **Non-Divisional*** | *142.25%* | (6) | *135.88%* |
| CSS (Common Site Support) | *Fully Loaded **Weekly** Time Worked - **Non-Divisional*** | *150.03%* | (6) | *143.46%* |
| | | | | |
| TSCS (Technical and Scientific Common Support) | *(applied to HEP funded accounts only, excludes Users)* | 2.7% | | |
| | *Effective Rate - Non-Divisional (TSCS & G&A)* | 16.82% | | |
| | | | | |
| G&A (General and Administrative) | | 13.75% | | |
| | | | | |
| Pass-Through | | 1.5% | | |

The above rates have been submitted to the U. S. Department of Energy (DOE) and have been neither approved nor disapproved. The Laboratory's current Cost Accounting Standards Disclosure Statement has been approved by DOE. All the above rates are subject to adjustment to actual at least once per year in September. History of adjustments available on Accounting website.

**Shop Chargeback Rates \***

| | | |
|---|---|---|
| | Machine Shop | $75 |
| | FESS Construction Engineering | $96 |

\* The chargeback rates are subject to the CSS, TSCS and G&A indirect rates. Machine Shop is also subject to the TD PS rate.

Approved: 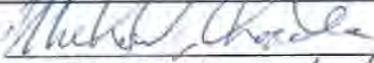 Approved: 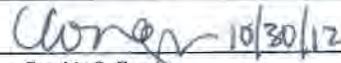 10/30/12

| Michael F. Rhoades | 10/29/12 | Cynthia S. Conger |
|---|---|---|
| Chief Accounting Officer | | Chief Financial Officer |

(1) - Effective rate on Time Worked after fringe rate applied to time worked + vacation + OPTO.
(2) - Effective rate on total labor cost (see (1) above) after PS, CSS, TSCS & G&A rate applied. Shop charges are also subject to this effective rate.
(3) - Effective rate on Time Worked after PS, CSS, TSCS & G&A rate applied to total labor cost (see (1) above).
(4) - Effective rate on M&S costs after MSA, TSCS & G&A rate applied.
(5) - Effective rate on total labor cost (see (1) above) after CSS, TSCS & G&A rate applied. Shop charges are also subject to this effective rate.
(6) - Effective rate on Time Worked after CSS, TSCS & G&A rate applied to total labor cost (see (1) above).

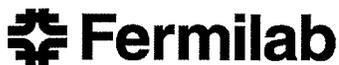

# Fermi National Accelerator Laboratory
# Indirect Burden Allocation Policy and Methodology

## I.     General Policy

It is the policy of the Laboratory to allocate indirect expenses to all final cost objectives. The types of indirect expenses allocated are Program Support (PS), Material/Service Acquisition (MSA), Common Site Support (CSS), Technical and Scientific Common Support (TSCS), and General and Administrative Expenses (G&A). All costs incurred are subject to the indirect burden allocation (as defined in Section III of this document) including the total cost of goods and services procured through Fermilab by user institutions or other external entities. This policy is consistent with the requirements of our Prime Contract with DOE to comply with Cost Accounting Standards (CAS), as well as the requirements of DOE Order 522.1, "Pricing of Departmental Materials and Services".

The Budget Office and the Accounting Department develop provisional indirect burden rates jointly. Normally provisional rates are based upon the Laboratory's current year budget of direct and indirect costs. However, in the case where the approved budget for the upcoming year has not been finalized, or in the case of a new methodology with which the Lab has little or no experience, the rates may be based on prior year(s) actual rates or projections.

The provisional rates will go into effect on the first day of the fiscal year, with a retroactive adjustment to actual (variance distribution) at least annually at year end. Major changes in funding, budgetary allocations, or the Lab's indirect burden allocation methodology could necessitate a rate change (increase or decrease) and/or a variance distribution during the year; such rate changes and variance distributions are subject to CFO approval. CAS requires rate changes to be retroactive to the beginning of the fiscal year.

The Accounting Department performs a monthly analysis of indirect allocations and monitors the rates and accumulated variances.

## II.     Exceptions

The indirect burden allocation rates are generally applied to all final cost objectives with the following exceptions as stated below:

A.      When Laboratory personnel provide a service (labor cost) while away from the Laboratory for a period of more than 180 days, the Common Site Support rate will not be charged upon request of applicable Division/Section management. The labor must be segregated into a task to be established, not subject to CSS.

B.      When special conditions or extenuating circumstances exist, the amount or applicability of the indirect charge to a collaborating institution may be negotiated with the Laboratory, with approval of the CFO. When an adjustment is granted, said adjustment should be included in the "Agreement for an Experiment", Memorandum of Understanding, or other formal document as appropriate, and such document should be provided to the Accounting Department before charges are incurred. If costs billed to a user institution will not include indirect charges, those charges must be transferred to the benefiting Laboratory B&R code, which can be determined by the purpose stated in the Agreement. The transfer is completed as part of the month-end burdening process where indirect costs are mirrored to the appropriate lab B&R. It should be noted that the use of this exception may cause difficulty in explaining the nature of these charges to DOE program managers and/or other reviewers of the benefiting B&R.

C.      Large Cap Purchase Order Indirect Cost Limitation:
Purchase orders with a single procurement action of greater than $500,000 will have indirect costs capped on $500,000 of costs. This exception recognizes that the causal-beneficial relationship between the procurement effort and procurement output is not linear (i.e. a typical $2 million order does not require four times the effort to place as a typical

$500,000 order).

All purchase orders entitled to the Indirect Cost Cap should be written against an exempt expenditure type as shown below.

Exempt – Civil Construction
Exempt – Equipment Rental
Exempt – Equipment
Exempt – Fabrication Procurement

Exempt – Material Purchases
Exempt – Subcontract Services Pass Thru
Exempt – Subcontract Services
Exempt – Temporary Help

The exempt expenditure types are excluded from the Indirect Burden Allocation process that is run at month-end; they do not receive a system generated indirect charge. Instead, the indirect charge for a purchase order written against any of these expenditure types is manually computed on the first $500,000 and recorded in Oracle Projects during the month that the first vendor invoice against the order is processed. Allocation of the indirect cost across Project/Task, in the case of split-coded orders, is based upon the P.O. distribution reported in Oracle eBS at the time of receipt of the first invoice.

The following table summarizes conditions under which a procurement action may or may not qualify for treatment under the Large Cap exception.

| 1. | Initial P.O.'s > $500K qualify. |
|----|---------------------------------|
| 2. | Change orders > $500K qualify. Total order amount is subject to the cap. |
| 3. | Change orders < $500K qualify if the original P.O. is > $500K. Total order amount is subject to the cap. |
| 4. | Total orders that are over $500K as a result of original P.O. and subsequent change orders under $500K do not qualify. Orders approved as Pass-Through orders (see below) are not subject to this requirement. |
| 5. | Original total orders that are over $500K as a result of multiple requisitions < $500K placed on one purchase order qualify. |
| 6. | Master Contract releases > $500K qualify but releases < $500K do not qualify. |

D.      Pass Through Orders:

Purchase orders that meet the definition of a Pass-Through purchase order are eligible for a reduced indirect rate of 1.5%. Pass-through orders are agreements that transfer funds to a program's/project's collaborating institution to carry out agreed-upon work. The reduced rate reflects the reduced indirect effort associated with such a procurement action. The reduced effort is primarily due to the preexisting documented agreement between the institutions, including deliverables, milestones, and due dates, most commonly through a Memorandum of Understanding.

A request for pass-through action must have CFO approval. All purchase requisitions representing approved pass-through actions must be recorded against expenditure type Subcontract Services Pass-Thru or Exempt - Subcontract Services Pass-Thru (further discussed below). Requisitions requesting pass-through treatment must include evidence of the collaborative character of the work with respect to the program/project and must generally meet the condition of requiring reduced indirect effort.

Application of the pass-through rate is subject to the $500,000 ceiling like any other large cap order. Purchase requisitions representing a pass-through action in excess of $500,000 should be recorded against the exempt expenditure type, Exempt - Subcontract Services Pass-Thru, instead of expenditure type Subcontract Services Pass-Thru.

E.    Restricted/Exempt Burden Schedules:

There are tasks that fall under specific Service Types that use a restricted or exempt burden schedule due to special circumstances. Examples include Inventory and certain costs unallowable on the contract.

# III.    Indirect Burden Rate Calculation

The rates for the provisional PS, MSA, CSS, TSCS, and G&A burden allocations will be based on the following formulas, where the numerator is the pool, and the denominator is the distribution base of the pool:

## A.    Program Support Burden Allocation

Each of our four Divisions (Accelerator, Computing, Particle Physics, and Technical) has a Program Support allocation to cover the costs associated with central division administration, departmental management, central computing support, certain travel and training, and related costs, as well as similar costs that exist in the Centers that are associated with those Divisions.

1. **Accelerator Division – (Includes APC)**
   B&R Code:  EC0101070 – AD Program Support
   Service Type:  OP-BURDEN-PS AD
   Expenditure Type: AD Program Support Allocation

2. **Computing Division**
   B&R Code: EC0101080 – CD Program Support
   Service Type:  OP-BURDEN-PS CD
   Expenditure Type:  CD Program Support Allocation

3. **Particle Physics Division – (Includes FCPA & CMS Center)**
   B&R Code: EC0101090 – PPD Program Support
   Service Type:  OP-BURDEN-PS PPD
   Expenditure Type: PPD Program Support Allocation

4. **Technical Division**
   B&R Code: EC0101010 – TD Program Support
   Service Type:  OP-BURDEN-PS TD
   Expenditure Type: TD Program Support Allocation

**Pool** – Each Division's Pool contains costs from the following activities as applicable:

| | |
|---|---|
| D/S/C Administration | Project Oversight |
| Document Workflow & Information Systems | Transportation, Distribution, Fleet |
| Financial Management | Quality Assurance |
| Human Resources | Safety Services |
| IT Governance and Oversight | |

The Centers have pool costs that are primarily D/S/C Administration.

**Base** – The Program Support burden allocation will be applied to the following expenditure types of the respective division:

| | |
|---|---|
| Accounting Transfers – Labor | Overtime |
| Construction Engineering | Service Organization Distribution |
| EOM Wage Accrual – Monthly | Special Compensation |
| EOM Wage Accrual – Summer/Temp | Summer/Temp Emp. Monthly |
| EOM Wage Accrual – Weekly | Summer/Temp Emp. Weekly |
| Fringe – Special | Time Worked – Monthly |
| Machine Shop | Time Worked – Weekly |

Vacation, OPTO, and Fringe allocations are also included in the base.
The base excludes costs charged to the PS burden pools themselves (i.e. to tasks with the service types of OP-BURDEN-PS AD, OP-BURDEN-PS CD, OP-BURDEN-PS PPD, or OP-BURDEN-PS TD.)

## B.     MSA Burden Allocation

The Material/Service Acquisition (MSA) burden pool represents the cost of purchasing services and materials. This includes a range of costs, from the cost of negotiation and execution of a contract to the cost of processing an invoice.

B&R Code: EC0101040 – Materials/Service and Acquisition
Service Type:  OP-BURDEN-MSA
Expenditure Type:  MSA Allocation

*Pool* – The MSA pool consists of all the costs from the following activities:

| | |
|---|---|
| Purchasing | Stock Room |
| Accounts Payable | Inventory Variances |
| Shipping/Receiving | |

*Base* –The MSA burden allocation will be applied to the following expenditure types:

| | |
|---|---|
| Civil Construction | Material Purchases |
| Computer Maintenance | Office Machine Maintenance |
| Computers, Desktop | ProCard Purchases |
| Computers, Hardware Maintenance | Professional Services |
| Computers, Software Licenses | Purchased Services |
| Computers. Software Maintenance | Spare parts/Other Issues |
| Computer Services Distribution | Special Process Spares Issues |
| Demurrage/Container Rental | Stores Issues |
| Equipment | Sub Contract Services |
| Equipment Rental | T&M Construction Services |
| Fabrication Procurement | T&M Electrical Services |
| Facility Rental | T&M Pipe Fitters |
| Freight | T&M Rigging Services |

| Gases/ Cryogenic Fluids | Telephone Expense |
|---|---|
| Honoraria | Telephone Expense Distribution |
| Insurance Premium | Temporary Help |

The base excludes costs charged to the MSA pool and burden pools previously allocated (i.e. to tasks with the service types of OP-BURDEN-PS AD, OP-BURDEN-PS CD, OP-BURDEN-PS PPD, or OP-BURDEN-PS TD, or OP-BURDEN-MSA.)

## C.    CSS Burden Allocation
The Common Site Support (CSS) burden represents the cost of running the physical facility and infrastructure (power, facility maintenance and management, computing infrastructure, telecommunications, mail service, etc.).

B&R Code:  EC0101050 – Common Site Support
Service Type:  OP-BURDEN-CSS
Expenditure Type:  CSS Allocation

*Pool* – The CSS pool consists of costs from the following activities:

Building Maintenance and Management Facilities
Engineering Services Section
ES&H Section
Accommodations
Computing Support
Farm Income
Food Service
Information Resources
Mail Operations
Non-programmatic building maintenance and utilities

Non-programmatic Power
Property Management
Telecommunications
Transportation
Travel Office
Vehicle Maintenance
Audio-Visual/Duplicating/Photo Services
Variances from the Service Centers
PS and MSA allocations on the costs in the CSS pool

Cyber Security other than:
Assignment of an Information Systems Security Manager
Assignment of a Certification Agent(s) – includes forensics analyst
Assignment of a full time Information System Security Officer (s)
SC HQ-mandated Security Implementations

*Base* – The CSS burden allocation will be applied to the expenditure types listed within the PS section of this document.

Vacation, OPTO, and Fringe labor burdens, as well as PS burden allocations, are also included in the base.

The base excludes costs charged to the CSS pool and pools previously allocated (i.e. to tasks with the service types of OP-BURDEN-PS AD, OP-BURDEN-PS CD, OP-BURDEN-PS PPD, or OP-BURDEN-PS TD, OP-BURDEN-MSA, or OP-BURDEN-CSS).

## D.    Technical and Scientific Common Support Allocation
The Technical and Scientific Common Support (TSCS) burden represents costs incurred by the Accelerator, Particle Physics and Technical Divisions associated with providing technical and scientific services necessary to support the high energy physics program at the laboratory. These include costs for Engineering Support, Vacuum and Fluids, the Silicon Detector, Carbon Fiber, Extrusion and Cryogenics facilities, Magnet Systems and Magnet Test facilities, and Meson Detector Building Test Facility.

B&R Code: EC-01-01-020
Service Type – OP-BURDEN TSCS
Expenditure Type: TSCS Burden Allocation

*Pool* – Each division above has pool costs from the following activities when associated with the support of technical and scientific needs:

| | |
|---|---|
| Accelerator Controls & Instrumentation | Technical Facilities Operations/Improvements |
| Accelerator Operations | PS, MSA and CSS allocations on the costs in the |
| Technical Centers Management | TSCS pool |

*Base* – The TSCS burden allocation will be applied to the expenditure types listed within the PS and MSA sections of this document.

Vacation, OPTO, and Fringe labor burdens, as well as PS, MSA and CSS burden allocations, are also included in the base.

Additionally, the base includes the following expenditure types:

| | |
|---|---|
| Accounting Transfers | Relocation |
| Donated Funds | Relocation – Temporary |
| Duplicating Services | Special Events |
| Educational Expense | Stipend Education |
| Fees | Stores & Spares loss / OBS |
| Housing Costs | Training |
| On-Site Travel Reimbursement | Travel, Domestic, Lab Employee |
| Other Costs and Credits | Travel, Domestic, Non Employee |
| Other Utilities | Travel, Foreign, Lab Employee |
| Photo/Graphic Services | Travel, Foreign, Non Employee |
| Physical Inventory Adj | Vehicle Maintenance |
| Postage and Related Costs | Video/Streaming |
| Proceeds from Personal Property Sales | Visitor Subsistence over 1 year |
| Recruitment | Visitor Subsistence < 1 year |

The base excludes costs charged to the TSCS pool and pools previously allocated (i.e. to tasks with the service types of OP-BURDEN-PS AD, OP-BURDEN-PS CD, OP-BURDEN-PS PPD, or OP-BURDEN-PS TD, OP-BURDEN-MSA, OP-BURDEN-CSS, or OP-BURDEN-TSCS). Additionally, costs charged to tasks in Non-HEP B&R's and in service type OP-R&D-USERS are also excluded.

Costs on project construction funds are not included in the base.

## E.    G & A Burden Allocation

The General and Administrative (G & A) burden pool represents the costs of the management and administration of the laboratory as a whole.

B&R Code:  EC0101060 – General and Administrative
Service Type:  OP-BURDEN-G&A
Expenditure Type:  G&A Allocation

*Pool* – The G & A pool consists of all the costs from the following activities.

Business Services Section Office (the portion not included within the MSA or CSS pools)
Computing Division Information Systems
Conference Office
Contractor Fee
Directorate
Education Office
Finance Section Office (the portion not included within the MSA pool)

Human Resources
Institutional General Plant Projects (IGPP)
Legal Office
Miscellaneous revenues
Offset from Pass-through Rate
Office of Communications
Payroll

Service on reviews of projects/programs/activities, and on lab wide committees not specific to mission work.
PS, MSA and CSS allocations on the costs in the G&A pool

*Base* – The G&A burden allocation will be applied to all expenditure types listed within the PS, MSA & TSCS sections of this document.

Vacation, OPTO, and Fringe labor burdens, as well as PS, MSA, CSS and TSCS burden allocations, are also included in the base.

The base excludes costs charged to the G&A pool and pools previously allocated (i.e. to tasks with the service types of OP-BURDEN-PS AD, OP-BURDEN-PS CD, OP-BURDEN-PS PPD, or OP-BURDEN-PS TD, OP-BURDEN-MSA, OP-BURDEN-CSS, OP-BURDEN-TSCS or OP-BURDEN-G&A).

## IV.    Current Year Rates

The following link will advise the current year's rates.
Provisional Labor, Indirect, and Shop Rates

Updated:  February 1, 2013

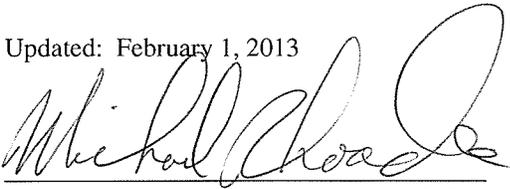

______________________________
Michael Rhoades – Chief Accounting Officer

2/8/13

______________________________
Date

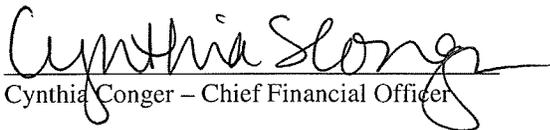

______________________________
Cynthia Conger – Chief Financial Officer

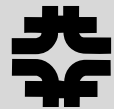

APPENDIX B nuSTORM Conventional Facilities

## Preliminary Drawings

- PDR-01   Location Plans and List of Drawings
- PDR-02   Existing – Site Plan
- PDR-03   Underground – Site Plan
- PDR-04   Surface Buildings – Site Plan
- PDR-05   Primary Beamline
- PDR-06   Main Injector and Primary Beamline Sections
- PDR-07   Enclosure Plan at Target Hall
- PDR-08   Target Hall Building Plan Sheet 1
- PDR-09   Target Hall Building Plan Sheet 2
- PDR-10   Target Hall Building Sections Sheet 1
- PDR-11   Target Hall Building Sections Sheet 2
- PDR-12   Target Hall Building Sections Sheet 3
- PDR-13   Enlarged Plans and Sections
- PDR-14   Muon Decay Ring Plan
- PDR-15   Muon Decay Ring Section
- PDR-16   Muon Decay Ring Building Plan
- PDR-17   Cryogenic Building Plan
- PDR-18   Muon Decay Ring Building Section
- PDR-19   Near Detector Hall Floor Plan Sheet 1
- PDR-20   Near Detector Hall Floor Plan Sheet 2
- PDR-21   Near Detector Hall Floor Plan Sheet 3
- PDR-22   Building and Enclosure Section Sheet 1
- PDR-23   Building and Enclosure Section Sheet 2
- PDR-24   Far Detector Floor Plan
- PDR-25   Far Detector Floor Plan
- PDR-26   Building Section

# nuSTORM

## PROJECT NUMBER 6-13-1

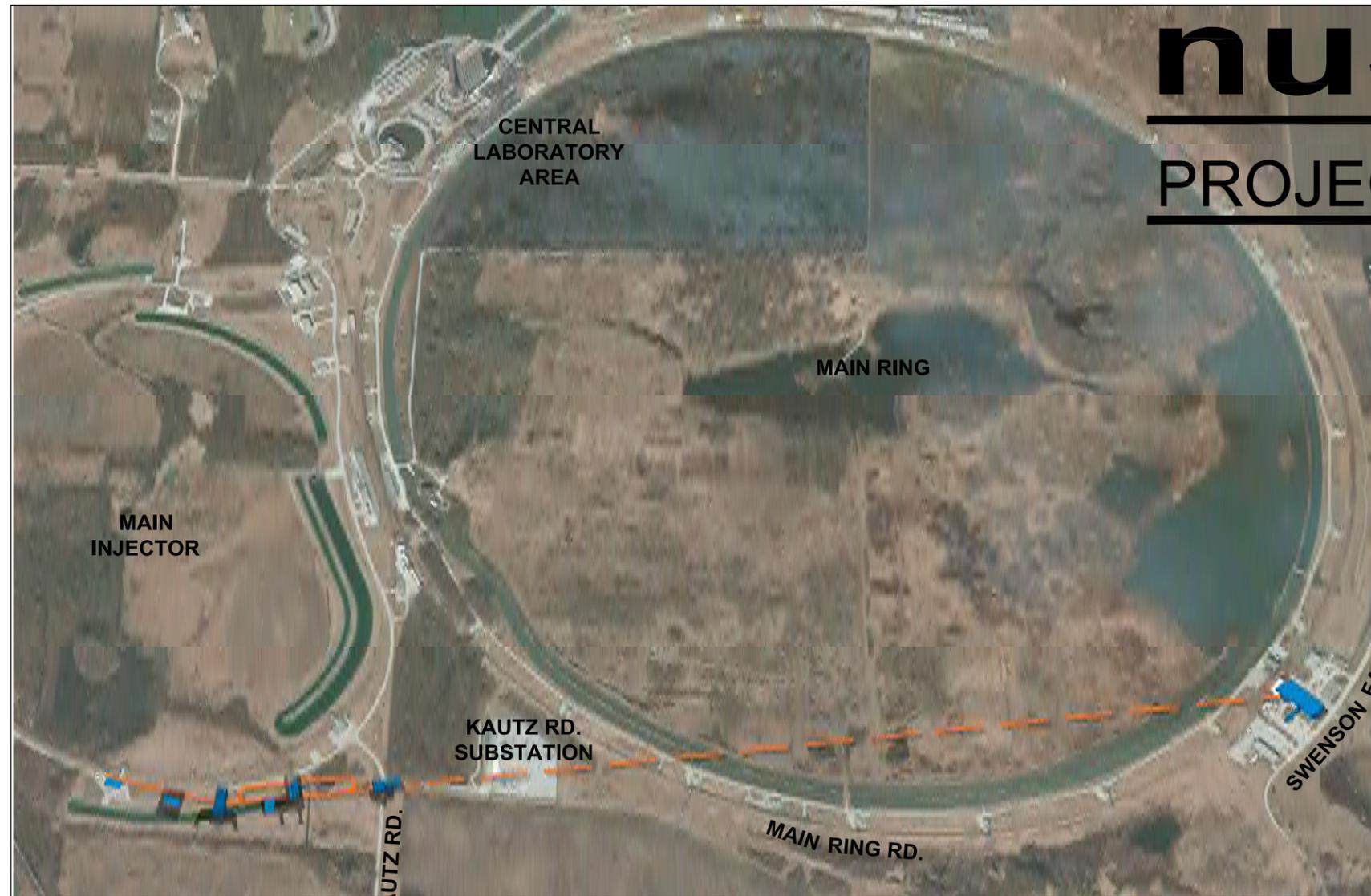

### LIST OF DRAWINGS

## VICINITY PLAN
SCALE: N.T.S.

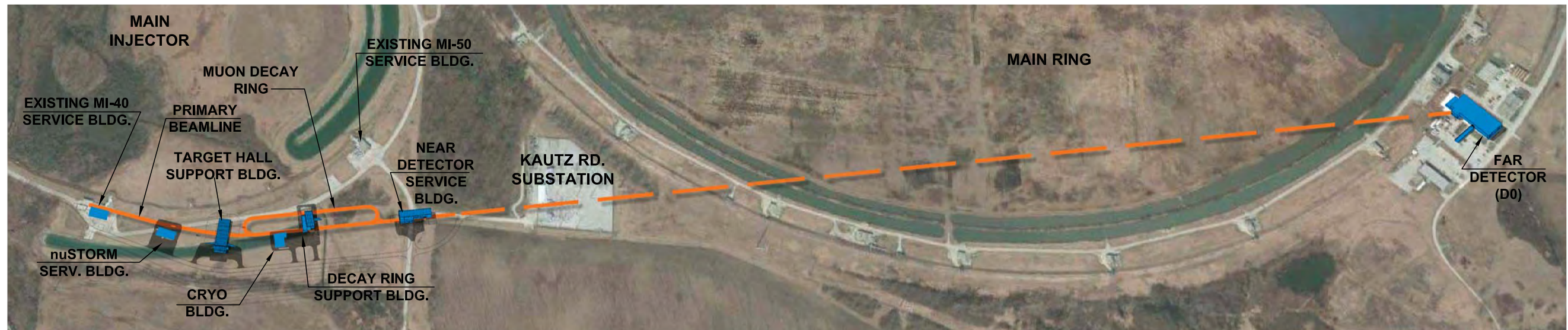

## LOCATION PLAN
SCALE: N.T.S.

SCALE:

PROJECT NORTH

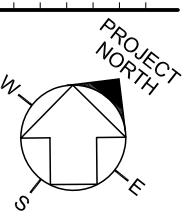

nuSTORM

LOCATION PLANS AND LIST OF DRAWINGS

PDR

vSTORM

⚛ Fermilab

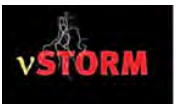

U.S. DEPARTMENT OF ENERGY   Office of Science

DATE
**08 MAY, 2013**

PROJECT NO.
**6-13-1**

DRAWING NO.
**PDR-01**

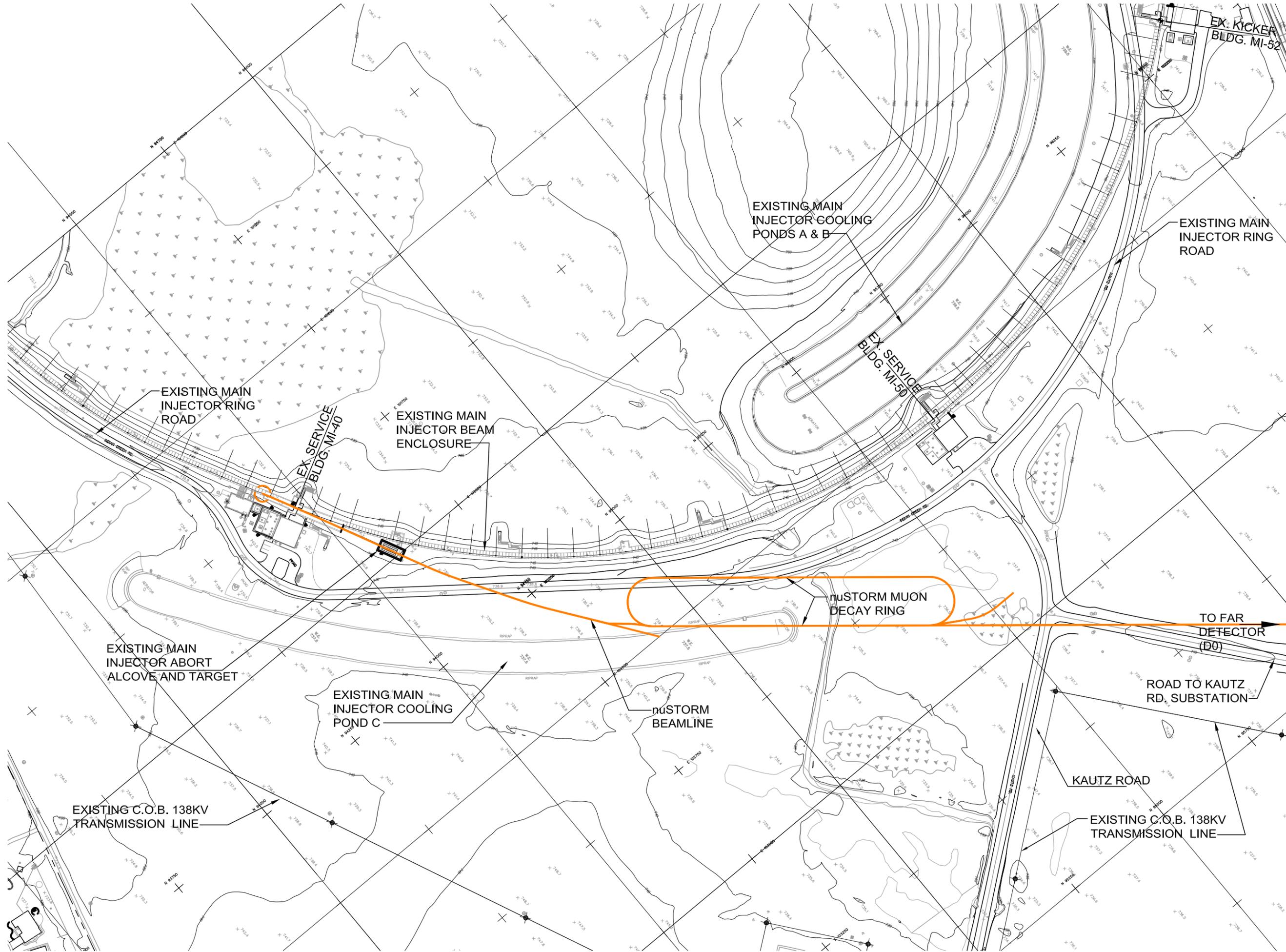

EX. KICKER
BLDG. MI-52

EXISTING MAIN
INJECTOR COOLING
PONDS A & B

EXISTING MAIN
INJECTOR RING
ROAD

EX. SERVICE
BLDG. MI-60

EXISTING MAIN
INJECTOR RING
ROAD

EXISTING MAIN
INJECTOR BEAM
ENCLOSURE

EX. SERVICE
BLDG. MI-40

nuSTORM MUON
DECAY RING

TO FAR
DETECTOR
(D0)

EXISTING MAIN
INJECTOR ABORT
ALCOVE AND TARGET

EXISTING MAIN
INJECTOR COOLING
POND C

nuSTORM
BEAMLINE

ROAD TO KAUTZ
RD. SUBSTATION

KAUTZ ROAD

EXISTING C.O.B. 138KV
TRANSMISSION LINE

EXISTING C.O.B. 138KV
TRANSMISSION LINE

SCALE:

1" = 640'

PROJECT NORTH

**nuSTORM**

EXISTING - SITE PLAN

PDR

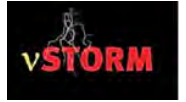

**Fermilab**

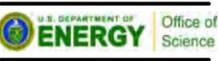
ENERGY | Office of Science

DATE
**08 MAY, 2013**

PROJECT NO.
**6-13-1**

DRAWING NO.
**PDR-02**

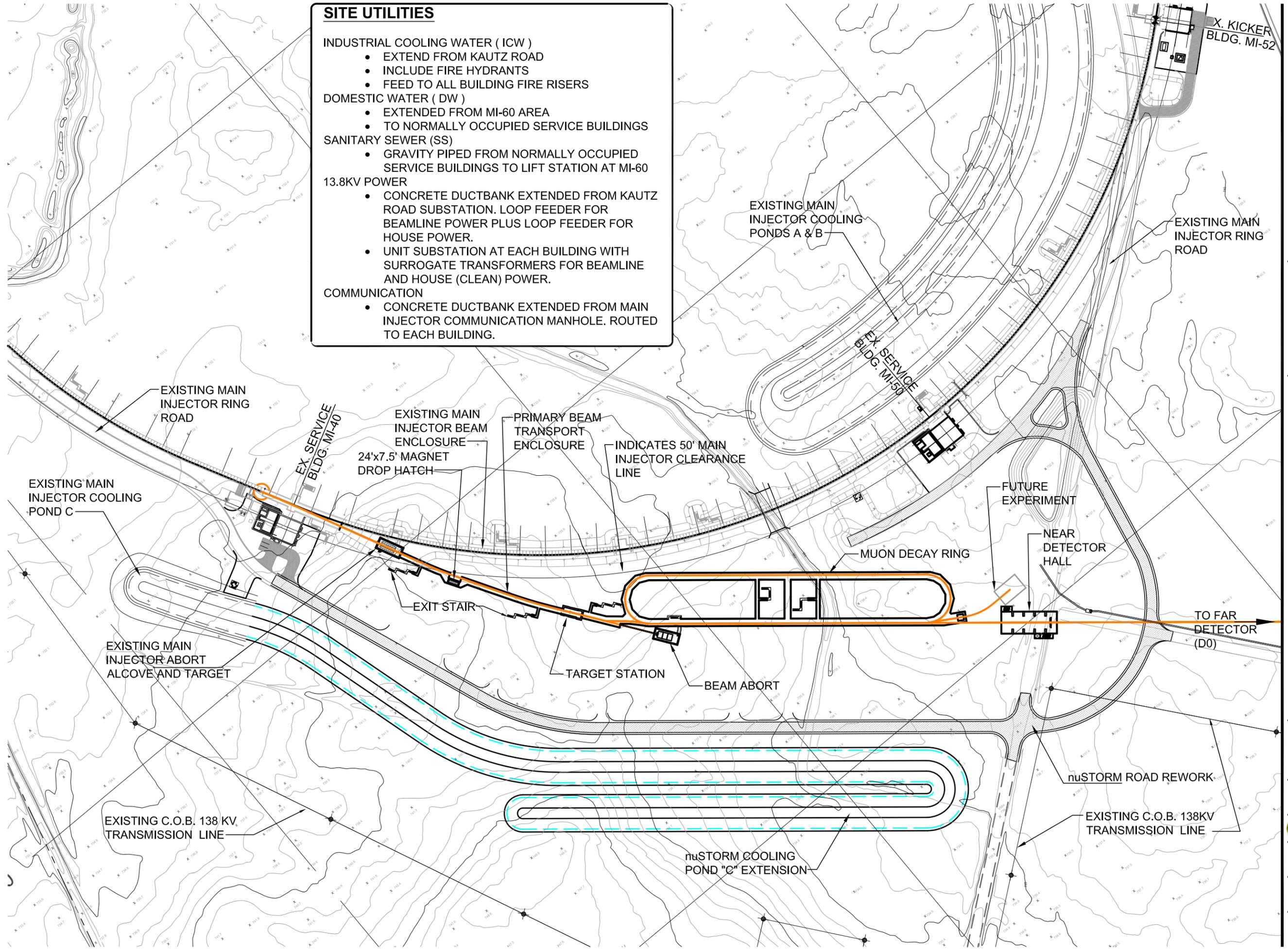

# SITE UTILITIES

**INDUSTRIAL COOLING WATER ( ICW )**
- EXTEND FROM KAUTZ ROAD
- INCLUDE FIRE HYDRANTS
- FEED TO ALL BUILDING FIRE RISERS

**DOMESTIC WATER ( DW )**
- EXTENDED FROM MI-60 AREA
- TO NORMALLY OCCUPIED SERVICE BUILDINGS

**SANITARY SEWER (SS)**
- GRAVITY PIPED FROM NORMALLY OCCUPIED SERVICE BUILDINGS TO LIFT STATION AT MI-60

**13.8KV POWER**
- CONCRETE DUCTBANK EXTENDED FROM KAUTZ ROAD SUBSTATION. LOOP FEEDER FOR BEAMLINE POWER PLUS LOOP FEEDER FOR HOUSE POWER.
- UNIT SUBSTATION AT EACH BUILDING WITH SURROGATE TRANSFORMERS FOR BEAMLINE AND HOUSE (CLEAN) POWER.

**COMMUNICATION**
- CONCRETE DUCTBANK EXTENDED FROM MAIN INJECTOR COMMUNICATION MANHOLE. ROUTED TO EACH BUILDING.

EX. KICKER BLDG. MI-52

EXISTING MAIN INJECTOR COOLING PONDS A & B

EXISTING MAIN INJECTOR RING ROAD

EX. SERVICE BLDG. MI-39

EXISTING MAIN INJECTOR RING ROAD

EX. SERVICE BLDG. MI-40

EXISTING MAIN INJECTOR BEAM ENCLOSURE

24'x7.5' MAGNET DROP HATCH

PRIMARY BEAM TRANSPORT ENCLOSURE

INDICATES 50' MAIN INJECTOR CLEARANCE LINE

EXISTING MAIN INJECTOR COOLING POND C

EXIT STAIR

MUON DECAY RING

FUTURE EXPERIMENT

NEAR DETECTOR HALL

EXISTING MAIN INJECTOR ABORT ALCOVE AND TARGET

TARGET STATION

BEAM ABORT

TO FAR DETECTOR (D0)

EXISTING C.O.B. 138 KV TRANSMISSION LINE

nuSTORM COOLING POND "C" EXTENSION

nuSTORM ROAD REWORK

EXISTING C.O.B. 138KV TRANSMISSION LINE

## SCALE:

1" = 60'-0"

# nuSTORM
## UNDERGROUND - SITE PLAN

PDR

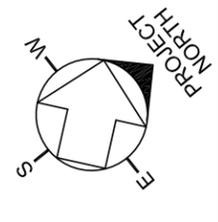

vSTORM

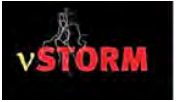

**Fermilab**

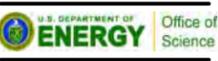

U.S. DEPARTMENT OF ENERGY · Office of Science

DATE
**08 MAY, 2013**

PROJECT NO.
**6-13-1**

DRAWING NO.
**PDR-3**

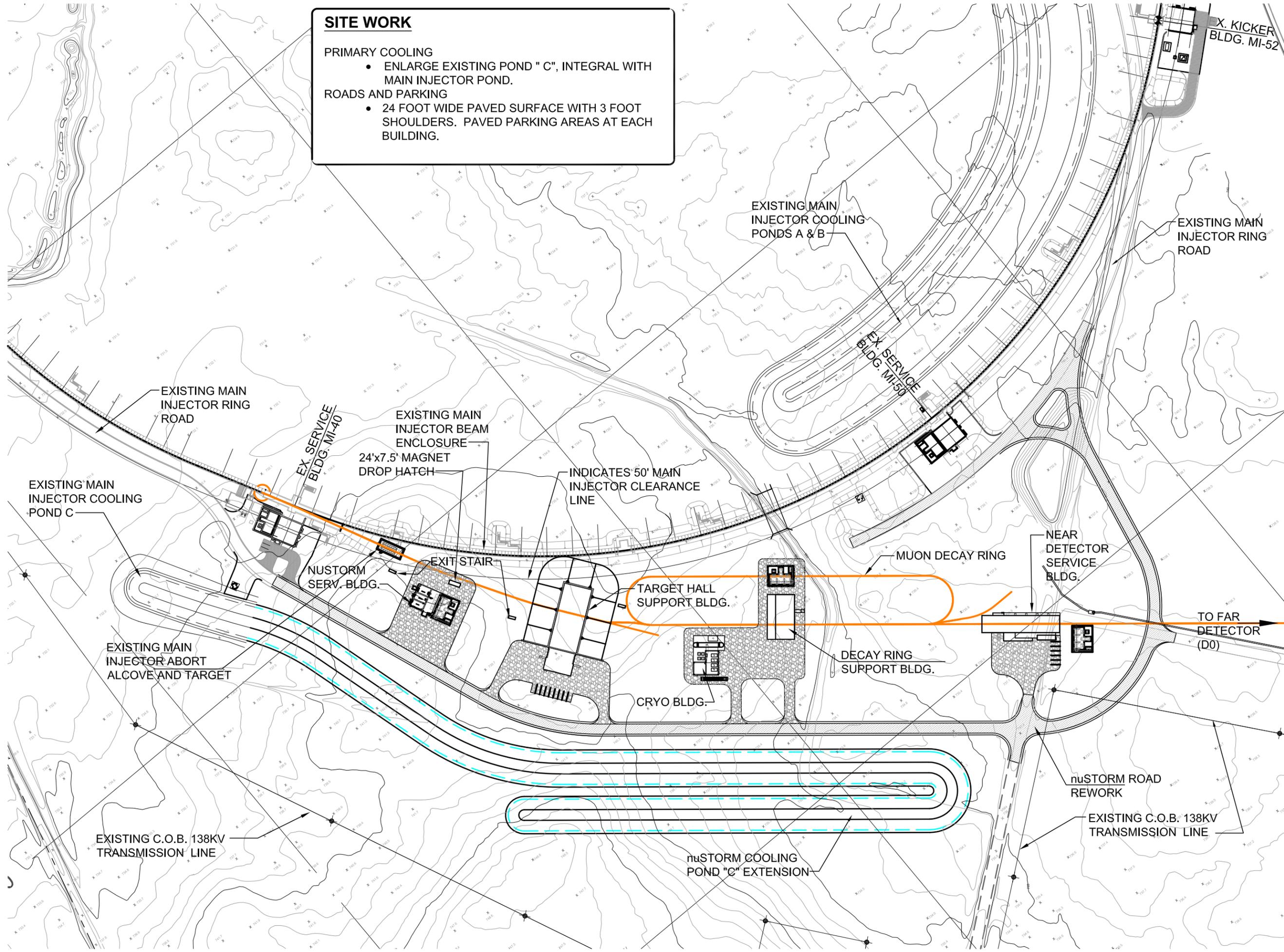
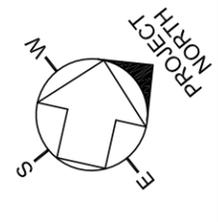
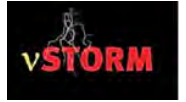

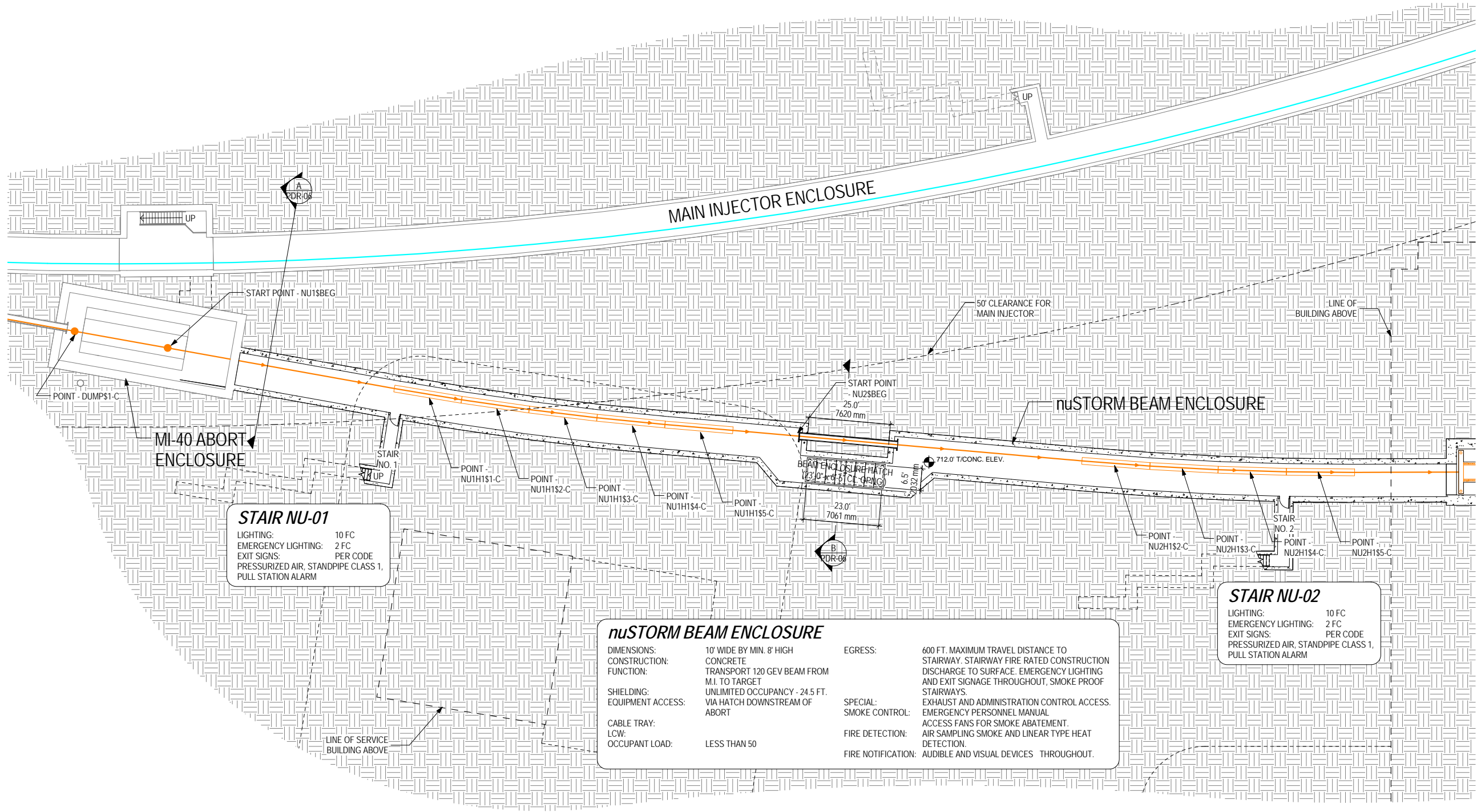

MAIN INJECTOR ENCLOSURE

A
FDR-06

UP

START POINT - NU1SBEG

50' CLEARANCE FOR
MAIN INJECTOR

LINE OF
BUILDING ABOVE

POINT - DUMPS1-C

MI-40 ABORT
ENCLOSURE

STAIR
NO. 1
UP

nuSTORM BEAM ENCLOSURE

START POINT -
NU2SBEG
25.0'
7620 mm

BEAM ENCLOSURE HATCH
(23'-0" x 6'-6") CC. OPENING

712.0' T/CONC. ELEV.

6.5'
1981 mm

23.0'
7001 mm

POINT
NU1H1S1-C

POINT
NU1H1S2-C

POINT
NU1H1S3-C

POINT
NU1H1S4-C

POINT
NU1H1S5-C

B
FDR-06

POINT -
NU2H1S2-C

POINT -
NU2H1S3-C

STAIR
NO. 2

POINT
NU2H1S4-C

POINT -
NU2H1S5-C

### STAIR NU-01
LIGHTING:   10 FC
EMERGENCY LIGHTING:   2 FC
EXIT SIGNS:   PER CODE
PRESSURIZED AIR, STANDPIPE CLASS 1,
PULL STATION ALARM

### STAIR NU-02
LIGHTING:   10 FC
EMERGENCY LIGHTING:   2 FC
EXIT SIGNS:   PER CODE
PRESSURIZED AIR, STANDPIPE CLASS 1,
PULL STATION ALARM

### nuSTORM BEAM ENCLOSURE
DIMENSIONS: 10' WIDE BY MIN. 8' HIGH
CONSTRUCTION: CONCRETE
FUNCTION: TRANSPORT 120 GEV BEAM FROM MI1 TO TARGET
SHIELDING: UNLIMITED OCCUPANCY - 24.5 FT.
EQUIPMENT ACCESS: VIA HATCH DOWNSTREAM OF ABORT
CABLE TRAY: LOW
OCCUPANT LOAD: LESS THAN 50

EGRESS: 600 FT. MAXIMUM TRAVEL DISTANCE TO STAIRWAY. STAIRWAY FIRE RATED CONSTRUCTION DISCHARGE TO SURFACE. EMERGENCY LIGHTING AND EXIT SIGNAGE THROUGHOUT. SMOKE PROOF STAIRWAYS.
SPECIAL: EXHAUST AND ADMINISTRATION CONTROL ACCESS.
SMOKE CONTROL: EMERGENCY PERSONNEL MANUAL ACCESS FANS FOR SMOKE ABATEMENT.
FIRE DETECTION: AIR SAMPLING SMOKE AND LINEAR TYPE HEAT DETECTION.
FIRE NOTIFICATION: AUDIBLE AND VISUAL DEVICES THROUGHOUT.

LINE OF SERVICE
BUILDING ABOVE.

PRIMARY BEAM LINE
SCALE: 1/16" = 1'-0"

PRIMARY BEAMLINE

SCALE:

1/16" = 1'-0"

SCALE

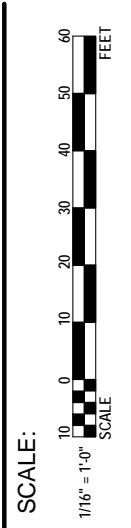

FEET

PROJECT
NORTH

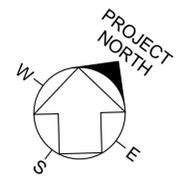

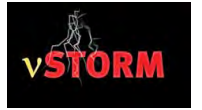
νSTORM

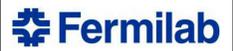
Fermilab

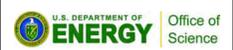
ENERGY   Office of Science

DATE
**08 MAY 2013**
PROJECT NO.
**6-13-1**
DRAWING NO.
**PDR-05**

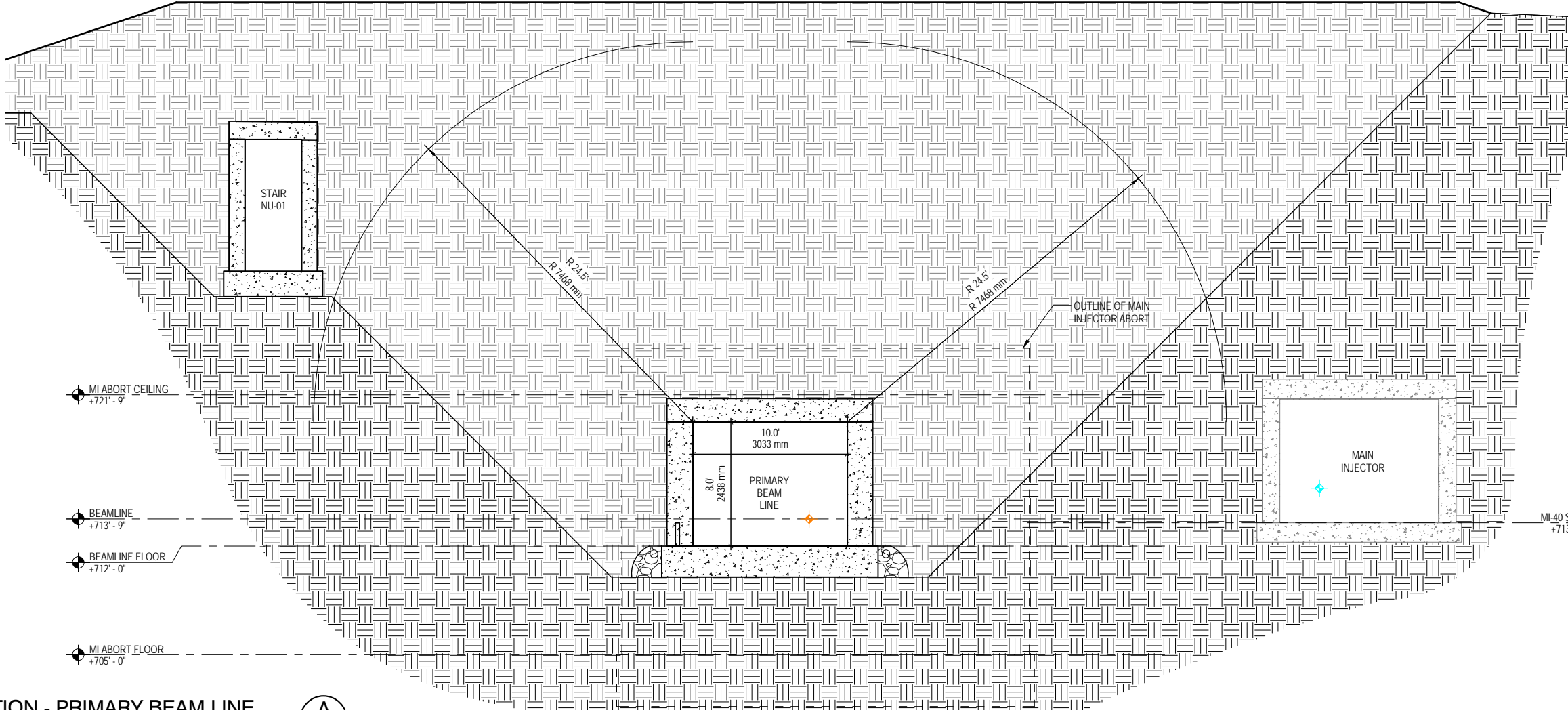

**SECTION - PRIMARY BEAM LINE** A
SCALE: 1/4" = 1'-0"

MI ABORT CEILING
+721' - 9"

BEAMLINE
+713' - 9"

BEAMLINE FLOOR
+712' - 0"

MI ABORT FLOOR
+705' - 0"

STAIR
NU-01

R-24.5'
R-7468 mm

R-24.5'
R-7468 mm

OUTLINE OF MAIN
INJECTOR ABORT

10.0'
3033 mm

8.0'
2438 mm

PRIMARY
BEAM
LINE

MAIN
INJECTOR

MI-40 SLAB
+713' - 6"

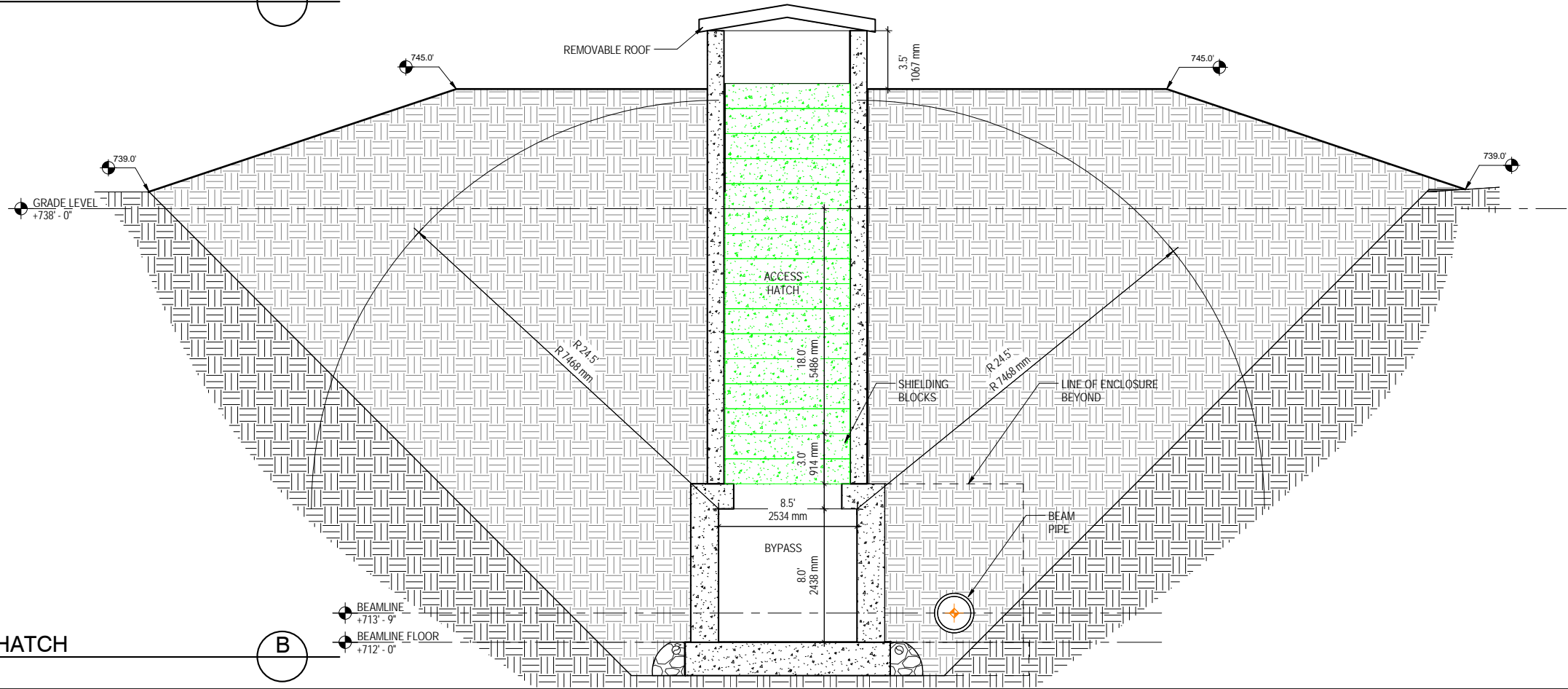

**SECTION - HATCH** B
SCALE: 1/4" = 1'-0"

REMOVABLE ROOF

3.5'
1067 mm

745.0'

745.0'

739.0'

GRADE LEVEL
+738' - 0"

ACCESS
HATCH

18.0'
5486 mm

3.0'
914 mm

R-24.5'
R-7468 mm

R-24.5'
R-7468 mm

SHIELDING
BLOCKS

LINE OF ENCLOSURE
BEYOND

8.5'
2534 mm

8.0'
2438 mm

BYPASS

BEAM
PIPE

BEAMLINE
+713' - 9"

BEAMLINE FLOOR
+712' - 0"

5/29/2013 10:41:00 AM    C:\Users\Matt_B\Documents\1523818-ARCH TARGET_Matt_B.rvt

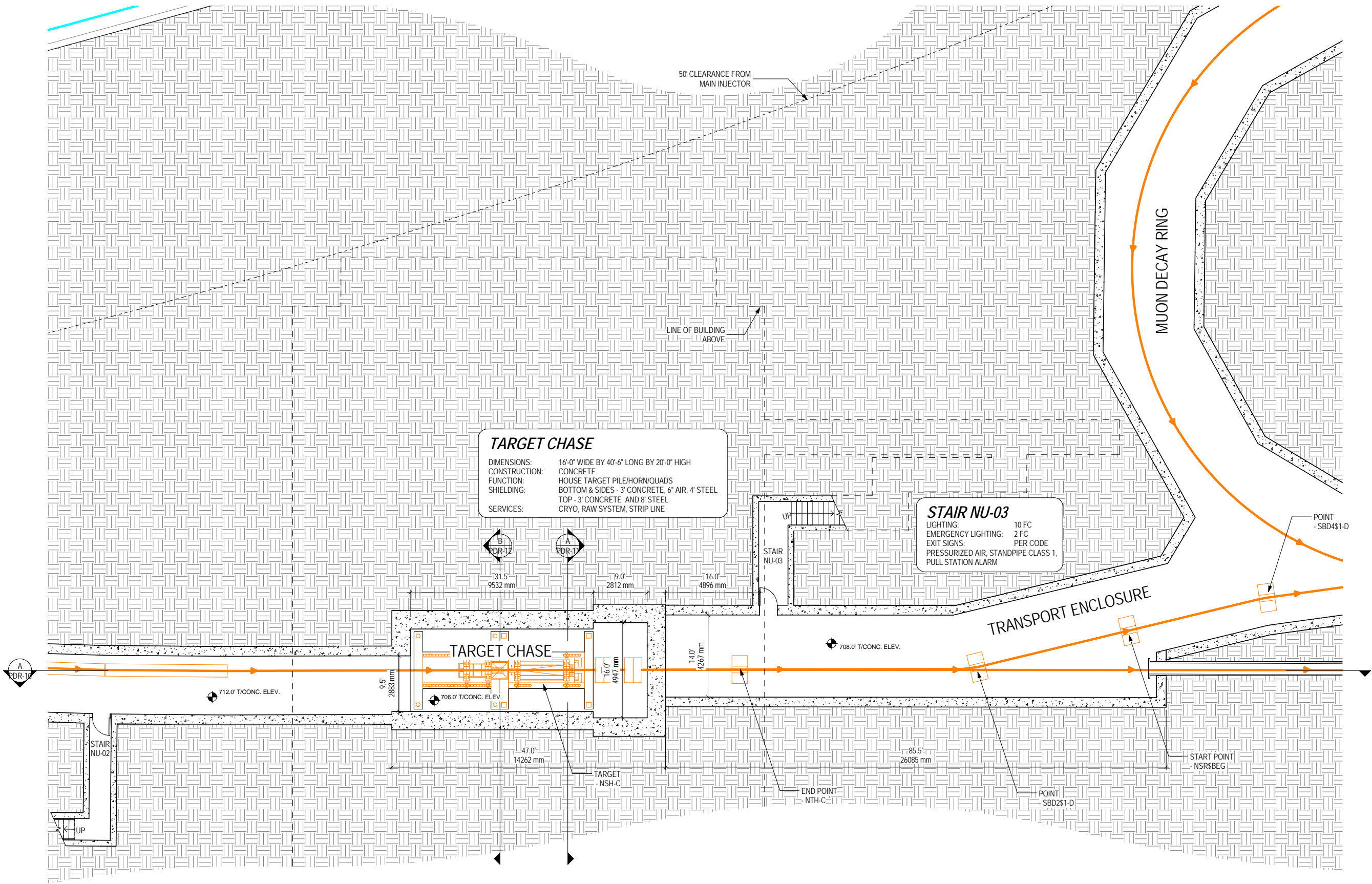

50' CLEARANCE FROM
MAIN INJECTOR

MUON DECAY RING

LINE OF BUILDING
ABOVE

**TARGET CHASE**

DIMENSIONS: 16'-0" WIDE BY 40'-6" LONG BY 20'-0" HIGH
CONSTRUCTION: CONCRETE
FUNCTION: HOUSE TARGET PILE/HORN/QUADS
SHIELDING: BOTTOM & SIDES - 3' CONCRETE, 6" AIR, 4' STEEL
TOP - 3' CONCRETE  AND 8' STEEL
SERVICES: CRYO, RAW SYSTEM, STRIP LINE

**STAIR NU-03**

LIGHTING: 10 FC
EMERGENCY LIGHTING: 2 FC
EXIT SIGNS: PER CODE
PRESSURIZED AIR, STANDPIPE CLASS 1,
PULL STATION ALARM

STAIR
NU-03

POINT
- SBD481-D

TRANSPORT ENCLOSURE

31.5'
9532 mm

9.0'
2812 mm

16.0'
4896 mm

TARGET CHASE

9.5'
2883 mm

16.0'
494 mm

14.0'
426 mm

708.0' T/CONC. ELEV.

POINT
- SBD261-D

START POINT
- NSRSBEG

712.0' T/CONC. ELEV.

706.0' T/CONC. ELEV.

47.0'
14262 mm

TARGET
- NSH-C

END POINT
- NTH-C

85.5'
26085 mm

STAIR
NU-02

UP

**ENCLOSURE AT TARGET CHASE**

SCALE:   1/8" = 1'-0"

SCALE:

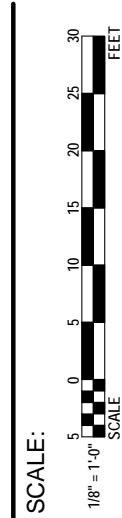

PROJECT
NORTH

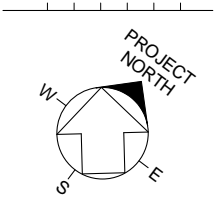

W        E
SW      SE
S

ENCLOSURE PLAN AT TARGET HALL

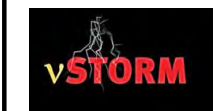
vSTORM

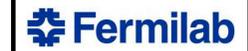
Fermilab

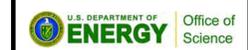
ENERGY  Office of Science

DATE
**08 MAY 2013**

PROJECT NO.
**6-13-1**

DRAWING NO.
**PDR-07**

C:\Users\Matt_B\Documents\152389B-ARCH-TARGET_Matt_B.rvt

5/29/2013 10:47:02 AM

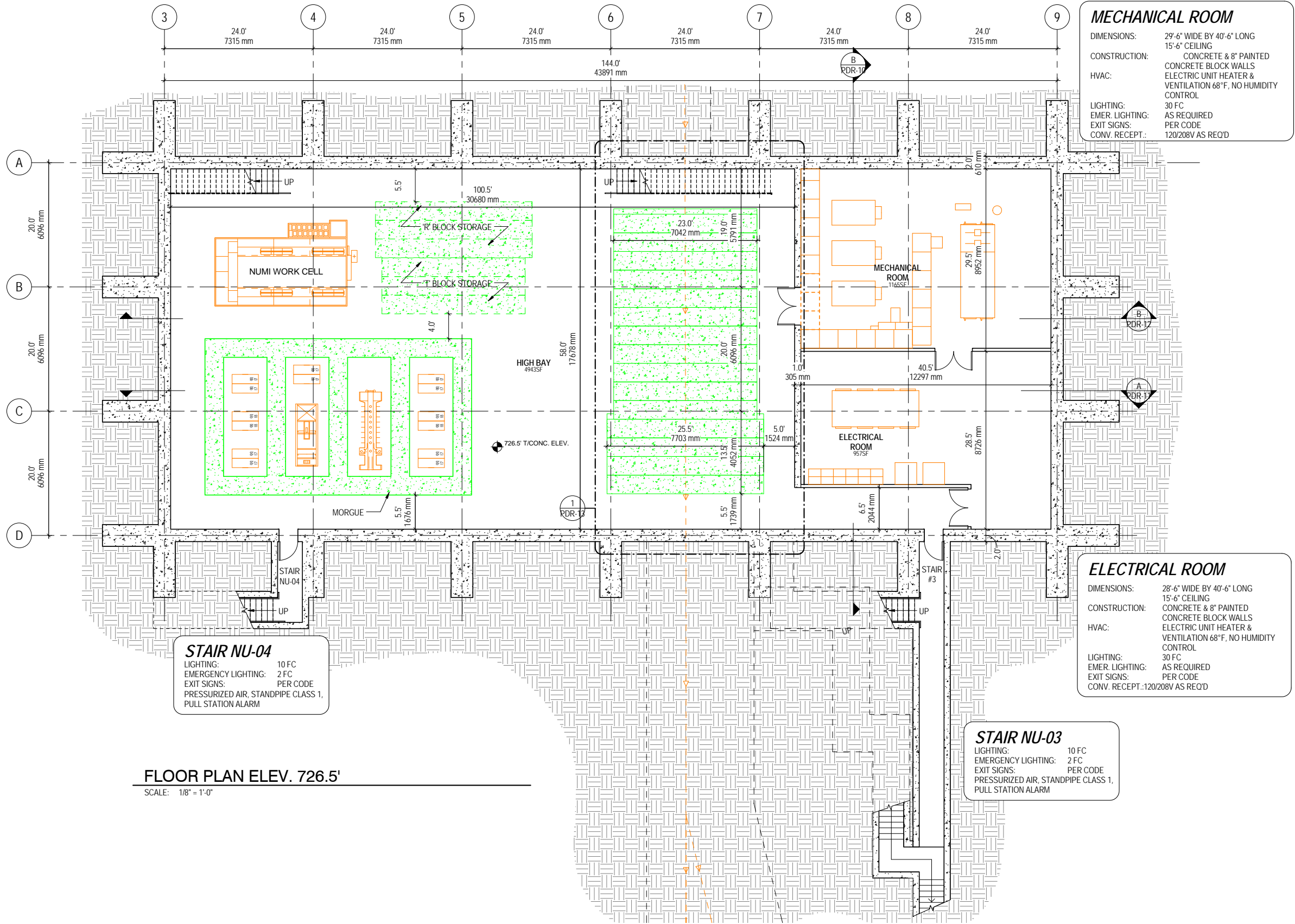

DIMENSIONS: 29'-6" WIDE BY 40'-6" LONG
15'-6" CEILING
CONSTRUCTION: CONCRETE & 8" PAINTED
CONCRETE BLOCK WALLS
HVAC: ELECTRIC UNIT HEATER &
VENTILATION 68°F, NO HUMIDITY
CONTROL
LIGHTING: 30 FC
EMER. LIGHTING: AS REQUIRED
EXIT SIGNS: PER CODE
CONV. RECEPT: 120/208V AS REQ'D

## ELECTRICAL ROOM

DIMENSIONS: 28'-6" WIDE BY 40'-6" LONG
15'-6" CEILING
CONSTRUCTION: CONCRETE & 8" PAINTED
CONCRETE BLOCK WALLS
HVAC: ELECTRIC UNIT HEATER &
VENTILATION 68°F, NO HUMIDITY
CONTROL
LIGHTING: 30 FC
EMER. LIGHTING: AS REQUIRED
EXIT SIGNS: PER CODE
CONV. RECEPT: 120/208V AS REQ'D

### STAIR NU-04
LIGHTING: 10 FC
EMERGENCY LIGHTING: 2 FC
EXIT SIGNS: PER CODE
PRESSURIZED AIR, STANDPIPE CLASS 1,
PULL STATION ALARM

### STAIR NU-03
LIGHTING: 10 FC
EMERGENCY LIGHTING: 2 FC
EXIT SIGNS: PER CODE
PRESSURIZED AIR, STANDPIPE CLASS 1,
PULL STATION ALARM

FLOOR PLAN ELEV. 726.5'
SCALE: 1/8" = 1'-0"

NUMI WORK CELL

8" BLOCK STORAGE

1' BLOCK STORAGE

HIGH BAY

MECHANICAL ROOM

ELECTRICAL ROOM

MORGUE

726.5' T/CONC. ELEV.

STAIR NU-04

STAIR #3

SCALE:

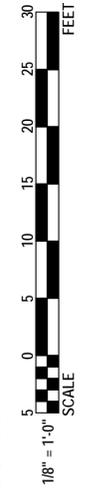

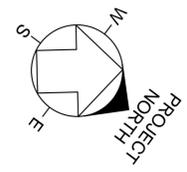

TARGET HALL BUILDING PLAN SHEET 1

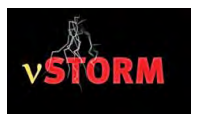

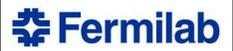

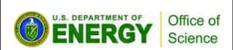

DATE
08 MAY 2013
PROJECT NO.
6-13-1
DRAWING NO.
PDR-08

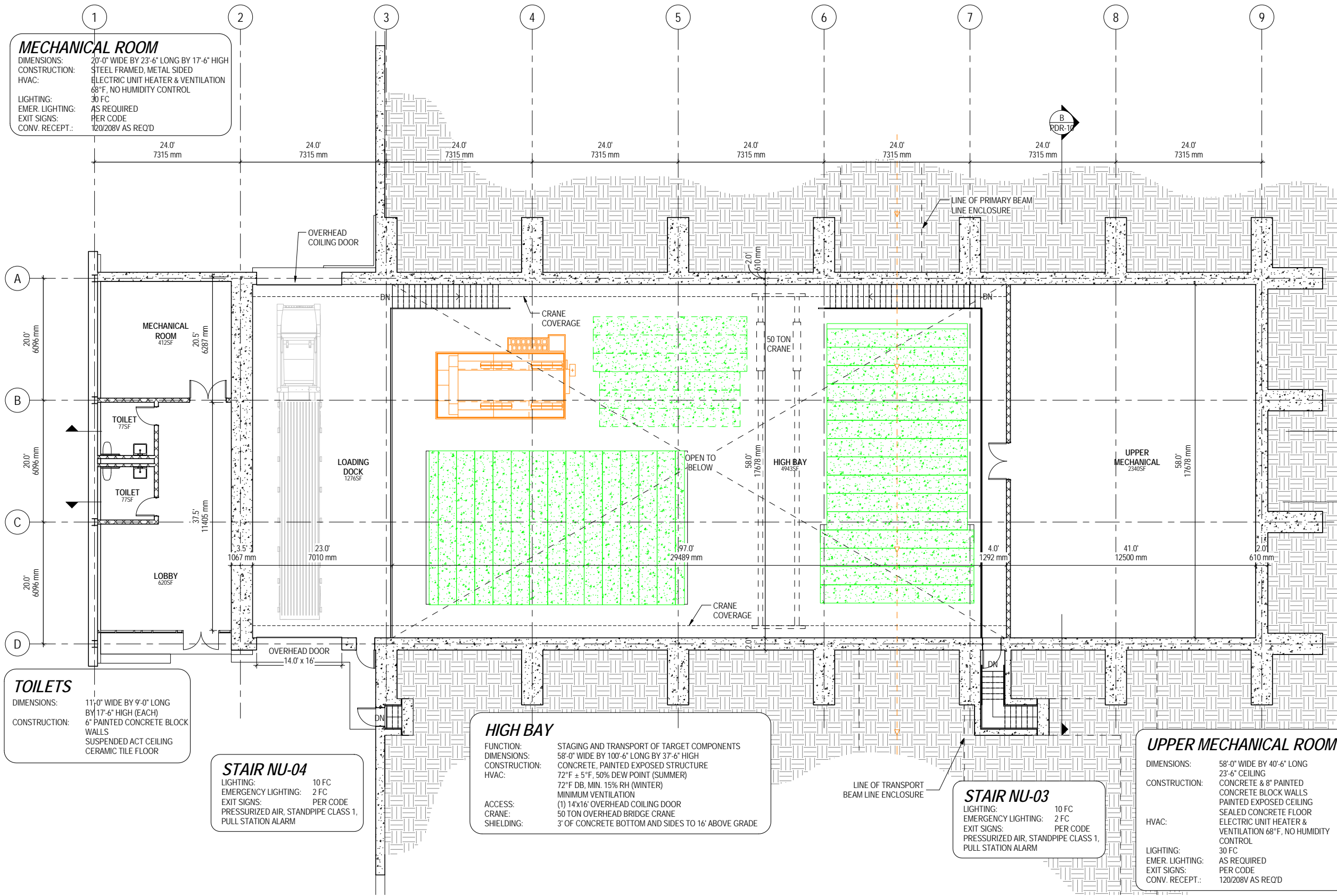

**MECHANICAL ROOM**
DIMENSIONS: 20'-0" WIDE BY 23'-6" LONG BY 17'-6" HIGH
CONSTRUCTION: STEEL FRAMED, METAL SIDED
HVAC: ELECTRIC UNIT HEATER & VENTILATION
68°F, NO HUMIDITY CONTROL
LIGHTING: 30 FC
EMER. LIGHTING: AS REQUIRED
EXIT SIGNS: PER CODE
CONV. RECEPT.: 120/208V AS REQ'D

OVERHEAD COILING DOOR

CRANE COVERAGE

LINE OF PRIMARY BEAM
LINE ENCLOSURE

MECHANICAL ROOM
470SF

TOILET
77SF

TOILET
77SF

LOBBY
630SF

LOADING DOCK
1276SF

OPEN TO BELOW

50 TON CRANE

HIGH BAY
4502SF

UPPER MECHANICAL
2945SF

CRANE COVERAGE

OVERHEAD DOOR
14'-0" x 16'

**TOILETS**
DIMENSIONS: 11'-0" WIDE BY 9'-0" LONG
BY 17'-6" HIGH (EACH)
CONSTRUCTION: 6" PAINTED CONCRETE BLOCK
WALLS
SUSPENDED ACT CEILING
CERAMIC TILE FLOOR

**STAIR NU-04**
LIGHTING: 10 FC
EMERGENCY LIGHTING: 2 FC
EXIT SIGNS: PER CODE
PRESSURIZED AIR, STANDPIPE CLASS 1,
PULL STATION ALARM

**HIGH BAY**
FUNCTION: STAGING AND TRANSPORT OF TARGET COMPONENTS
DIMENSIONS: 58'-0" WIDE BY 100'-6" LONG BY 37'-6" HIGH
CONSTRUCTION: CONCRETE, PAINTED EXPOSED STRUCTURE
HVAC: 72°F ± 5°F, 50% DEW POINT (SUMMER)
72°F DB, MIN. 15% RH (WINTER)
MINIMUM VENTILATION
ACCESS: (1) 14'x16' OVERHEAD COILING DOOR
CRANE: 50 TON OVERHEAD BRIDGE CRANE
SHIELDING: 3'-0" OF CONCRETE BOTTOM AND SIDES TO 16' ABOVE GRADE

**STAIR NU-03**
LIGHTING: 10 FC
EMERGENCY LIGHTING: 2 FC
EXIT SIGNS: PER CODE
PRESSURIZED AIR, STANDPIPE CLASS 1,
PULL STATION ALARM

LINE OF TRANSPORT
BEAM LINE ENCLOSURE

**UPPER MECHANICAL ROOM**
DIMENSIONS: 58'-0" WIDE BY 40'-6" LONG
23'-6" CEILING
CONSTRUCTION: CONCRETE & 8" PAINTED
CONCRETE BLOCK WALLS
PAINTED EXPOSED CEILING
SEALED CONCRETE FLOOR
HVAC: ELECTRIC UNIT HEATER &
VENTILATION 68°F, NO HUMIDITY
CONTROL
LIGHTING: 30 FC
EMER. LIGHTING: AS REQUIRED
EXIT SIGNS: PER CODE
CONV. RECEPT.: 120/208V AS REQ'D

**FLOOR PLAN ELEV. 738.00' & 742.50'**
SCALE: 1/8" = 1'-0"

TARGET HALL BUILDING PLAN SHEET 2

SCALE:

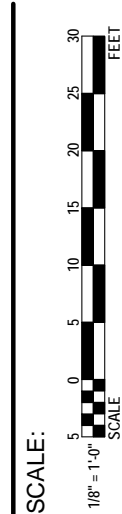

C:\Users\Matt_B\Documents\52381B-ARCH-TARGET_Matt_B.rvt
5/29/2013 10:47:06 AM

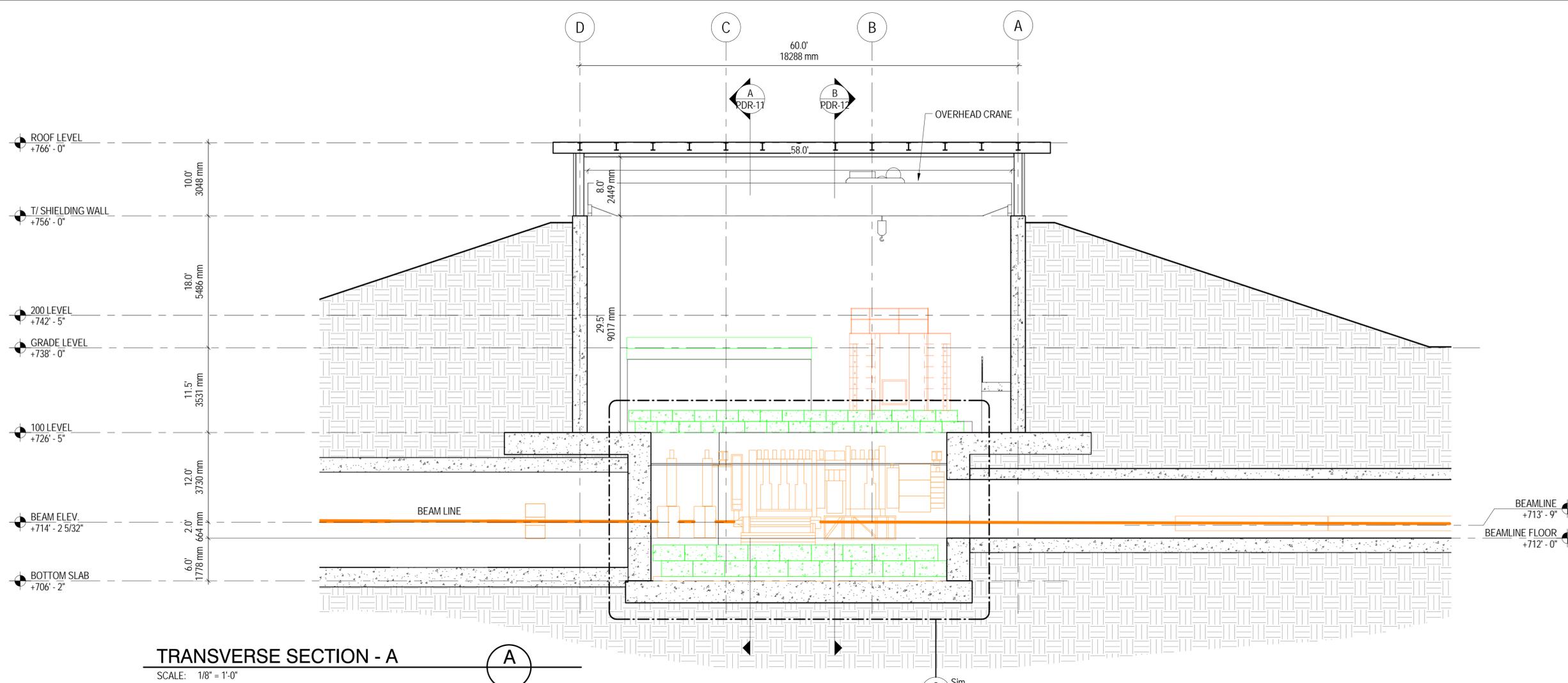

## TRANSVERSE SECTION - A
SCALE: 1/8" = 1'-0"

A

- ROOF LEVEL +766'-0"
- T/ SHIELDING WALL +756'-0"
- 200 LEVEL +742'-5"
- GRADE LEVEL +738'-0"
- 100 LEVEL +726'-5"
- BEAM ELEV +714'-2 5/32"
- BOTTOM SLAB +706'-2"

10'-0" 3048 mm
18'-0" 5486 mm
11'-5" 3531 mm
12'-0" 3730 mm
2'-0" 664 mm
6'-0" 1778 mm

8'-0" 2449 mm
29'-6" 9017 mm

60'-0" 18288 mm

D   C   B   A

A PDR-11
B PDR-11

OVERHEAD CRANE

58'-0"

BEAM LINE

BEAMLINE +713'-9"
BEAMLINE FLOOR +712'-0"

2 Sim
PDR-13 opp

## TRANSVERSE SECTION - B
SCALE: 1/8" = 1'-0"

B

A   B   C   D

B PDR-12
A PDR-11

- ROOF LEVEL +766'-0"
- T/ SHIELDING WALL +756'-0"
- 200 LEVEL +742'-5"
- GRADE LEVEL +738'-0"
- 100 LEVEL +726'-5"

755'-0"

13'-5" 4146 mm
16'-0" 4877 mm

UPPER MECHANICAL

MECHANICAL ROOM

ELECTRICAL ROOM

5/29/2013 10:41:07 AM   C:\Users\Matt_B\Documents\1523818-ARCH-TARGET_Matt_B.rvt

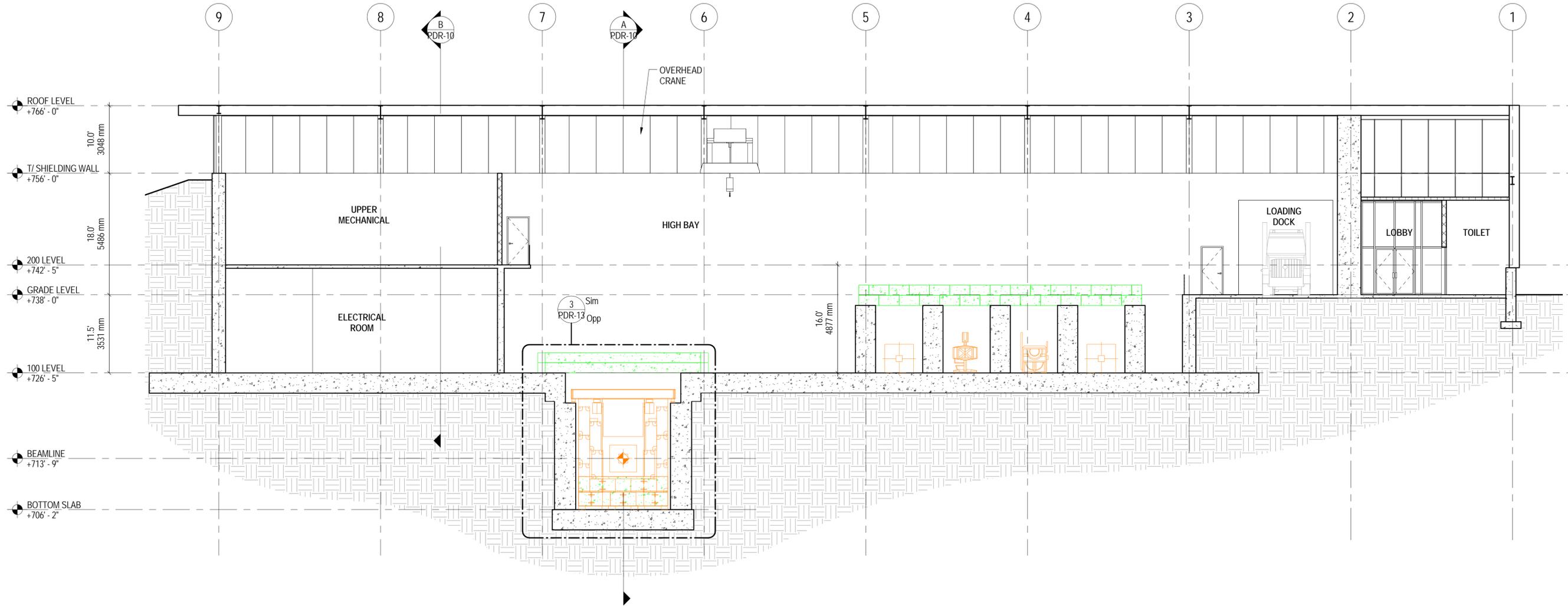

LONGITUDINAL SECTION - A
SCALE: 1/8" = 1'-0"  (A)

OVERHEAD CRANE

ROOF LEVEL
+766' - 0"

7' SHIELDING WALL
+756' - 0"

200 LEVEL
+742' - 5"

GRADE LEVEL
+738' - 0"

100 LEVEL
+726' - 5"

BEAMLINE
+713' - 9"

BOTTOM SLAB
+706' - 2"

UPPER MECHANICAL

HIGH BAY

LOADING DOCK

LOBBY

TOILET

ELECTRICAL ROOM

Sim
PDR-13 Opp

C:\Users\Matt_B\Documents\1523\BIB-ARCH-TARGET_Matt_B.rvt

5/29/2013 10:47:08 AM

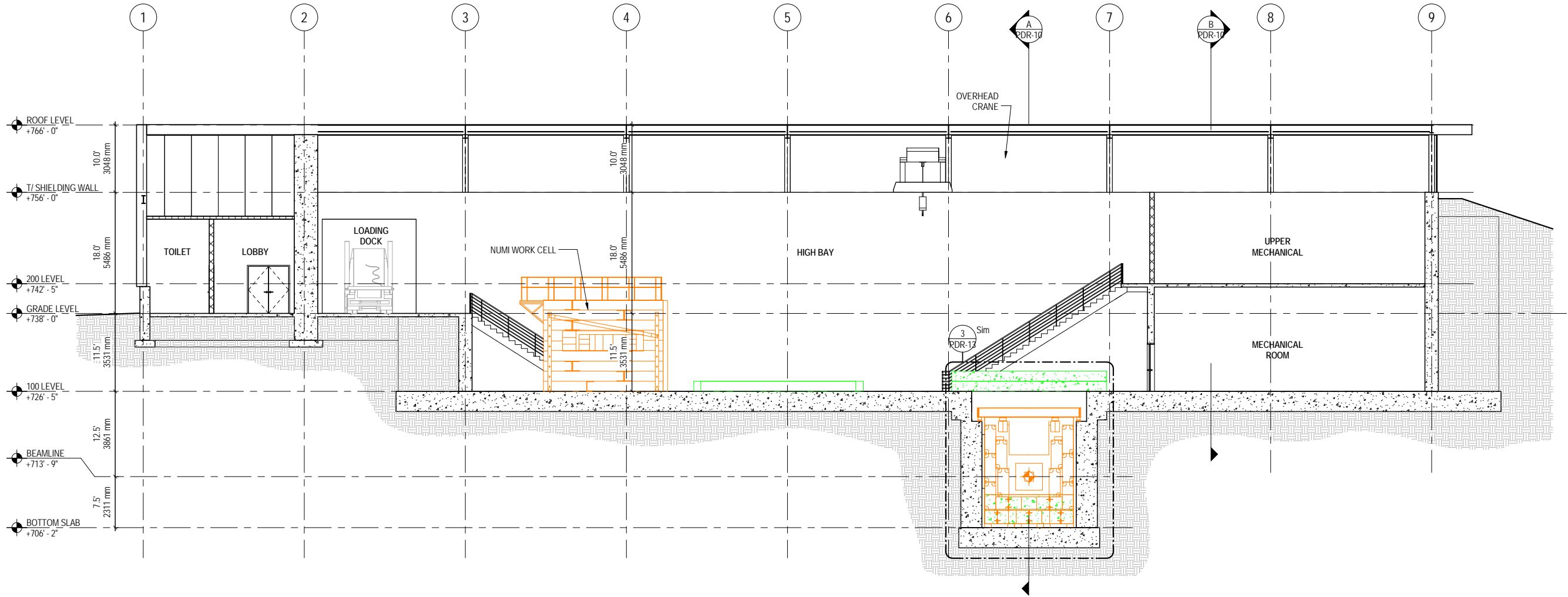

ROOF LEVEL
+766' - 0"

T/ SHIELDING WALL
+756' - 0"

200 LEVEL
+742' - 5"

GRADE LEVEL
+738' - 0"

100 LEVEL
+726' - 5"

BEAMLINE
+713' - 9"

BOTTOM SLAB
+706' - 2"

10'0"
3048 mm

18'0"
5486 mm

11'5"
3531 mm

12'5"
3861 mm

7'5"
2311 mm

10'0"
3048 mm

18'0"
5486 mm

11'5"
3531 mm

TOILET

LOBBY

LOADING DOCK

NUMI WORK CELL

HIGH BAY

OVERHEAD CRANE

UPPER MECHANICAL

MECHANICAL ROOM

3 Sim
PDR-13

A
PDR-10

B
PDR-9

LONGITUDINAL SECTION - B

SCALE: 1/8" = 1'-0"

B

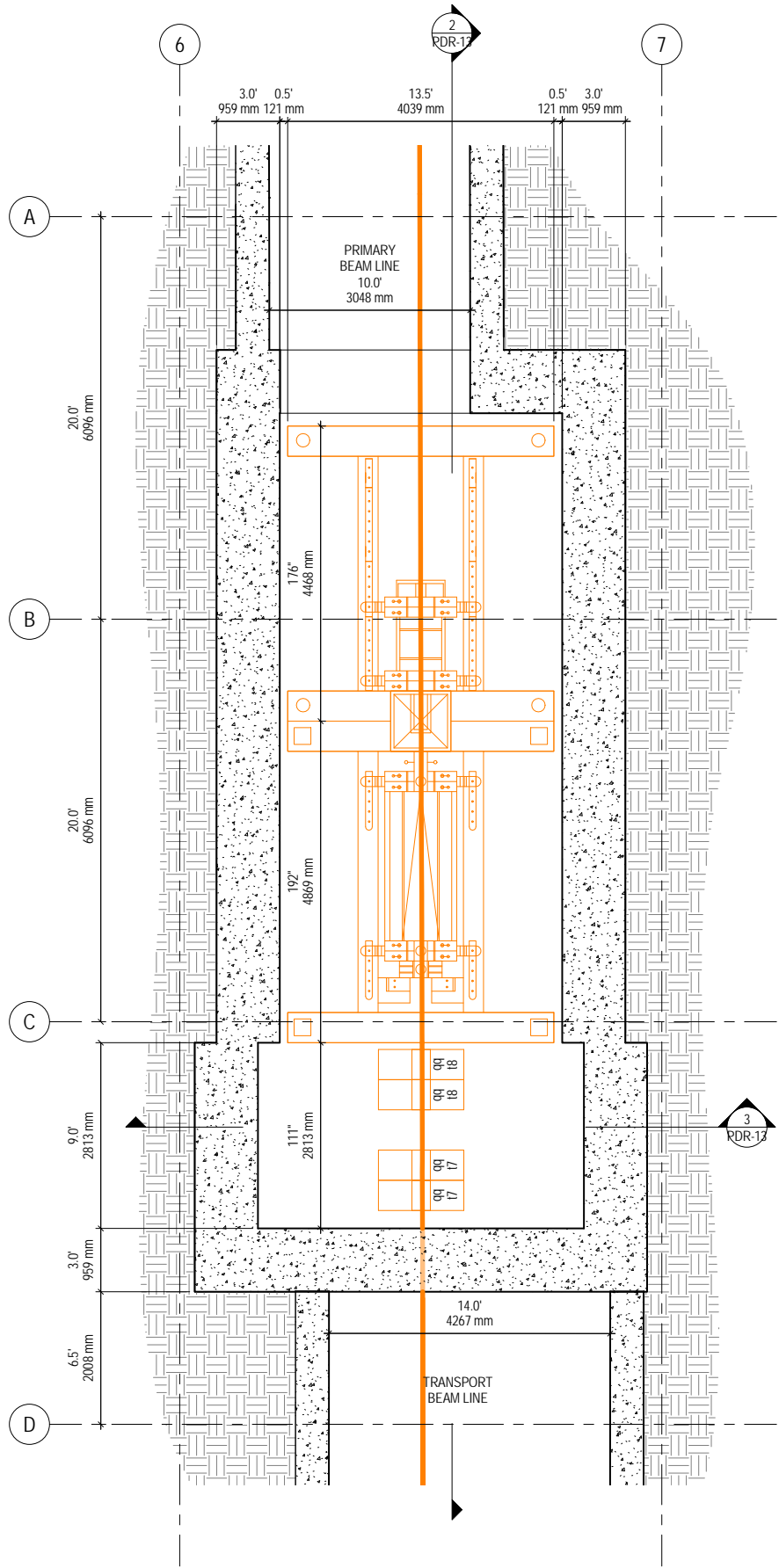

**TARGET HALL - 100 LEVEL ENLARGED PLAN**
SCALE: 1/4" = 1'-0"

PRIMARY
BEAM LINE
10.0'
3048 mm

TRANSPORT
BEAM LINE

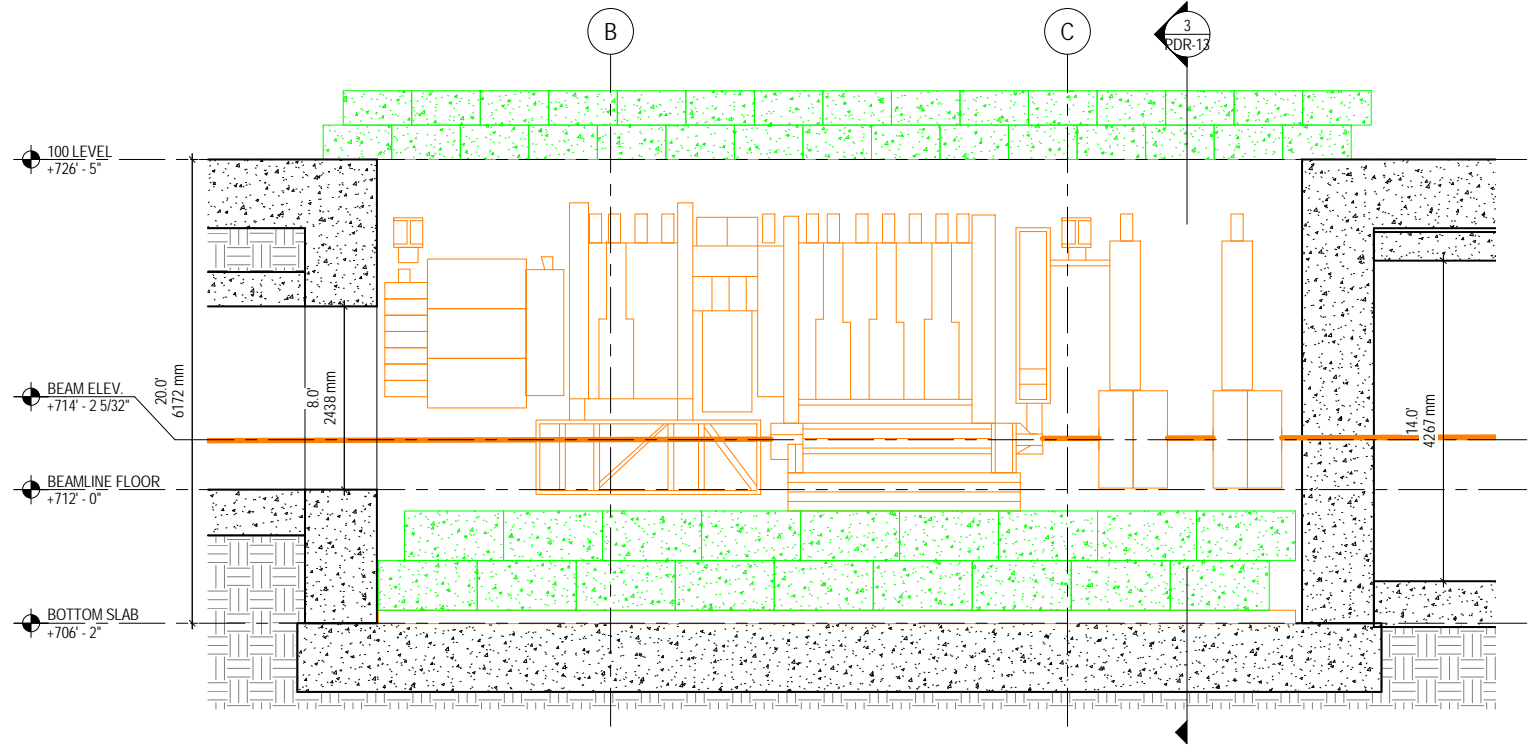

**LONGITUDINAL SECTION DETAIL - TARGET**
SCALE: 1/4" = 1'-0"

100 LEVEL
+726' - 5"

BEAM ELEV
+714' - 2 5/32"

BEAMLINE FLOOR
+712' - 0"

BOTTOM SLAB
+706' - 2"

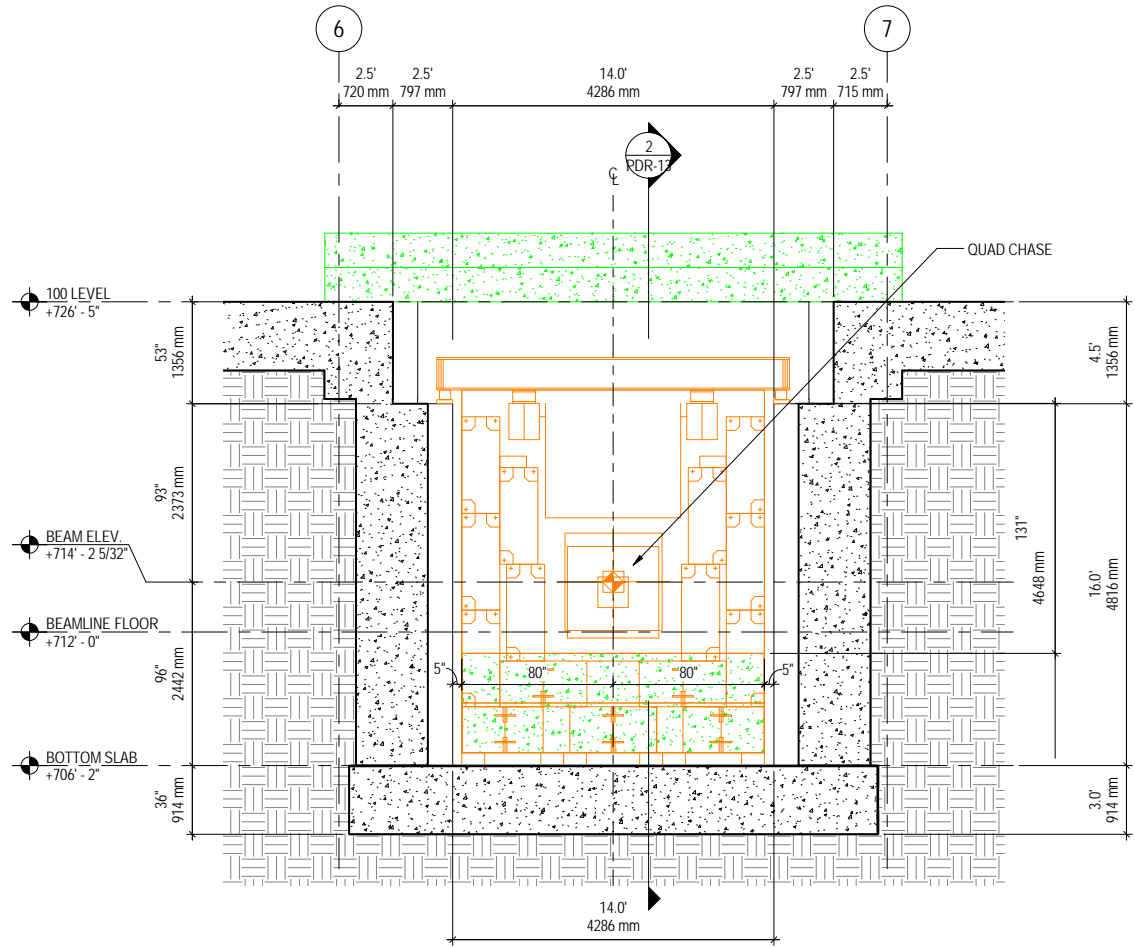

**TRANSVERSE SECTION DETAIL - TARGET**
SCALE: 1/4" = 1'-0"

100 LEVEL
+726' - 5"

BEAM ELEV
+714' - 2 5/32"

BEAMLINE FLOOR
+712' - 0"

BOTTOM SLAB
+706' - 2"

QUAD CHASE

C:\Users\Matt_B\Documents\1528IB-ARCH-TARGET_Matt_B.rvt

5/29/2013 10:41:11 AM

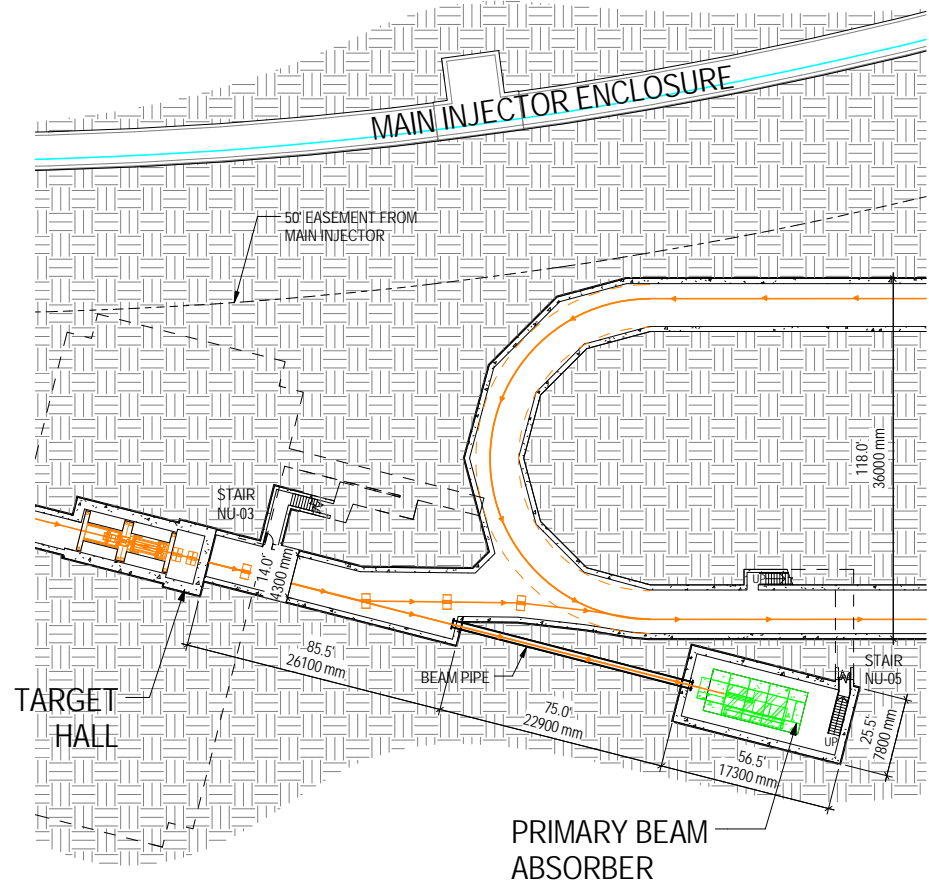

MAIN INJECTOR ENCLOSURE

50' EASEMENT FROM
MAIN INJECTOR

STAIR
NU-03

TARGET
HALL

118.0'
36000 mm

7.4'
2300 mm

85.5'
26100 mm

BEAM PIPE

75.0'
22900 mm

56.5'
17300 mm

25.5'
7800 mm

STAIR
NU-05

PRIMARY BEAM
ABSORBER

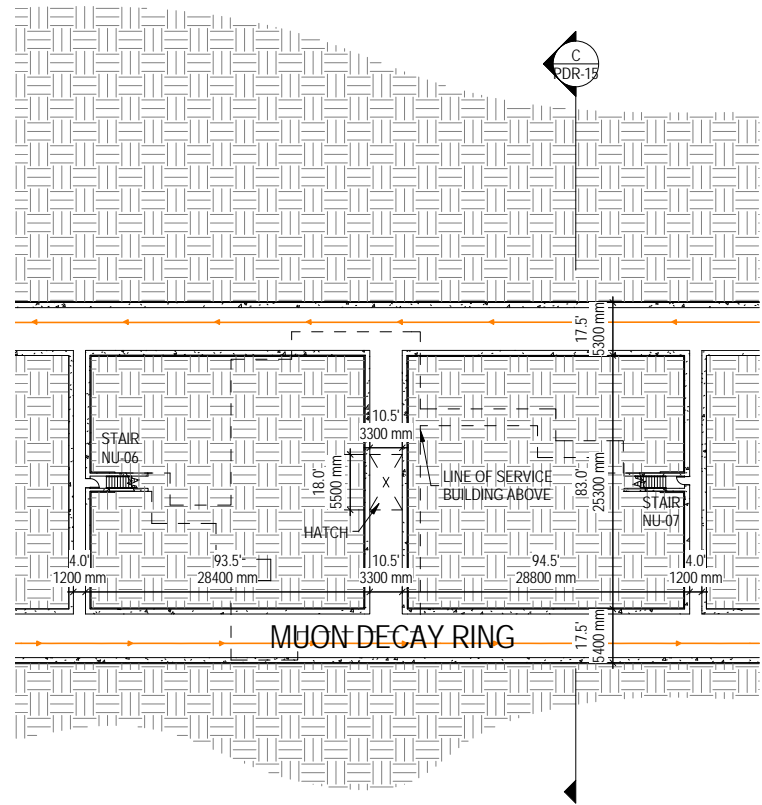

CDR-18

STAIR
NU-06

4.0'
1200 mm

93.5'
28400 mm

HATCH

10.5'
3300 mm

10.5'
3300 mm

18.0'
5500 mm

LINE OF SERVICE
BUILDING ABOVE

94.5'
28800 mm

83.0'
25300 mm

17.5'
5300 mm

STAIR
NU-07

4.0'
1200 mm

MUON DECAY RING

17.5'
5400 mm

FLOOR PLAN ELEV. 708.00'
SCALE:   1" = 30'-0"

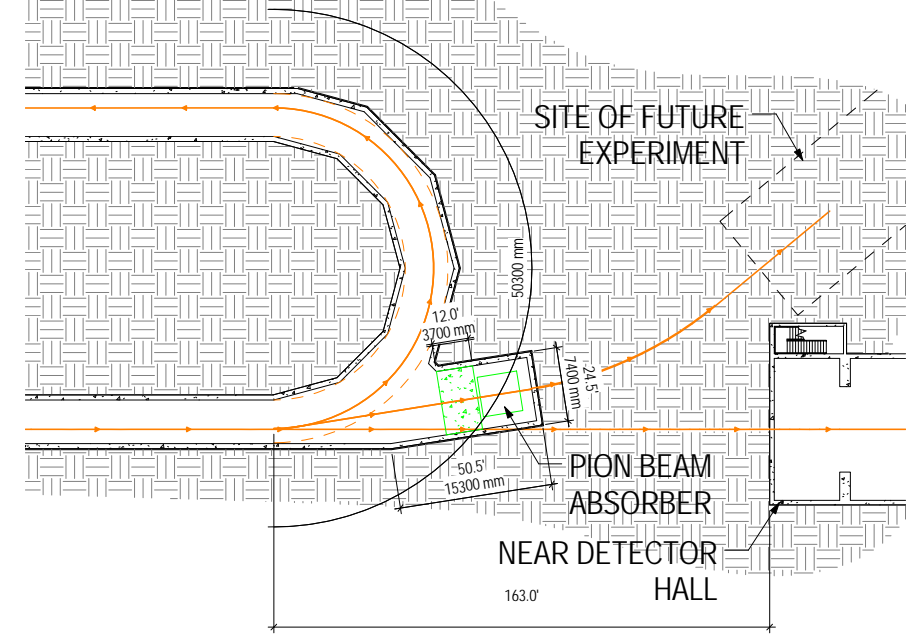

SITE OF FUTURE
EXPERIMENT

50.0'
15300 mm

12.0'
3700 mm

29.5'
9000 mm

21.0'
6400 mm

50.5'
15300 mm

163.0'

PION BEAM
ABSORBER

NEAR DETECTOR
HALL

SCALE:

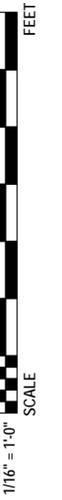

1/16" = 1'-0"        SCALE

PROJECT
NORTH

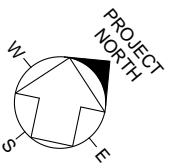

W    N    E
S

MUON DECAY RING PLAN

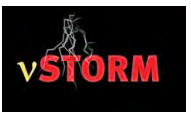
νSTORM

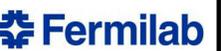
Fermilab

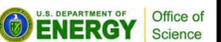
ENERGY  Office of Science

DATE
08 MAY 2013

PROJECT NO.
6-13-1

DRAWING NO.
PDR-14

5/29/2013 11:27:16 AM    C:\Users\Matt_B\Documents\1523818-ARCH-NEAR_Matt_B.rvt

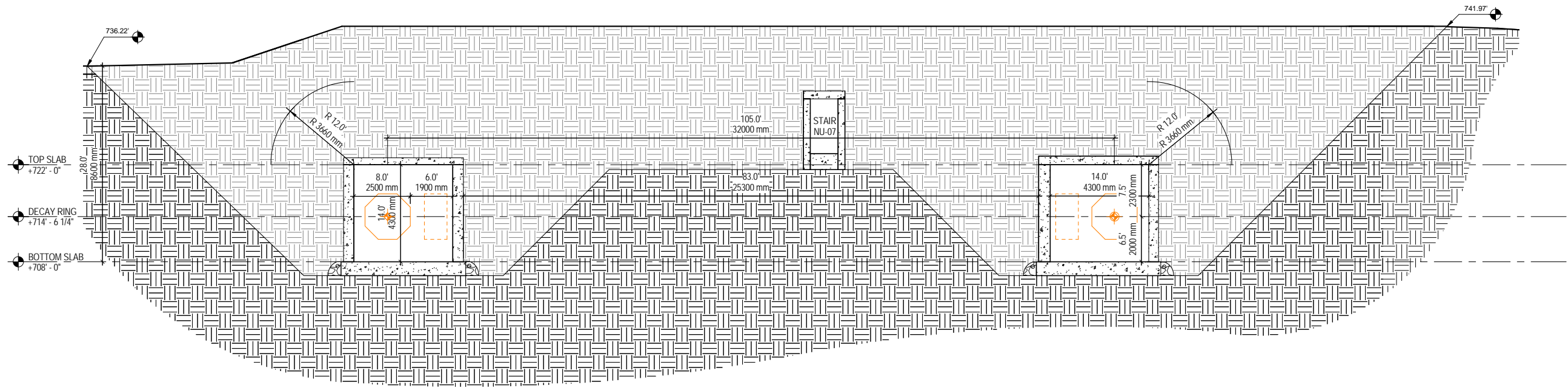

SECTION   ⟨ C ⟩

SCALE:   1/8" = 1'-0"

Section labels:

TOP SLAB
+722' - 0"

DECAY RING
+714' - 6 1/4"

BOTTOM SLAB
+708' - 0"

736.22

741.97

STAIR
NU-07

R 12.0'
8' 3660 mm

R 12.0'
8' 3660 mm

8.0'
2500 mm

6.0'
1900 mm

14.0'
4300 mm

16.0'
4300 mm

105.0'
32000 mm

83.0'
25300 mm

6.5'
2000 mm

7.3'
2200 mm

28.0'
8600 mm

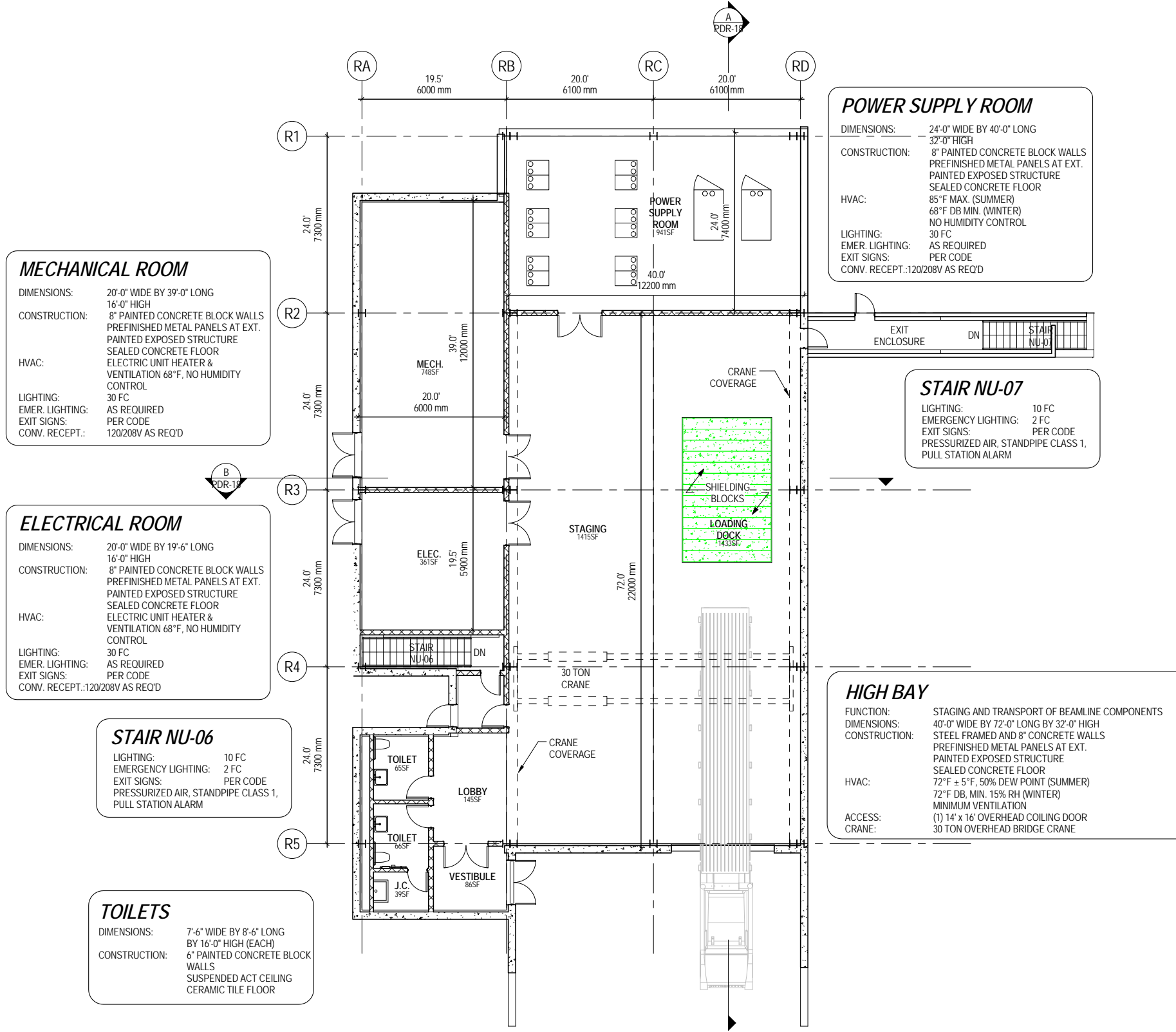

## POWER SUPPLY ROOM

| | |
|---|---|
| DIMENSIONS: | 24'-0" WIDE BY 40'-0" LONG |
| | 32'-0" HIGH |
| CONSTRUCTION: | 8" PAINTED CONCRETE BLOCK WALLS |
| | PREFINISHED METAL PANELS AT EXT. |
| | PAINTED EXPOSED STRUCTURE |
| | SEALED CONCRETE FLOOR |
| HVAC: | 85°F MAX (SUMMER) |
| | 68°F DB MIN. (WINTER) |
| | NO HUMIDITY CONTROL |
| LIGHTING: | 30 FC |
| EMER. LIGHTING: | AS REQUIRED |
| EXIT SIGNS: | PER CODE |
| CONV. RECEPT.: | 120/208V AS REQ'D |

## MECHANICAL ROOM

| | |
|---|---|
| DIMENSIONS: | 20'-0" WIDE BY 39'-0" LONG |
| | 16'-0" HIGH |
| CONSTRUCTION: | 8" PAINTED CONCRETE BLOCK WALLS |
| | PREFINISHED METAL PANELS AT EXT. |
| | PAINTED EXPOSED STRUCTURE |
| | SEALED CONCRETE FLOOR |
| HVAC: | ELECTRIC UNIT HEATER & |
| | VENTILATION 68°F, NO HUMIDITY |
| | CONTROL |
| LIGHTING: | 30 FC |
| EMER. LIGHTING: | AS REQUIRED |
| EXIT SIGNS: | PER CODE |
| CONV. RECEPT.: | 120/208V AS REQ'D |

## STAIR NU-07

| | |
|---|---|
| LIGHTING: | 10 FC |
| EMERGENCY LIGHTING: | 2 FC |
| EXIT SIGNS: | PER CODE |
| PRESSURIZED AIR, STANDPIPE CLASS 1, |
| PULL STATION ALARM |

## ELECTRICAL ROOM

| | |
|---|---|
| DIMENSIONS: | 20'-0" WIDE BY 19'-6" LONG |
| | 16'-0" HIGH |
| CONSTRUCTION: | 8" PAINTED CONCRETE BLOCK WALLS |
| | PREFINISHED METAL PANELS AT EXT. |
| | PAINTED EXPOSED STRUCTURE |
| | SEALED CONCRETE FLOOR |
| HVAC: | ELECTRIC UNIT HEATER & |
| | VENTILATION 68°F, NO HUMIDITY |
| | CONTROL |
| LIGHTING: | 30 FC |
| EMER. LIGHTING: | AS REQUIRED |
| EXIT SIGNS: | PER CODE |
| CONV. RECEPT.: | 120/208V AS REQ'D |

## STAIR NU-06

| | |
|---|---|
| LIGHTING: | 10 FC |
| EMERGENCY LIGHTING: | 2 FC |
| EXIT SIGNS: | PER CODE |
| PRESSURIZED AIR, STANDPIPE CLASS 1, |
| PULL STATION ALARM |

## HIGH BAY

| | |
|---|---|
| FUNCTION: | STAGING AND TRANSPORT OF BEAMLINE COMPONENTS |
| DIMENSIONS: | 40'-0" WIDE BY 72'-0" LONG BY 32'-0" HIGH |
| CONSTRUCTION: | STEEL FRAMED AND 8" CONCRETE WALLS |
| | PREFINISHED METAL PANELS AT EXT. |
| | PAINTED EXPOSED STRUCTURE |
| | SEALED CONCRETE FLOOR |
| HVAC: | 72°F ± 5°F, 50% DEW POINT (SUMMER) |
| | 72°F DB, MIN. 15% RH (WINTER) |
| | MINIMUM VENTILATION |
| ACCESS: | (1) 14' x 16' OVERHEAD COILING DOOR |
| CRANE: | 30 TON OVERHEAD BRIDGE CRANE |

## TOILETS

| | |
|---|---|
| DIMENSIONS: | 7'-6" WIDE BY 8'-6" LONG |
| | BY 16'-0" HIGH (EACH) |
| CONSTRUCTION: | 6" PAINTED CONCRETE BLOCK |
| | WALLS |
| | SUSPENDED ACT CEILING |
| | CERAMIC TILE FLOOR |

FLOOR PLAN ELEV. 737.50'
SCALE: 1/8" = 1'-0"

C:\Users\Matt_B\Documents\523818-ARCH-NEAR_Matt_B.rvt
5/29/2013 11:27:19 AM

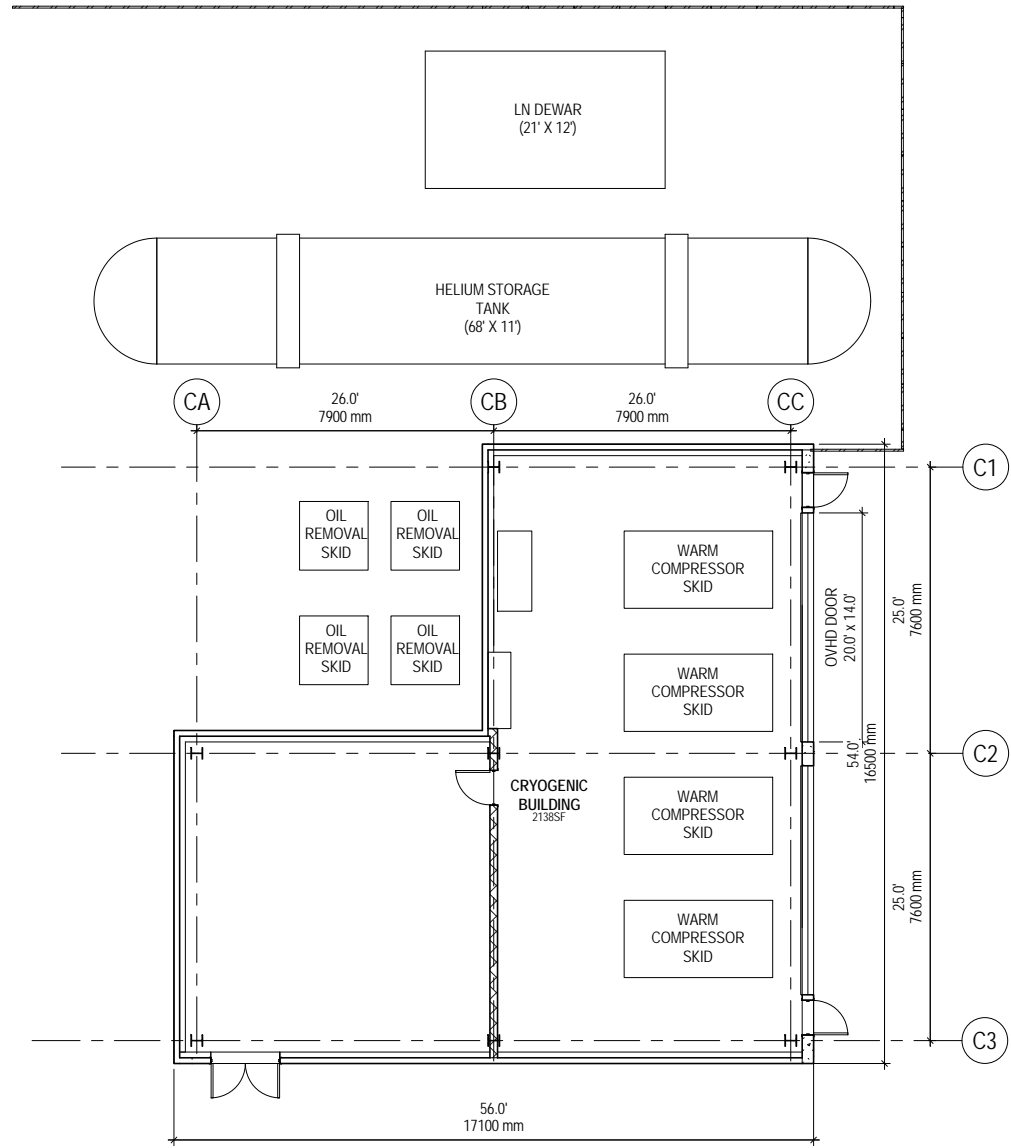

**FLOOR PLAN ELEV. 737.50'**
SCALE: 1/8" = 1'-0"

LN DEWAR
(21' X 12')

HELIUM STORAGE TANK
(68' X 11')

CA — 26.0' / 7900 mm — CB — 26.0' / 7900 mm — CC

C1
C2
C3

OIL REMOVAL SKID
OIL REMOVAL SKID
OIL REMOVAL SKID
OIL REMOVAL SKID

WARM COMPRESSOR SKID
WARM COMPRESSOR SKID
WARM COMPRESSOR SKID
WARM COMPRESSOR SKID

CRYOGENIC BUILDING
2138SF

OVHD DOOR
20.0' X 14.0'

25.0' / 7600 mm
54.0' / 16500 mm
25.0' / 7600 mm

56.0' / 17100 mm

### CRYOGENIC BUILDING

FUNCTION: HOUSES CRYO EQUIPMENT TO PROVIDE SEPARATION FROM SENSITIVE ELECTRONIC EQUIPMENT.
DIMENSIONS: 54'-0" WIDE BY 56'-0" LONG
CONSTRUCTION: METAL PANELS ON BRACED STEEL FRAME: BUILT UP ROOF ON INSULATION AND METAL DECK
ACCESS: (2) 20' x 14' OVERHEAD COILING DOORS FOR EQUIPMENT ACCESS
HANDLING: FORKLIFT
HVAC: VENTILATION AND HEATING ONLY MIN. 68°F, NO MIN. RH
PROCESS WATER: ICW FOR WARM COMPRESSOR
PURGE (ODH) VENT.: REQUIRED
EGRESS: 400 FT MAXIMUM TRAVEL DISTANCE TO STAIRWAY. STAIRWAY FIRE RATED CONSTRUCTION DISCHARGE TO SURFACE. EMERGENCY LIGHTING AND EXIT SIGNAGE THROUGHOUT. MANUAL PULLS STATIONS AT EXIT DOORS
ERRKWCP: O2 MONITORING SENSORS THROUGHOUT, EXHAUST PURGE FANS AND ADMINISTRATION CONTROL ACCESS PROVIDED.
SMOKE CONTROL: EMERGENCY PERSONAL MANUAL ACCESS FANS FOR SMOKE ABATEMENT.
FIRE DETECTION: AIR SAMPLING SMOKE & LINEAR TYPE HEAT DETECTION.
FIRE NOTIFICATION: AUDIBLE & VISUAL DEVICES THROUGHOUT.
FIRE SUPPRESSION: PREACTION (DRY PIPE) AUTOMATIC SPRINKLER SYSTEM (DUAL INTERLOCK)

SCALE:

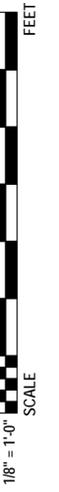

1/8" = 1'-0"

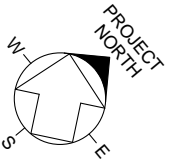

PROJECT NORTH

**CRYOGENIC BUILDING PLAN**

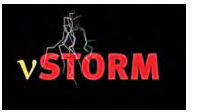

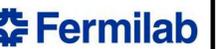

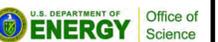

DATE
**08 MAY 2013**

PROJECT NO.
**6-13-1**

DRAWING NO.
**PDR-17**

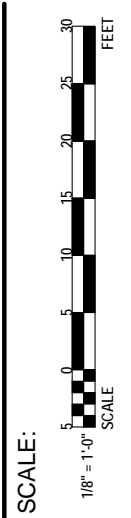

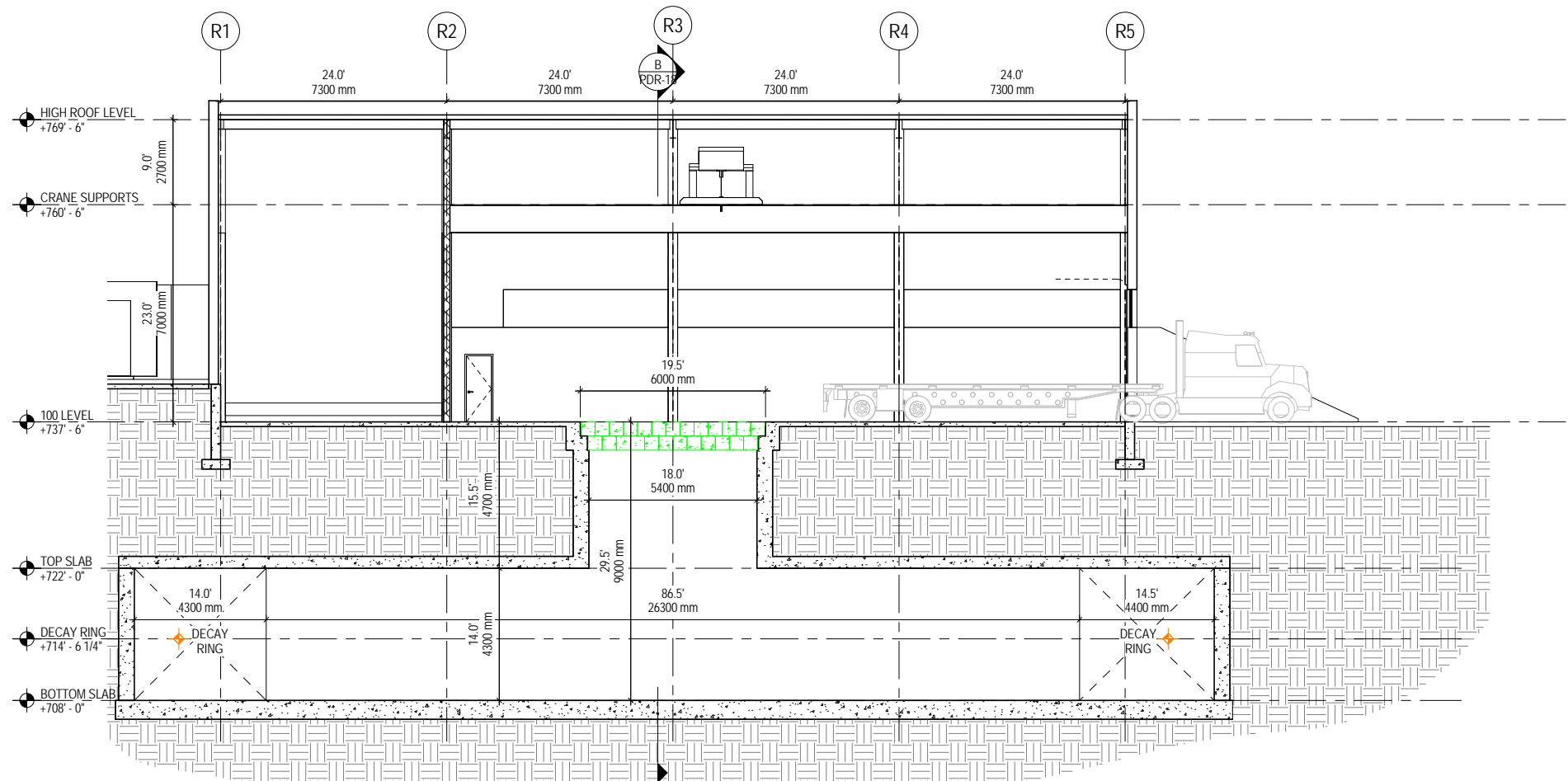

**LONGITUDINAL SECTION** ⟨A⟩
SCALE: 1/8" = 1'-0"

Grid lines: R1 · R2 · R3 · R4 · R5

PDR-18

24.0' 7300 mm · 24.0' 7300 mm · 24.0' 7300 mm · 24.0' 7300 mm

- HIGH ROOF LEVEL +769' - 6"
- 9.0' 2700 mm
- CRANE SUPPORTS +760' - 6"
- 23.0' 7000 mm
- 100 LEVEL +737' - 6"
- TOP SLAB +722' - 0"
- DECAY RING +714' - 6 1/4"
- BOTTOM SLAB +708' - 0"

15.5' 4700 mm · 18.0' 5400 mm · 19.5' 6000 mm
29.5' 9000 mm
14.0' 4300 mm · 14.0' 4300 mm · 14.5' 4400 mm
86.5' 26300 mm

DECAY RING · DECAY RING

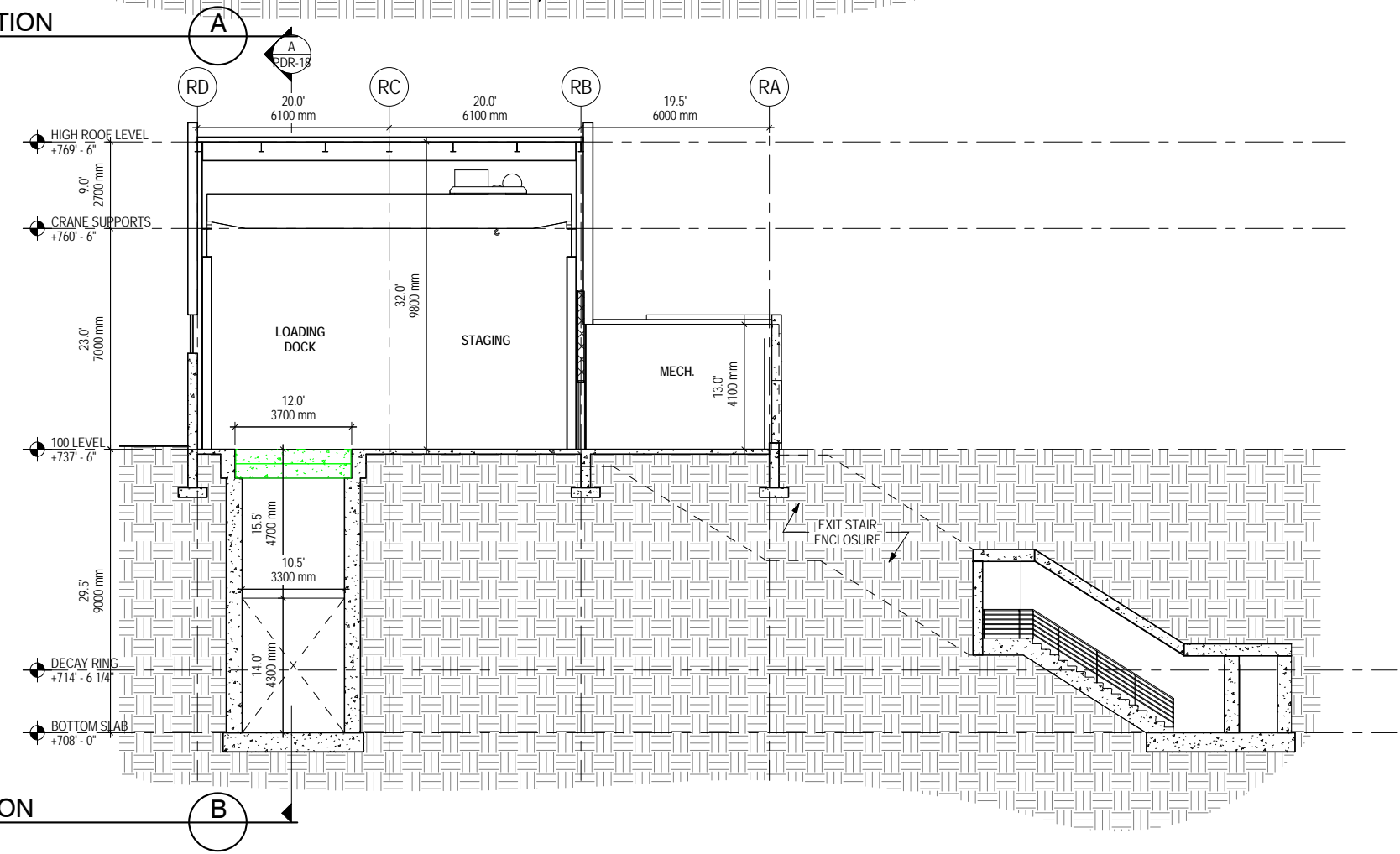

**TRANSVERSE SECTION** ⟨B⟩
SCALE: 1/8" = 1'-0"

Grid lines: RD · RC · RB · RA

PDR-18

20.0' 6100 mm · 20.0' 6100 mm · 19.5' 6000 mm

- HIGH ROOF LEVEL +769' - 6"
- 9.0' 2700 mm
- CRANE SUPPORTS +760' - 6"
- 23.0' 7000 mm
- 32.0' 9800 mm
- 100 LEVEL +737' - 6"
- 29.5' 9000 mm
- DECAY RING +714' - 6 1/4"
- BOTTOM SLAB +708' - 0"

LOADING DOCK · STAGING · MECH.

12.0' 3700 mm · 13.0' 4100 mm
15.5' 4700 mm
10.5' 3300 mm
14.0' 4300 mm

EXIT STAIR ENCLOSURE

C:\Users\Matt_B\Documents\1523818-ARCH-NEAR_Matt_B.rvt

5/29/2013 11:27:20 AM

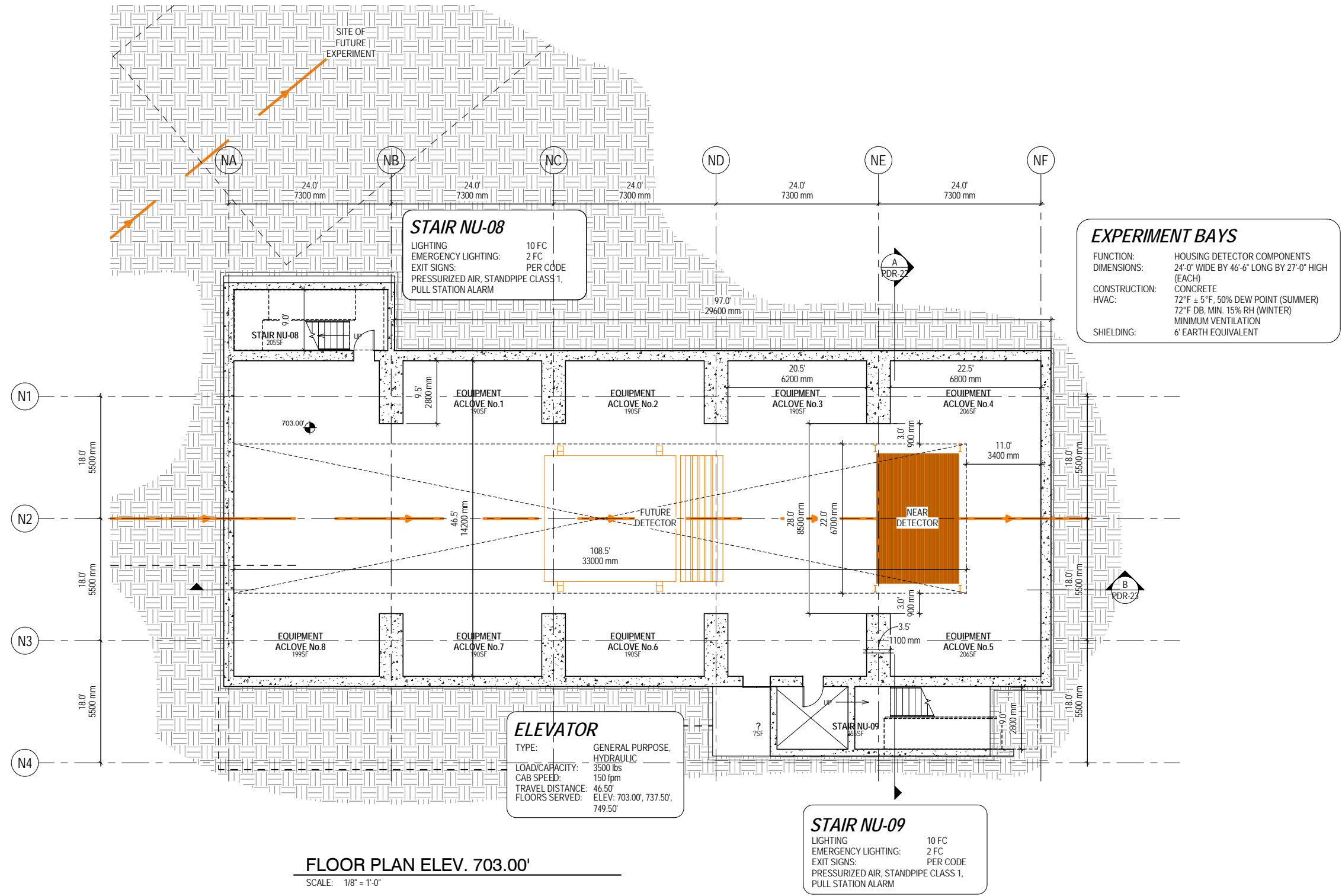

SITE OF FUTURE EXPERIMENT

NA  NB  NC  ND  NE  NF

24.0' / 7300 mm    24.0' / 7300 mm    24.0' / 7300 mm    24.0' / 7300 mm    24.0' / 7300 mm

### STAIR NU-08

LIGHTING 10 FC
EMERGENCY LIGHTING: 2 FC
EXIT SIGNS: PER CODE
PRESSURIZED AIR, STANDPIPE CLASS 1,
PULL STATION ALARM

### EXPERIMENT BAYS

FUNCTION: HOUSING DETECTOR COMPONENTS
DIMENSIONS: 24'-0" WIDE BY 46'-6" LONG BY 27'-0" HIGH (EACH)
CONSTRUCTION: CONCRETE
HVAC: 72°F ± 5°F, 50% DEW POINT (SUMMER)
72°F DB, MIN. 15% RH (WINTER)
MINIMUM VENTILATION
SHIELDING: 6' EARTH EQUIVALENT

A
PDR-22

97.0' / 29600 mm

STAIR NU-08
TYP.

N1   18.0' / 5500 mm

EQUIPMENT ALCOVE No.1

EQUIPMENT ALCOVE No.2

20.5' / 6200 mm
EQUIPMENT ALCOVE No.3

22.5' / 6800 mm
EQUIPMENT ALCOVE No.4

703.00'

11.0' / 3400 mm

N2   18.0' / 5500 mm

FUTURE DETECTOR

NEAR DETECTOR

108.5' / 33000 mm

22.0' / 6700 mm

B
PDR-22

N3   18.0' / 5500 mm

EQUIPMENT ALCOVE No.8

EQUIPMENT ALCOVE No.7

EQUIPMENT ALCOVE No.6

EQUIPMENT ALCOVE No.5

3.5' / 1100 mm

### ELEVATOR

TYPE: GENERAL PURPOSE, HYDRAULIC
LOAD/CAPACITY: 3500 lbs
CAB SPEED: 150 fpm
TRAVEL DISTANCE: 46.50'
FLOORS SERVED: ELEV. 703.00', 737.50', 749.50'

STAIR NU-09

N4   18.0' / 5500 mm

### STAIR NU-09

LIGHTING 10 FC
EMERGENCY LIGHTING: 2 FC
EXIT SIGNS: PER CODE
PRESSURIZED AIR, STANDPIPE CLASS 1,
PULL STATION ALARM

### FLOOR PLAN ELEV. 703.00'

SCALE: 1/8" = 1'-0"

C:\Users\Matt_B\Documents\S238\B-ARCH-NEAR_Matt_B.rvt

5/29/2013 11:27:21 AM

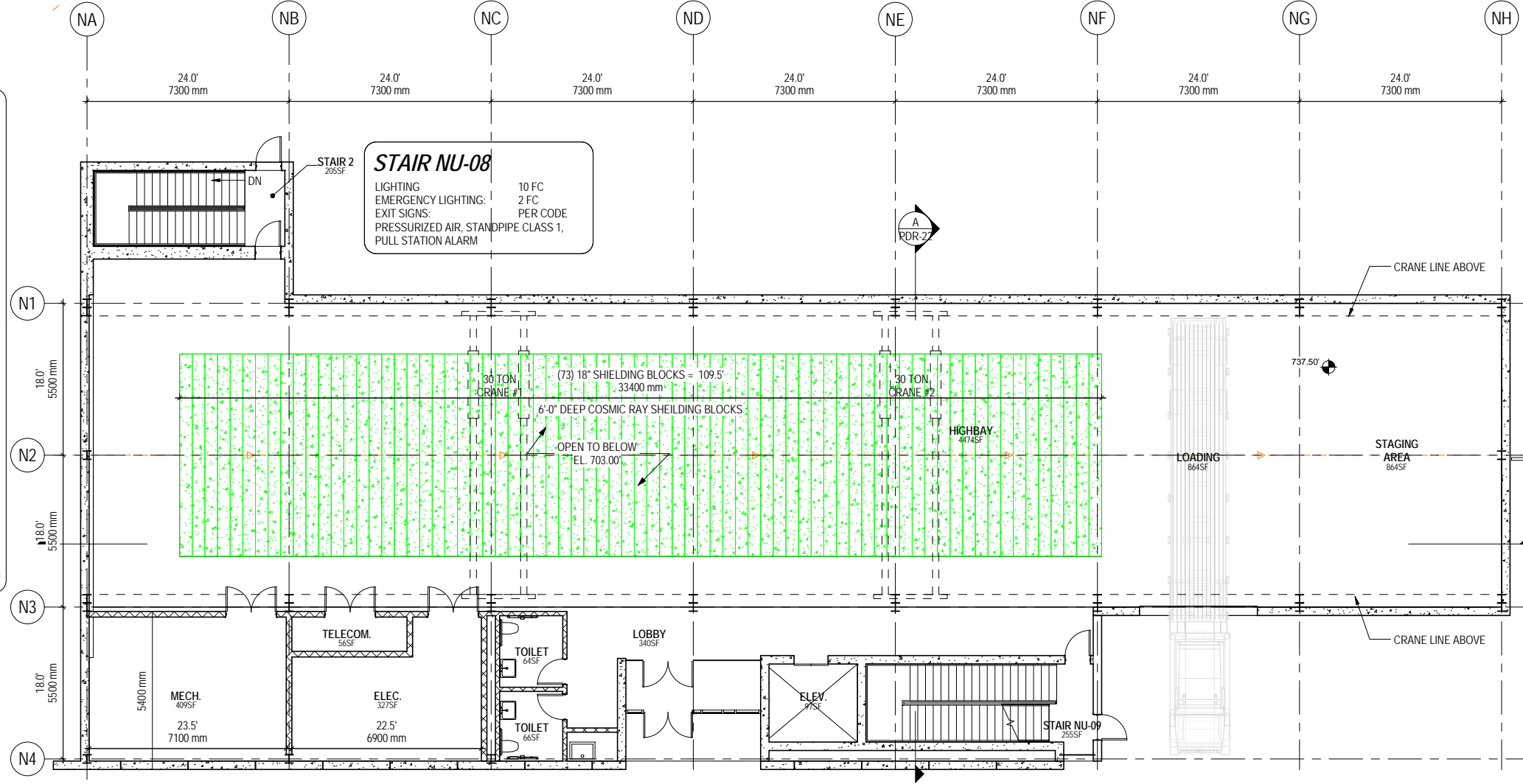

## HIGH BAY

| | |
|---|---|
| FUNCTION: | PROVIDE INSTALLATION AND ASSEMBLY OF nuSTORM NEAR DETECTOR AND PROVIDE SPACE FOR FUTURE DETECTORS |
| DIMENSIONS: | 317'-0" WIDE BY 169'-0" LONG BY 29'-6" HIGH |
| CONSTRUCTION: | CONCRETE |
| SHIELDING: | BOTTOM & SIDES: 1'-6" CONCRETE W/ 4'-6" EARTH TOP: 6'-0" CONC. (2 LAYERS OF 3'-0" CONC. SHIELDING BLOCKS) |
| EQUIPMENT ACCESS: | (3) 14' X 16' OVERHEAD ROLLUP DOOR FOR SEMI TRUCK ACCESS |
| CRANE: | (2) 30 TON OVERHEAD BRIDGE CRANES |
| HVAC: | SUMMER: 72°F (+/-5°F), 50°F DEW POINT WINTER: 72°F, MIN. 15% RH VENTILATION: TBD |
| EGRESS: | 400 FOOT MAXIMUM TRAVEL DISTANCE TO STAIRWAY. STAIRWAY TO BE OF FIRE RATED CONSTRUCTION. DISCHARGE TO SURFACE. EMERGENCY LIGHTING AND EXIT SIGNAGE THROUGHOUT MANUAL PULL STATIONS AT EXIT DOORS. |
| SMOKE CONTROL: | EMERGENCY PERSONAL MANUAL ACCESS FANS FOR SMOKE ABATEMENT |
| FIRE DETECTION: | AIR SAMPLING SMOKE & LINEAR TYPE HEAT DETECTION |
| FIRE NOTIFICATION: | AUDIBLE AND VISUAL DEVICES THROUGHOUT |
| FIRE SUPPRESSION: | PREACTION (DRY PIPE) AUTOMATIC SPRINKLER SYSTEM (DUAL INTERLOCK) |

## MECHANICAL ROOM

| | |
|---|---|
| DIMENSIONS: | 24'-0" WIDE BY 17'-6" LONG 16'-0" HIGH |
| CONSTRUCTION: | 8" PAINTED CONCRETE BLOCK WALLS PREFINISHED METAL PANELS AT EXT. PAINTED EXPOSED STRUCTURE SEALED CONCRETE FLOOR |
| HVAC: | ELECTRIC UNIT HEATER & VENTILATION 68°F, NO HUMIDITY CONTROL |
| LIGHTING: | 30 FC |
| EMER. LIGHTING: | AS REQUIRED |
| EXIT SIGNS: | PER CODE |
| CONV. RECEPT.: | 120/208V AS REQ'D |

## ELECTRICAL ROOM

| | |
|---|---|
| DIMENSIONS: | 22'-6" WIDE BY 17'-6" LONG 16'-0" HIGH |
| CONSTRUCTION: | 8" PAINTED CONCRETE BLOCK WALLS PREFINISHED METAL PANELS AT EXT. PAINTED EXPOSED STRUCTURE SEALED CONCRETE FLOOR |
| HVAC: | ELECTRIC UNIT HEATER & VENTILATION 68°F, NO HUMIDITY CONTROL |
| LIGHTING: | 30 FC |
| EMER. LIGHTING: | AS REQUIRED |
| EXIT SIGNS: | PER CODE |
| CONV. RECEPT.: | 120/208V AS REQ'D |

## TOILETS

| | |
|---|---|
| DIMENSIONS: | 7'-6" WIDE BY 8'-6" LONG BY 16'-0" HIGH (EACH) |
| CONSTRUCTION: | 6" PAINTED CONCRETE BLOCK WALLS SUSPENDED ACT CEILING CERAMIC TILE FLOOR |

## ELEVATOR

| | |
|---|---|
| TYPE: | GENERAL PURPOSE, HYDRAULIC |
| LOAD/CAPACITY: | 3500 lbs |
| CAB SPEED: | 150 fpm |
| TRAVEL DISTANCE: | 46.50' |
| FLOORS SERVED: | ELEV. 703.00', 737.50', 749.50' |

## STAIR NU-09

| | |
|---|---|
| LIGHTING | 10 FC |
| EMERGENCY LIGHTING: | 2 FC |
| EXIT SIGNS: | PER CODE |
| PRESSURIZED AIR, STANDPIPE CLASS 1, PULL STATION ALARM | |

**FLOOR PLAN ELEV. 737.50'**
SCALE: 1/8" = 1'-0"

STAIR 2
STAIR NU-08
LIGHTING     10 FC
EMERGENCY LIGHTING:   2 FC
EXIT SIGNS:   PER CODE
PRESSURIZED AIR, STANDPIPE CLASS 1,
PULL STATION ALARM

30 TON CRANE #1
(73) 18" SHIELDING BLOCKS = 109.5'  33400 mm
6'-0" DEEP COSMIC RAY SHIELDING BLOCKS
OPEN TO BELOW EL. 703.00'
30 TON CRANE #2
HIGHBAY
LOADING
STAGING AREA

CRANE LINE ABOVE

TELECOM.
MECH.
ELEC.
TOILET
LOBBY
TOILET
ELEV
STAIR NU-09

NA  NB  NC  ND  NE  NF  NG  NH

24.0' 7300 mm (×7)

N1  N2  N3  N4

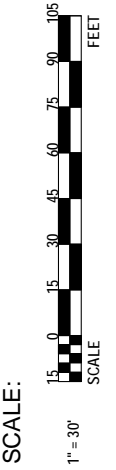

SCALE:
1" = 30'
FEET

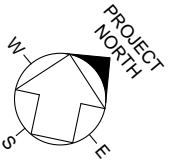

PROJECT NORTH

NEAR DETECTOR HALL FLOOR PLAN
SHEET 2

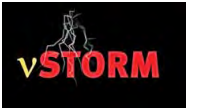
vSTORM

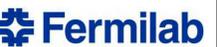
Fermilab

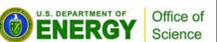
U.S. DEPARTMENT OF ENERGY   Office of Science

DATE
**08 MAY 2013**

PROJECT NO.
**6-13-1**

DRAWING NO.
**PDR-20**

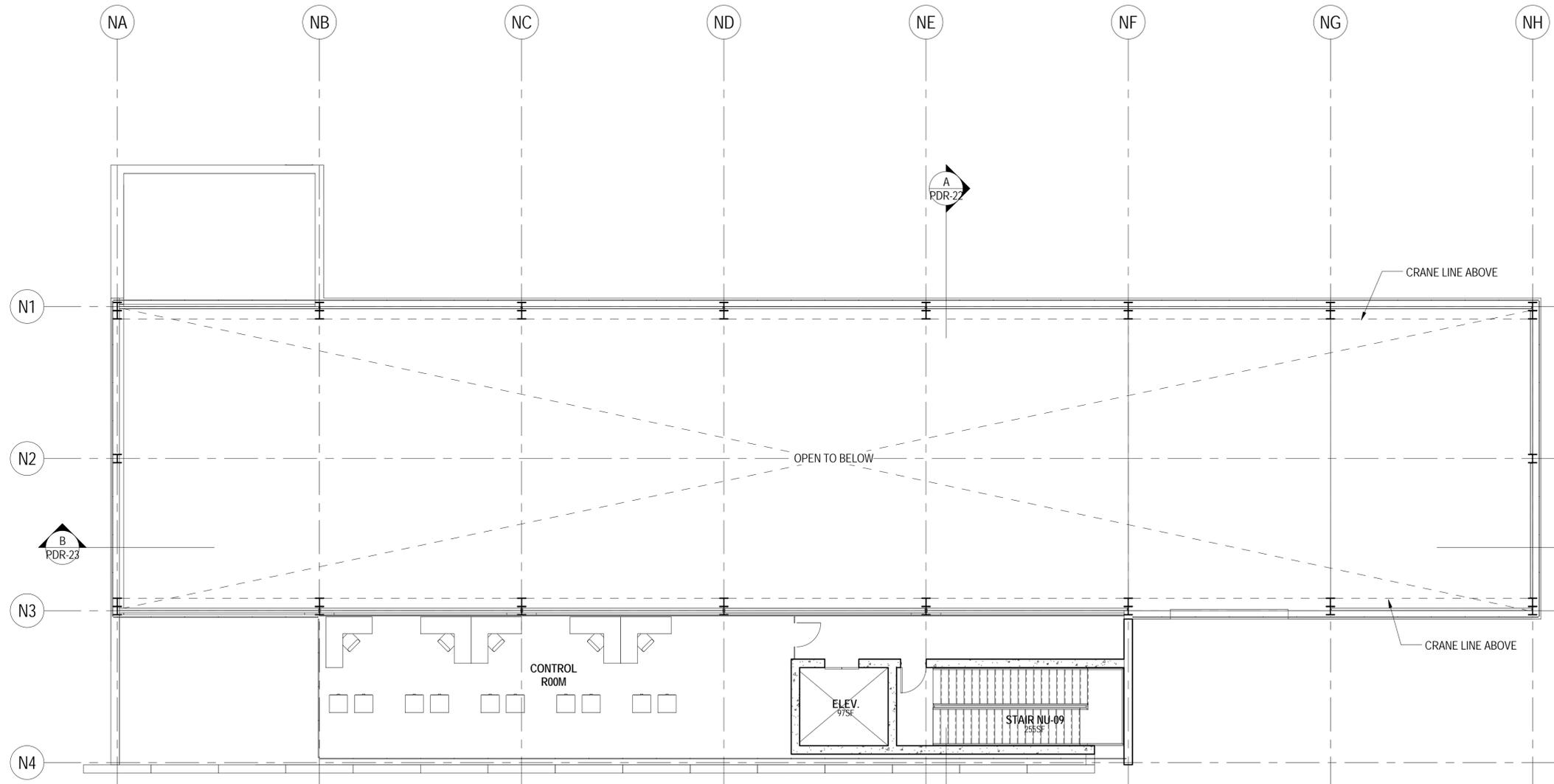

**CONTROL ROOM**

| | |
|---|---|
| HVAC: | 85°F AT 55% RH MAX.<br>21% RH MIN. |
| PROCESS WATER: | NONE |
| ODH VENTILATION: | NONE |
| SMOKE CONTROL: | EMERGENCY PERSONAL MANUAL |
| FIRE DETECTION: | AIR SAMPLING SMOKE DETECTION |
| FIRE NOTIFICATION: | AUDIBLE AND VISUAL DEVICES<br>THROUGHOUT |
| FIRE SUPPRESSION: | PREACTION (DRY) AUTOMATIC<br>SPRINKLER SYSTEM<br>(DUAL INTERLOCK) |

**ELEVATOR**

| | |
|---|---|
| TYPE: | GENERAL PURPOSE,<br>HYDRAULIC |
| LOAD/CAPACITY: | 3500 lbs |
| CAB SPEED: | 150 fpm |
| TRAVEL DISTANCE: | 46.50' |
| FLOORS SERVED: | ELEV. 703.00', 737.50',<br>749.50' |

**STAIR NU-09**

| | |
|---|---|
| LIGHTING: | 10 FC |
| EMERGENCY LIGHTING: | 2 FC |
| EXIT SIGNS: | PER CODE |
| PRESSURIZED AIR; STANDPIPE CLASS 1,<br>PULL STATION ALARM | |

FLOOR PLAN ELEV. 749.50'
SCALE: 1/8" = 1'-0"

NEAR DETECTOR HALL FLOOR PLAN
SHEET 3

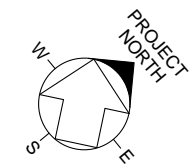

vSTORM

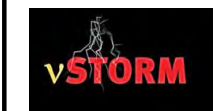
Fermilab

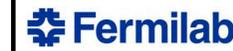
ENERGY  Office of Science

DATE
**08 MAY 2013**

PROJECT NO.
**6-13-1**

DRAWING NO.
**PDR-21**

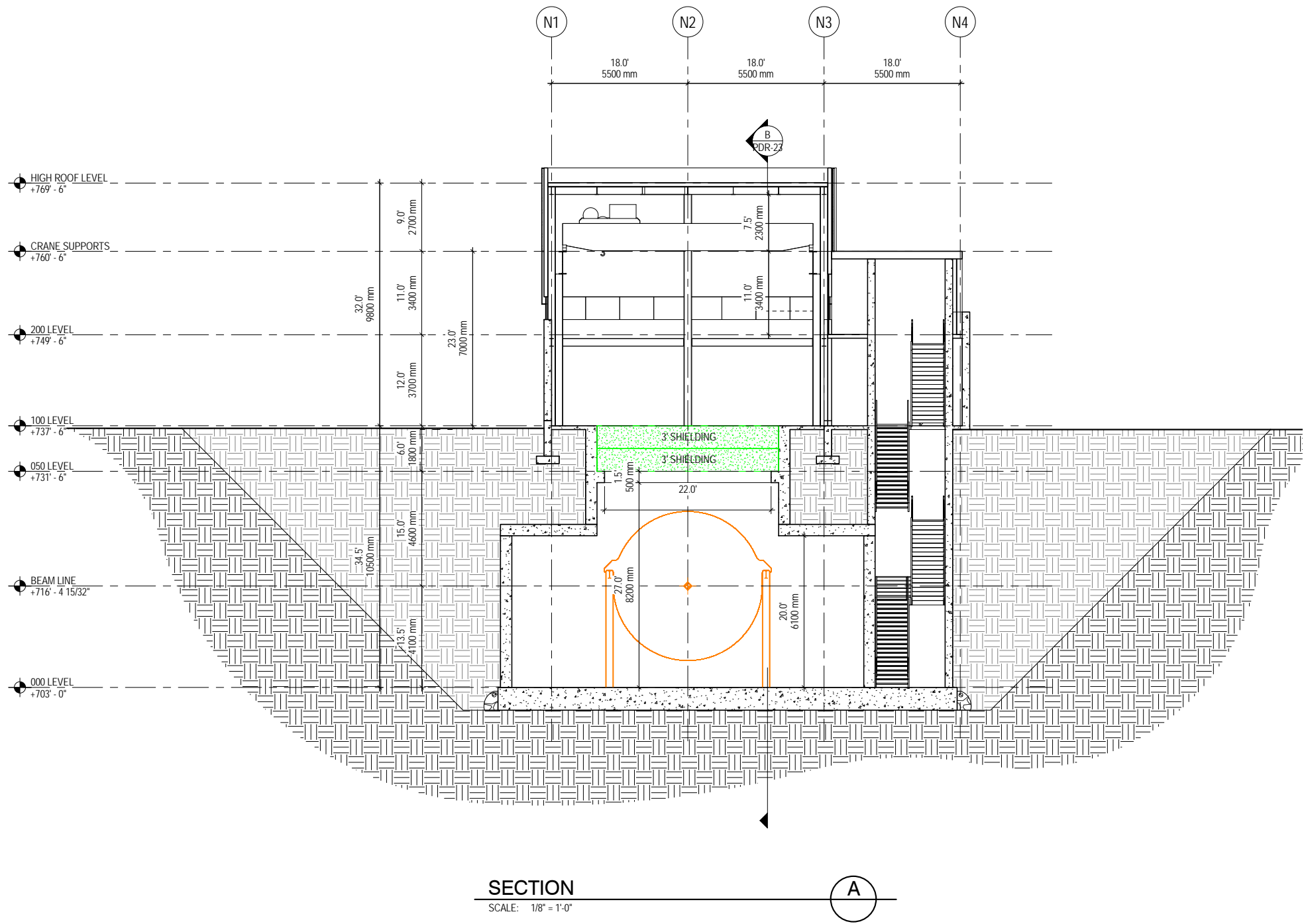

N1   N2   N3   N4

18.0'
5500 mm
18.0'
5500 mm
18.0'
5500 mm

PDR-23

HIGH ROOF LEVEL
+769' - 6"

CRANE SUPPORTS
+767' - 6"

200 LEVEL
+749' - 6"

100 LEVEL
+737' - 0"

050 LEVEL
+731' - 6"

BEAM LINE
+716' - 4 15/32"

000 LEVEL
+703' - 0"

9.0'
2700 mm

7.5'
2300 mm

11.0'
3400 mm

11.0'
3400 mm

32.0'
9800 mm

23.0'
7000 mm

12.0'
3700 mm

6.0'
1800 mm

3' SHIELDING

3' SHIELDING

1.5'
500 mm

22.0'

15.0'
4600 mm

34.5'
10500 mm

13.5'
4100 mm

22.0'
8000 mm

20.0'
6100 mm

SECTION
SCALE:  1/8" = 1'-0"

A

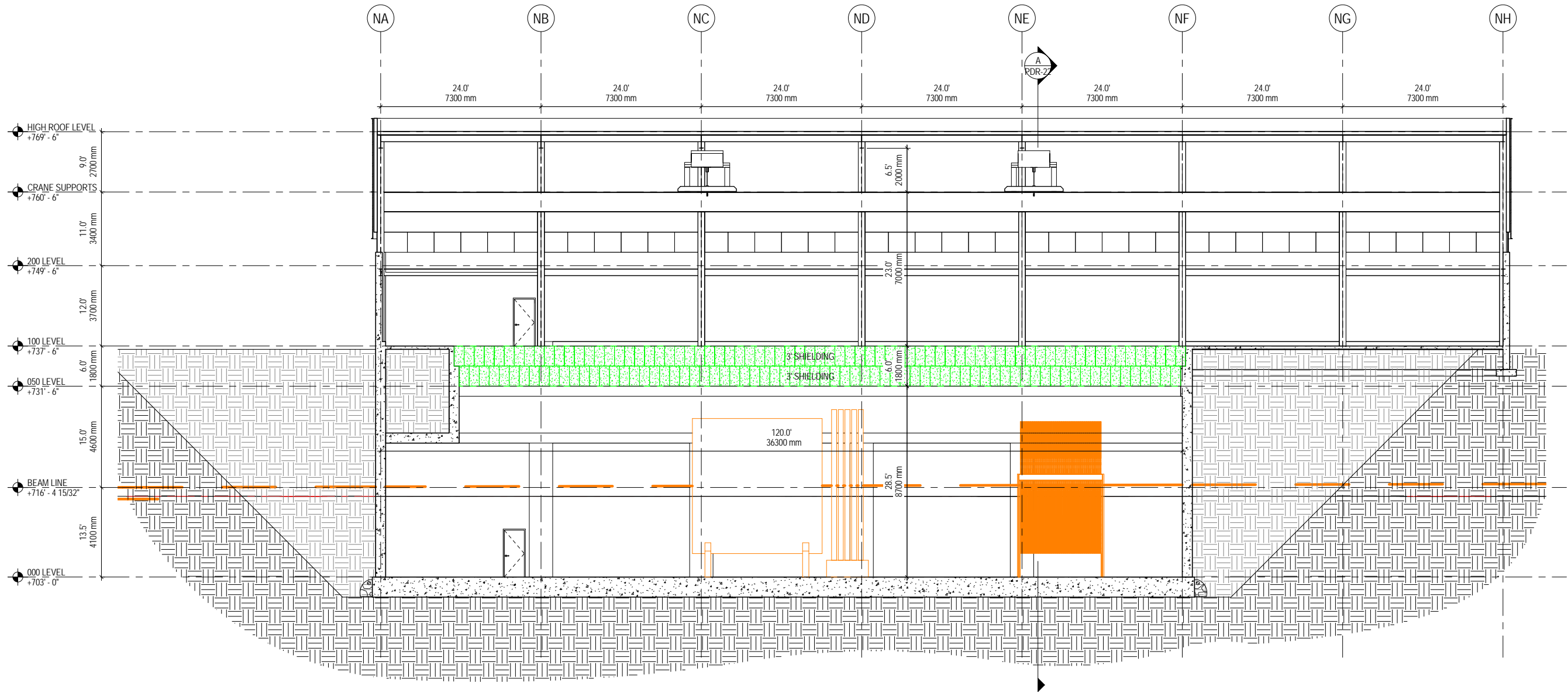

SECTION
SCALE: 1/8" = 1'-0"    (B)

SCALE:

1/8" = 1'-0"   SCALE

FEET:

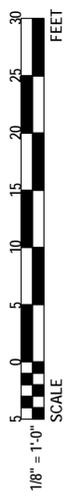

BUILDING & ENCLOSURE SECTION
SHEET 2

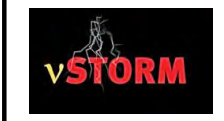
νSTORM

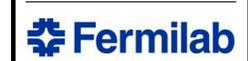
Fermilab

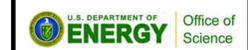
ENERGY    Office of Science

DATE
**08 MAY 2013**

PROJECT NO.
**6-13-1**

DRAWING NO.
**PDR-23**

HIGH ROOF LEVEL
+769' - 6"

CRANE SUPPORTS
+760' - 6"

200 LEVEL
+749' - 6"

100 LEVEL
+737' - 6"

050 LEVEL
+731' - 6"

BEAM LINE
+716' - 4 15/32"

000 LEVEL
+703' - 0"

9'0"
2700 mm

11'0"
3400 mm

12'0"
3700 mm

6'0"
1800 mm

15'0"
4600 mm

13'5"
4100 mm

3' SHIELDING
3' SHIELDING

120.0'
36300 mm

24.0'
7300 mm  (×7)

6.5'
2000 mm

20.0'
7900 mm

28.5'
8700 mm

NA  NB  NC  ND  NE  NF  NG  NH

A
PDR-22

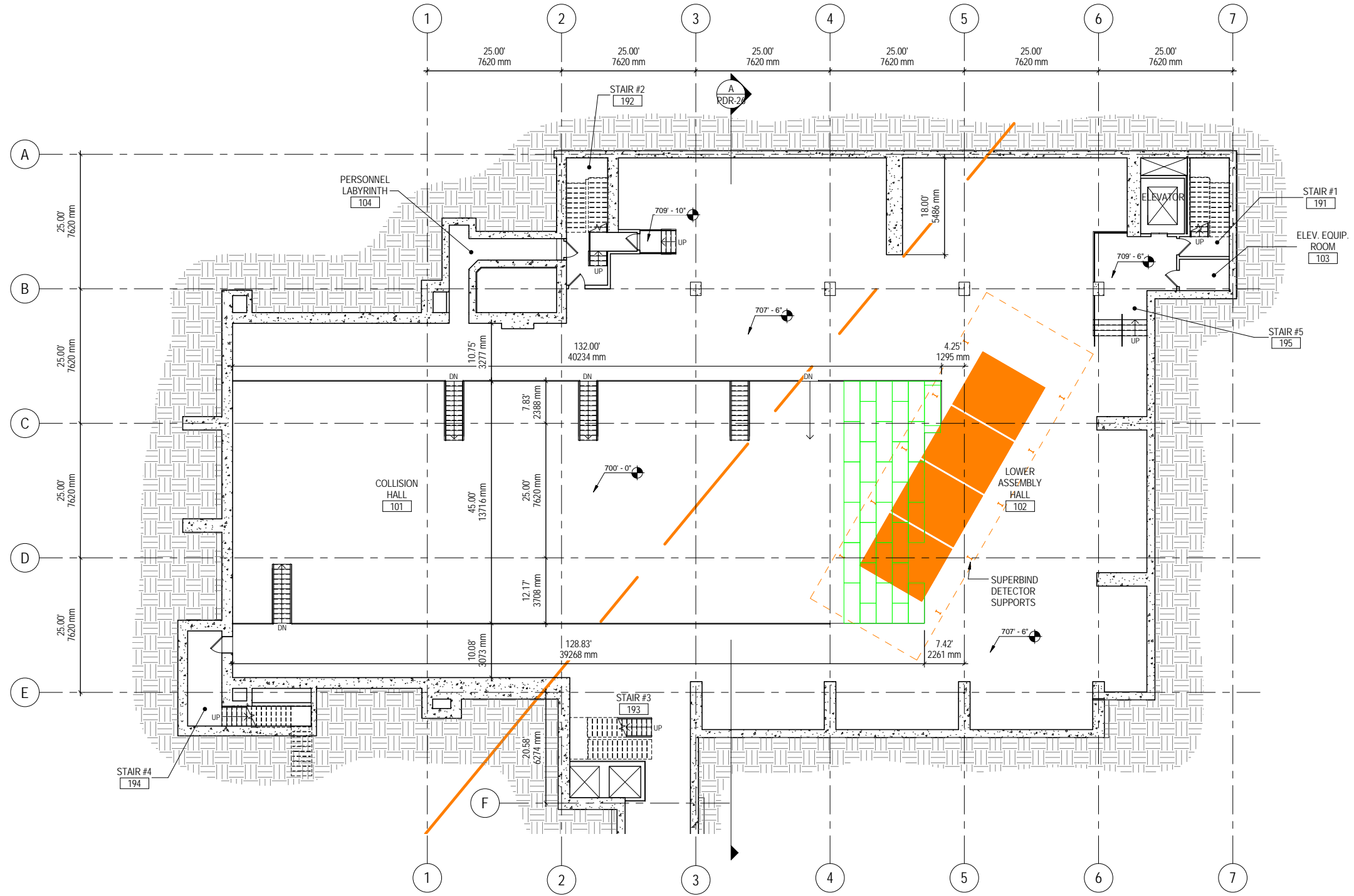

FLOOR PLAN ELEV. 700.00' & 707.50'
SCALE: 3/32" = 1'-0"

COLLISION HALL
101

LOWER ASSEMBLY HALL
102

PERSONNEL LABYRINTH
104

SUPERBIND DETECTOR SUPPORTS

STAIR #2
192

STAIR #1
191

STAIR #3
193

STAIR #4
194

STAIR #5
195

ELEVATOR

ELEV. EQUIP. ROOM
103

25.00'
7620 mm

25.00'
7620 mm

25.00'
7620 mm

25.00'
7620 mm

25.00'
7620 mm

25.00'
7620 mm

132.00'
40234 mm

45.00'
13716 mm

25.00'
7620 mm

10.75'
3277 mm

7.83'
2388 mm

12.17'
3708 mm

10.08'
3073 mm

128.83'
39268 mm

20.58'
6274 mm

4.25'
1295 mm

7.42'
2261 mm

18.00'
5486 mm

700' - 0"

709' - 10"

709' - 6"

707' - 6"

707' - 6"

A PDR-25

FAR DETECTOR FLOOR PLAN

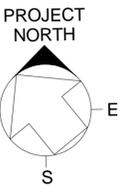

νSTORM

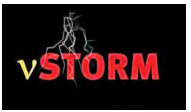
Fermilab

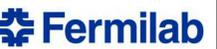
ENERGY   Office of Science

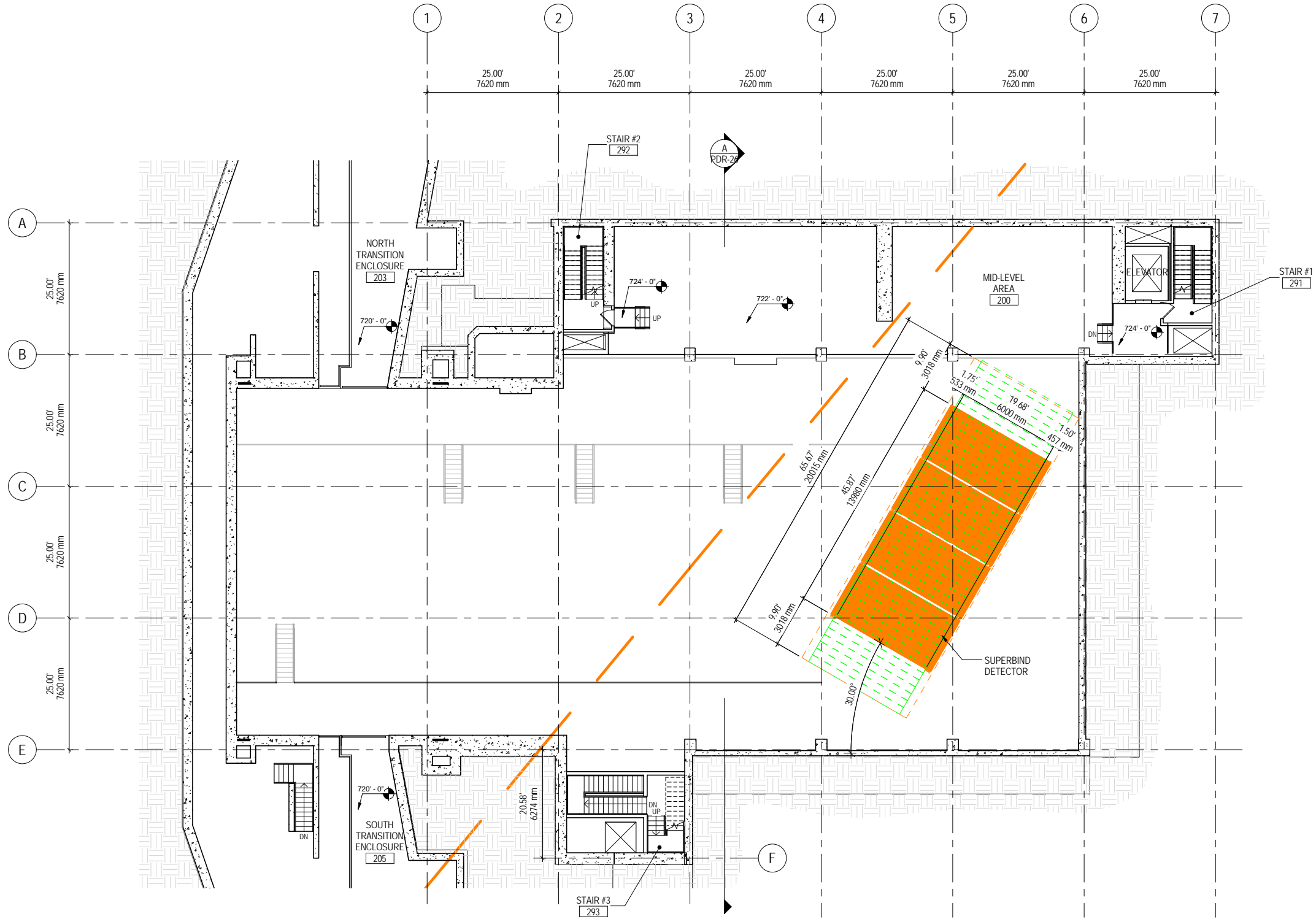

FLOOR PLAN ELEV. 722.00'

SCALE: 3/32" = 1'-0"

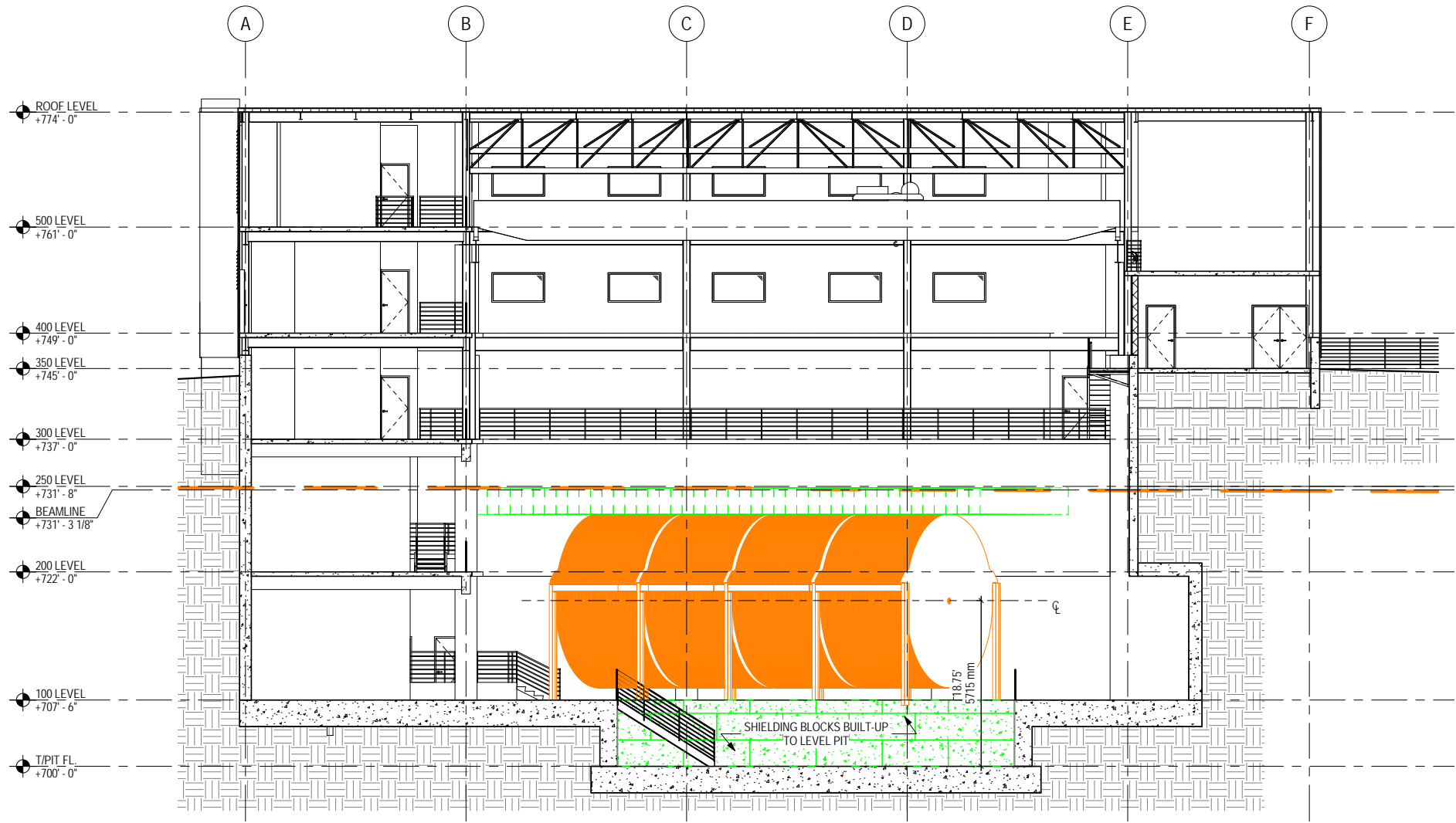

ROOF LEVEL
+774' - 0"

500 LEVEL
+761' - 0"

400 LEVEL
+749' - 0"

350 LEVEL
+745' - 0"

300 LEVEL
+737' - 0"

250 LEVEL
+731' - 8"

BEAMLINE
+731' - 3 1/8"

200 LEVEL
+722' - 0"

100 LEVEL
+707' - 6"

T/PIT FL
+700' - 0"

(A) (B) (C) (D) (E) (F)

118.75"
5715 mm

SHIELDING BLOCKS BUILT-UP
TO LEVEL PIT

TRANSVERSE SECTION    (A)
SCALE:   1/8" = 1'-0"

5/29/2013 11:45:02 AM    P:\15238.18 - Fermilab nu\STORM\01 Drawing Files\00-Revit Files\01-Project Files\01-Project Central File\15238.18-ARCH FAR-20.rvt